\documentclass{article} %
\usepackage[final]{colm2026_conference}

\usepackage{amsmath,amsfonts,bm}

\def\eqref#1{equation~\ref{#1}}

\def\1{\bm{1}}

\DeclareMathAlphabet{\mathsfit}{\encodingdefault}{\sfdefault}{m}{sl}
\SetMathAlphabet{\mathsfit}{bold}{\encodingdefault}{\sfdefault}{bx}{n}

\usepackage{microtype}
\usepackage{hyperref}
\usepackage{lineno}
\usepackage{url}

\usepackage[pdftex]{graphicx}
\usepackage{booktabs} %
\usepackage{comment} %
\usepackage{subcaption} %
\usepackage[most]{tcolorbox} %
\definecolor{darkgreen}{rgb}{0.0, 0.5, 0.0}  %
\usepackage{rotating} %

\newcommand{\twoimagegrid}[1]{%
    \begin{subfigure}[t]{0.47\textwidth}
        \includegraphics[width=\linewidth, keepaspectratio]{#1directional_feedback.pdf}
    \end{subfigure} &
    \begin{subfigure}[t]{0.47\textwidth}
        \includegraphics[width=\linewidth, keepaspectratio]{#1numerical_feedback.pdf}
    \end{subfigure}
}

\title{High Volatility and Action Bias Distinguish LLMs from \\ Humans in Group Coordination}

\author{Sahaj Singh Maini\textsuperscript{*†}, Robert L. Goldstone\textsuperscript{†‡} \& Zoran Tiganj\textsuperscript{*†‡} \\
\textsuperscript{*}Department of Computer Science, Indiana University Bloomington\\
\textsuperscript{†}Cognitive Science Program, Indiana University Bloomington\\
\textsuperscript{‡}Department of Psychological and Brain Sciences, Indiana University Bloomington \\
\texttt{\{sahmaini,rgoldsto,ztiganj\}@iu.edu} 
}

\begin{document}

\ifcolmsubmission
\linenumbers
\fi

\maketitle

\begingroup
\renewcommand{\thefootnote}{}
\footnotetext{\href{https://cogneuroai.github.io/Human-vs-LLM-Group-Coordination/}{https://cogneuroai.github.io/Human-vs-LLM-Group-Coordination}}
\endgroup

\begin{abstract}
Humans exhibit remarkable abilities to coordinate in groups. As large language models (LLMs) become more capable, it remains an open question whether they can demonstrate comparable adaptive coordination and whether they use the same strategies as humans. To better understand this, we compare LLM and human performance on a common-interest game with imperfect monitoring: Group Binary Search. In this $n$-player game, participants need to coordinate their actions to achieve a common objective. Players independently submit numerical values in an effort to collectively sum to a randomly assigned target number. Without direct communication, they rely on group feedback to iteratively adjust their submissions until they reach the target number. Our findings show that, unlike humans who adapt and stabilize their behavior over time, LLMs often fail to improve across games and exhibit excessive switching, which impairs group convergence. Moreover, richer feedback (e.g., numerical error magnitude) benefits humans substantially but has small effects on LLMs. Finally, we show that GRPO can be effective in reducing the excessive switching. Taken together, by grounding the analysis in human baselines and mechanism-level metrics, including reactivity scaling, switching dynamics, and learning across games, we point to differences in human and LLM groups and provide a behaviorally grounded diagnostic for closing the coordination gap.
\end{abstract}

\section{Introduction}
\label{sec:intro}
Coordination among agents is critical for effective group behavior in both natural and artificial systems. Humans demonstrate a remarkable capacity for adaptive group coordination, particularly in settings characterized by incomplete information and limited communication \citep{conradt2009group}. These capabilities are supported by dynamic strategies such as role differentiation, graded responsiveness to feedback, and self-consistency within groups, mechanisms that collectively enable convergence toward shared goals even without direct communication \citep{roberts2011adaptive}. Such adaptive coordination is crucial in numerous real-world scenarios, from team collaborations to organizational decision-making.

As large language models (LLMs) are increasingly deployed in multi-agent systems (MAS) to solve complex, collaborative tasks, their capacity to coordinate adaptively is a subject of intense investigation and debate. Recent advances have led to agentic frameworks in which multiple LLM-based agents perceive, learn, reason, and act collaboratively, leveraging collective intelligence to tackle problems beyond the scope of single agents \citep{tran2025multiagentcollaborationmechanismssurvey, du2023improving}. These MAS are being explored across diverse domains, including software engineering, scientific discovery, and social simulations (e.g., VIRSCI \citep{su-etal-2025-many} MetaGPT \citep{hong2024metagpt}, ChatDev \citep{qian2024chatdev}, HyperAgent \citep{phan2024hyperagent}, and frameworks like AutoHMA-LLM \citep{yang2025autohma}).

However, empirical results often show limited gains over single-agent baselines and frequent breakdowns in collaborative performance \citep{cemri2025why}.
Recent benchmarks and studies using behavioral game theory are beginning to probe the coordination and cooperation abilities of LLMs more directly \citep{huang2025competing, akata2025playing, fan2024can, duan2024gtbench, agashe-etal-2025-llm, sun2025collab}. Findings indicate that LLMs can perform well in certain interactive settings, particularly those aligning with self-interest, like the iterated Prisoner's Dilemma \citep{akata2025playing}. However, they often behave inefficiently in games requiring pure coordination, such as the Battle of the Sexes \citep{akata2025playing}, or complex Theory of Mind (ToM) reasoning, like Hanabi \citep{bard2020hanabi, agashe-etal-2025-llm}.
Nonetheless, some studies show LLMs matching or exceeding sophisticated reinforcement learning agents in coordination tasks like Overcooked-AI, which rely more on environmental understanding, and demonstrating robustness when paired with unseen partners (Zero-Shot Coordination) \citep{agashe-etal-2025-llm}. The potential for human-AI coordination is also being explored, leveraging the complementary strengths of human flexibility and contextual understanding with AI's data processing power \citep{patel2019humanai, NEURIPS2019_f5b1b89d}.

To dissect the mechanisms underpinning group coordination under imperfect monitoring in LLMs, we compare human and LLM performance on Group Binary Search (GBS). In this $n$-player game, participants must coordinate actions to collectively sum their independent guesses to a hidden target number. Without direct communication, they rely solely on shared group feedback - either directional (too high, too low, or just right) or numerical (e.g., too high by 25), to iteratively refine their choices. While existing research explores structured collaboration and game-theoretic interactions in LLMs, our work studies human and LLM behavior in a pure group coordination setting from a social psychology perspective, where success depends entirely on mutual adjustment and emergent alignment based on incomplete, group-level feedback. Prior work shows that human groups in this setting can adjust strategies, learn from experience, and develop stable heterogeneous adjustment patterns that reduce coordination costs \citep{roberts2011adaptive}. Recent work on group coordination in an imperfect monitoring setting \citep{riedl2026emergent} found that LLMs struggle with oscillations as group size increases and studied impact of prompt-induced persona differentiation. We focus on establishing direct human baselines in GBS to identify behavioral divergences in LLM groups, including persistent switching, overreactivity to feedback, and limited cross-game adaptation. 

We found systematic differences between LLM and human groups. Humans significantly benefited from numerical feedback, converging to targets faster than with directional feedback alone; LLM groups, however, showed little to no benefit from richer numerical feedback, performing worse than humans in most cases. Unlike humans who clearly improved across repeated games, LLMs exhibited minimal cross-game learning. These differences persisted across group sizes (2–17 players). Mechanistically, LLMs consistently displayed dramatically higher switching rates than humans, as well as greater overreactivity to feedback, and failed to reduce their adjustments as the group approached the target. Thus even when LLM performance reached human performance, the coordination was fundamentally different. Improving learning from feedback history and calibrating reactivity and switching could help reduce the observed gap. Along these lines, we investigated whether additional targeted fine-tuning could alter LLM switching behavior. Using reinforcement learning, we systematically increased stay probability, illustrating one pathway to reducing the observed gap between human and LLM groups.

\section{Methods}
\label{sec:methods}

\subsection{Group Binary Search}
\begin{figure}[ht!]
    \centering
    \includegraphics[width=0.7\textwidth]{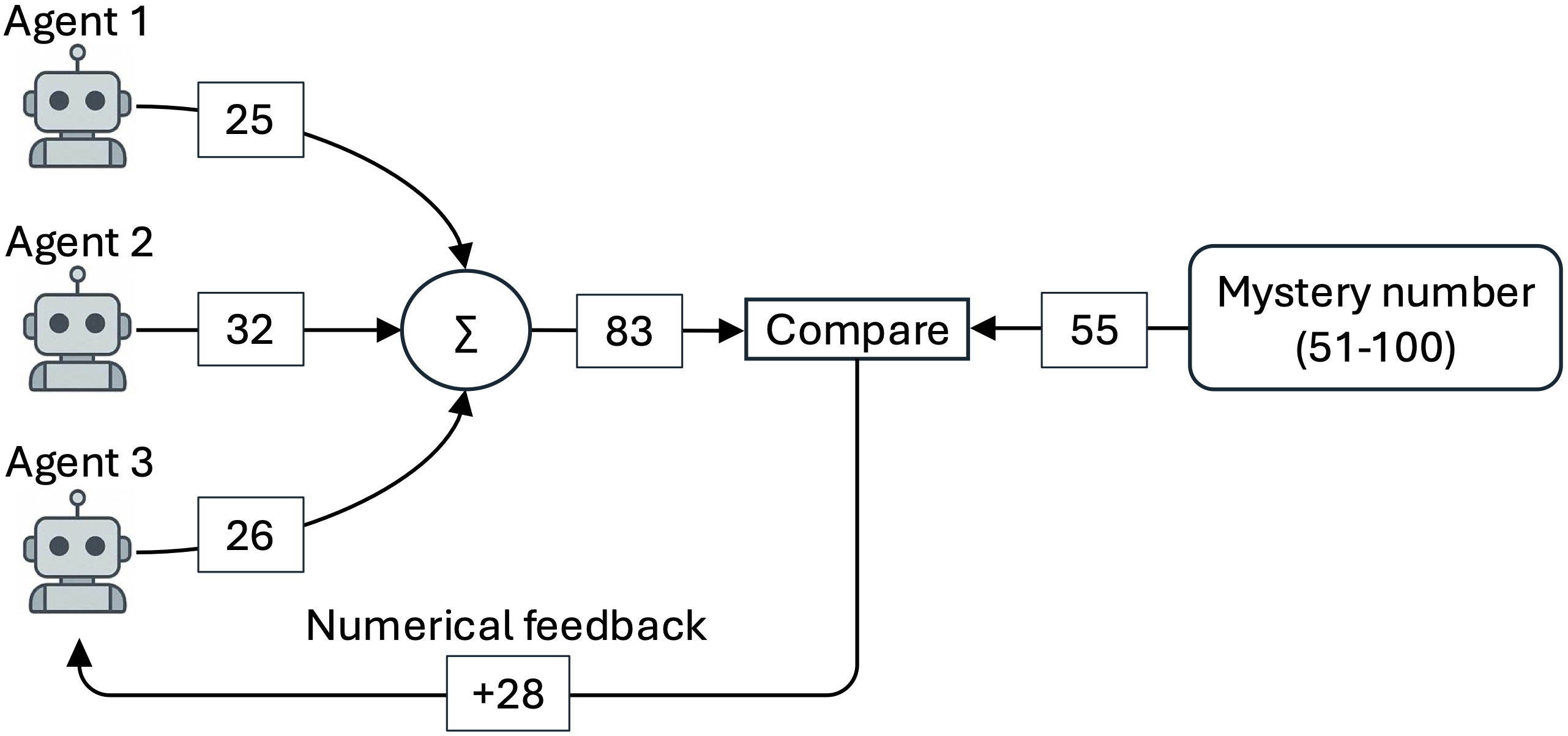}
    \caption{Schematic of a GBS game with three players and numerical feedback. The sum of guesses from each player is compared to the mystery number, and the players are provided feedback about the difference between the sum of their guesses and the mystery number. The players can then adjust their answer (without communicating with each other or knowing the guesses by other players), and the game continues until the sum of guesses matches the mystery number or until 15 rounds have been played.}
    \label{fig:schematic}
\end{figure}

We evaluate LLMs on a common-interest game with imperfect monitoring, GBS. In GBS, each participant makes a guess on each round and the sum of all of the guesses in the group is compared to a hidden mystery number at the end of every round. Participants are not provided the individual guesses of other players, and the mystery number remains fixed within a game. All players receive identical feedback, either directional feedback (too low, too high, or just right) or numerical error-magnitude feedback (too low by $k$, too high by $k$, or just right). The common goal of the group is to have the sum of all guesses exactly match the mystery number. The game ends when the group succeeds or reaches the maximum of 15 rounds (see Fig.~\ref{fig:schematic} for a schematic illustrating the game with numerical feedback). In our experiments, we closely follow the experiment structure in \citep{roberts2011adaptive}, whose data we use as a human baseline. Each experiment session contains 10 games played sequentially while alternating between directional and numerical feedback. To facilitate comparison with human data, we match the mystery number in each game and the number of players in each session to the human data. Similar to \citep{roberts2011adaptive}, the models were evaluated across 18 GBS experiments with varying sizes: six 2-player, three 3-player, three 4-player, one 6-player, one 7-player, one 10-player, one 16-player, and two 17-player games. The groups were categorized by size: 2-3 players (small groups), 4-7 players (medium groups), and 10-17 players (large groups).

\subsection{Prompts}

We used two different prompting conditions, Zero-Shot and Zero-Shot CoT \citep{kojima2022large} as shown in Figures \ref{fig: Zero-shot prompt} and \ref{fig: Zero-shot CoT prompt}. In both conditions, the context of LLMs keeps increasing as the experiment progresses. The results of all the previous rounds across all the previous games and the current game are maintained in the context. In the zero-shot CoT prompt condition, we parse the output of the model and retain only the numerical guesses of all the previous rounds in order to keep the number of tokens within desirable limits. 

\subsection{Models}

The evaluations were conducted using Deepseek-V3 \citep{liu2024deepseek}, Deepseek-V3.1-Terminus (referred to as Deepseek-V3.1-T throughout) \citep{liu2024deepseek}, Llama 3.3 \citep{grattafiori2024llama}, and Gemini 2.0 Flash \citep{blogIntroducingGemini}. On each experimental run, we set a different seed for each agent in the group for all models that expose a seed parameter. Temperature values were selected per model and prompting condition based on temperature ablation experiments in small groups (see Appendix Sec.~\ref{appendix:hyperparameters} for full details). All other generation parameters were left at provider defaults. We conducted additional analyses to assess robustness to temperature settings, temperature heterogeneity within groups, and alternative prompting strategies; these had minimal impact on coordination performance (see Appendix Sec.~\ref{appendix:robustness}). %

\subsection{Behavioral Intervention with GRPO}

We fine-tuned Qwen2.5-14B-Instruct with GRPO using \texttt{verl} \citep{sheng2024hybridflow} and make it easier for the model to learn the desired behavior, we divided the training into two phases. Phase 1 ran for 100 optimization steps on synthetic numerical-feedback GBS histories. Within each example, the model was told that it had made the same guess in every previous round of the current game; the repeated guess varied across games, while the round index (2--15) and group size (2--17) also varied. Phase 2 initialized from the Phase 1 step-100 model and ran for a further 100 steps on simulator-generated numerical-feedback states in which previous guesses could vary across rounds. Each prompt used 32 rollouts. Switches received zero reward, while stayed responses were eligible for positive reward only when GPT-4.1-mini judged the rationale coherent, consistent with the final guess, and grounded in group-level coordination; incoherent or action-inconsistent stay rationales were penalized. Checkpoints were saved every 10 Phase 2 steps. We evaluated the unmodified base model and Phase 2 checkpoints 50, 60, 70, 80, 90, and 100 in held-out GBS games. Checkpoint numbers in all figures therefore denote optimization steps within Phase 2.

\section{Results}
\label{sec:results}

We evaluated LLMs on their coordination performance in the GBS task, comparing them against human baselines. The evaluation varied across group sizes (small: 2-3 players, medium: 4-7 players, large: 10-17 players), feedback types (numerical vs. directional), and prompting strategies (Zero-Shot and Zero-Shot Chain-of-Thought (CoT)). Key results are summarized in Table~\ref{tab:group_sizes_feedback}. 

\begin{table}[ht!]
\centering
\footnotesize
\setlength{\tabcolsep}{4pt}
\begin{tabular}{@{}lcccccc@{}}
\toprule
& \multicolumn{2}{c}{\textbf{Small groups}} & \multicolumn{2}{c}{\textbf{Medium groups}} & \multicolumn{2}{c}{\textbf{Large groups}} \\
\cmidrule(lr){2-3} \cmidrule(lr){4-5} \cmidrule(lr){6-7}
\phantom{\textbf{Condition}} & \textit{Numerical} & \textit{Directional} & \textit{Numerical} & \textit{Directional} & \textit{Numerical} & \textit{Directional} \\
\midrule
\textbf{Human} & 4.34 (0.82) & 9.29 (1.99) & 8.72 (1.69) & 10.84 (2.32) & 11.05 (2.99) & 12.85 (1.89) \\
\addlinespace
\multicolumn{7}{@{}l}{\textbf{Zero-Shot}} \\
Deepseek-V3 & 8.46 (3.02) & 7.60 (2.02) & 10.16 (2.73) & 12.48 (0.63) & 14.65 (0.70) & 14.10 (1.04) \\
Gemini 2.0 Flash & 7.23 (2.16) & 8.85 (2.75) & 11.36 (3.94) & 11.40 (1.12) & 11.55 (2.88) & 12.40 (2.68) \\
Llama 3.3 & 11.33 (3.48) & 9.02 (2.75) & 15.00 (0.00) & 11.08 (2.23) & 15.00 (0.00) & 13.00 (1.36) \\
\addlinespace
\multicolumn{7}{@{}l}{\textbf{Zero-Shot CoT}} \\
Deepseek-V3 & 5.37 (1.47) & 6.72 (1.39) & 8.08 (1.57) & 7.60 (1.53) & 12.85 (1.86) & 13.65 (1.06) \\
Deepseek-V3.1-T & 5.02 (1.68) & 6.65 (1.19) & 8.40 (0.60) & 10.48 (2.89) & 12.00 (2.24) & 12.60 (2.85) \\
Gemini 2.0 Flash & 8.98 (2.08) & 8.09 (1.88) & 13.68 (2.34) & 11.60 (1.72) & 15.00 (0.00) & 14.60 (0.49) \\
Llama 3.3 & 10.34 (3.34) & 7.21 (1.03) & 14.64 (0.80) & 10.00 (2.63) & 14.75 (0.50) & 14.10 (1.80) \\
\addlinespace
\textbf{Mixed LLMs} & 13.82 (1.31) & 8.42 (1.86) & 15.00 (0.00) & 11.04 (2.92) & 15.00 (0.00) & 14.05 (0.74) \\
\bottomrule
\addlinespace 
\end{tabular}
\caption{Mean rounds to solution (games that remained unsolved at the 15-round limit assigned 15). Standard deviations are computed across session-level means, with each session-level mean averaging the five games with the corresponding feedback type. Success rates, rounds conditional on success, failure rates, and terminal errors are reported separately in Appendix Table~\ref{tab:success_failure_metrics}. See Table~\ref{tab:mixed_model_breakdown} for breakdown of the results for \textit{Mixed LLMs}.}
\label{tab:group_sizes_feedback}
\end{table}

Humans solved numerical-feedback games faster than directional-feedback games at every group size, whereas LLMs showed inconsistent benefits and often performed worse with numerical feedback. Both populations performed worse as group size increased, with ceiling effects for some LLMs in medium and large numerical-feedback games.

\begin{figure*}[h!]
\centering
\includegraphics[width=1.0\textwidth]{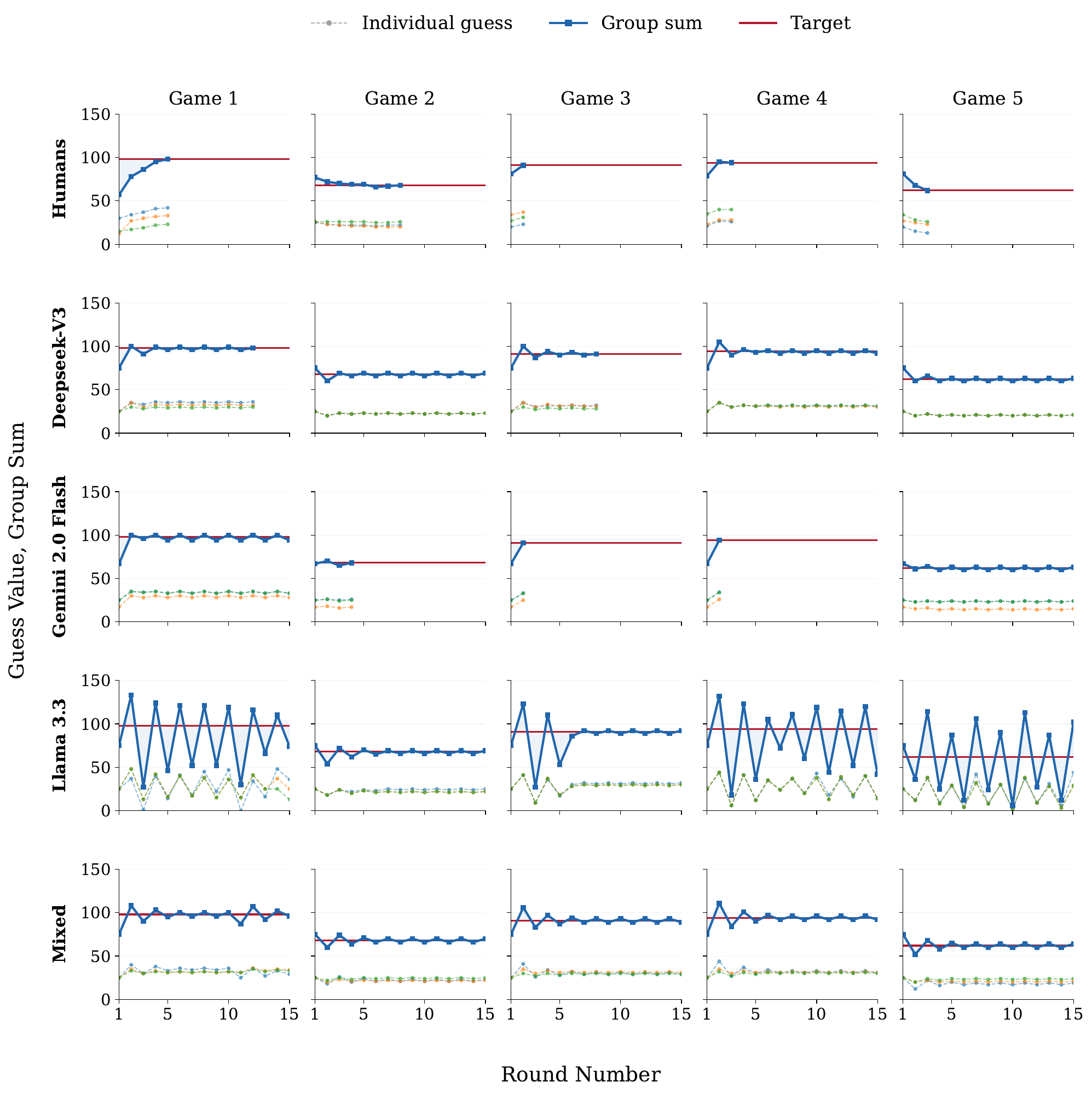}
\caption{Example of coordination in 3-player games with numerical feedback and zero-shot prompts. The solid horizontal line indicates the mystery number, the other solid line indicates the sum of the group, and the dashed line represents the decisions of each agent in the group.}
\label{fig:3-player-zero-shot-numerical-14}
\end{figure*}

To illustrate the dynamics of the GBS gameplay, we show in Fig~\ref{fig:3-player-zero-shot-numerical-14} responses for each player in each round for five consecutive three-player games with numerical feedback  for human and different LLM groups. Additional examples of gameplays are provided in Appendix Sec.~\ref{appendix:Gameplay}, and the rest are included in the Supplemental Material. 
It is noticeable that LLM groups often tended to overreact to feedback signals, leading to prolonged oscillations around the target number, whereas human groups showed calibrated, progressively refined adjustments leading to rapid convergence. We provided further quantification of this behavior in subsequent sections.

\subsection{Adaptation and Learning Over Games}
\begin{figure*}[h!]
\centering
\includegraphics[width=\textwidth]{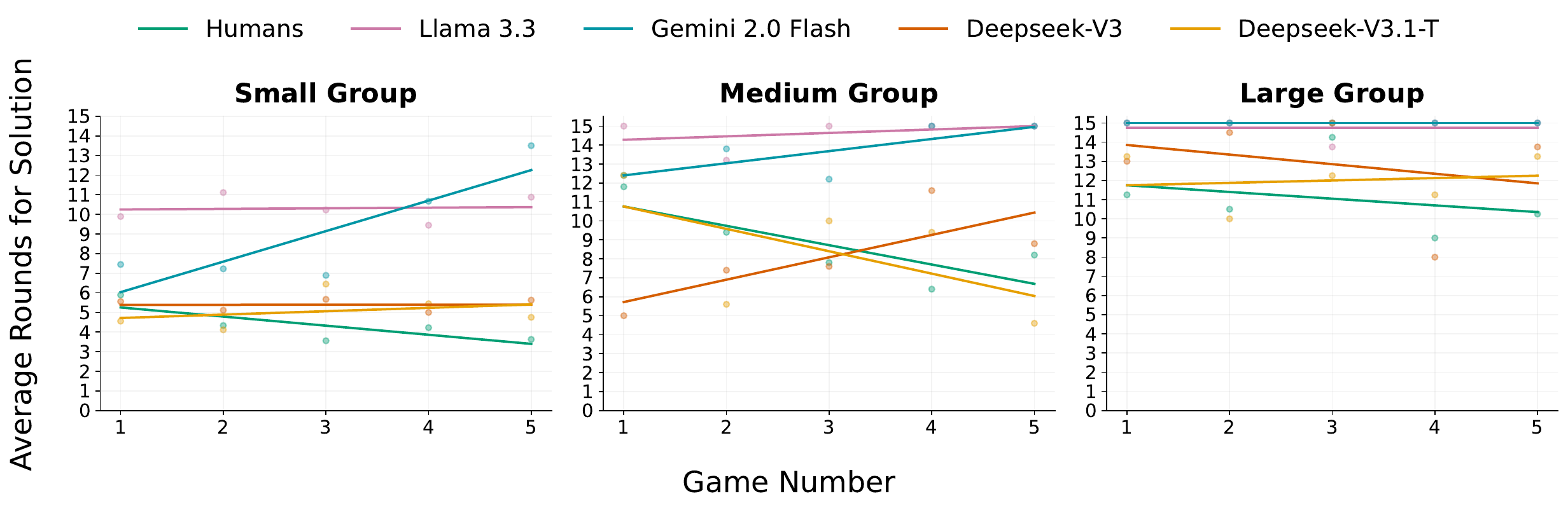}
\caption{Average number of rounds needed to finish the game with zero-shot CoT prompts under numerical feedback.}
\label{fig:rounds_to_solution_numerical_cot}
\end{figure*}

An important difference between human and LLM groups emerged in adaptation over successive games. The learning curves shown in Fig.~\ref{fig:rounds_to_solution_numerical_cot} and Fig.~\ref{fig:rounds_to_solution_rest} indicate that human groups generally improve more than LLM groups for both directional and numerical feedback across successive games. To quantify this, for each run and feedback type, we fit a linear slope relating rounds to solution and game index (1--5), where negative slopes indicate improvement across games. Across runs, human groups showed negative mean slopes for both directional feedback (mean slope $= -0.91$ rounds/game, 95\% bootstrap CI $[-1.34, -0.43]$; $78\%$ of runs negative) and numerical feedback (mean slope $= -0.57$, 95\% bootstrap CI $[-1.12, -0.04]$; $72\%$ of runs negative). Human slopes were negative on average in small, medium, and large groups for both feedback types. LLM adaptation was weaker and more model-dependent. Under directional feedback, model-specific mean slopes ranged from $-0.39$ for Deepseek-V3 zero-shot to $0.31$ for Deepseek-V3 zero-shot CoT, and all bootstrap intervals included zero. Under numerical feedback, most LLM conditions were again close to flat, with the clearest worsening trend observed for Gemini 2.0 Flash with zero-shot CoT prompting (mean slope $= 0.89$, 95\% bootstrap CI $[0.33, 1.52]$; $89\%$ of runs non-negative). All per-model slope statistics for zero-shot and zero-shot CoT conditions are reported in Tables~\ref{slope_rounds_to_solution_zs} and \ref{slope_rounds_to_solution_zs_cot}, respectively. Overall, the learning curves point to a consistent pattern: human groups improved more reliably across repeated games, whereas LLM adaptation was limited and depended strongly on model and prompting condition. This could reflect limitations in how current LLMs manage evolving context or refine behavioral strategies across temporally extended tasks.

\subsection{Group Reaction Strategies}
\begin{figure*}[h!]
\centering
\includegraphics[width=\textwidth]{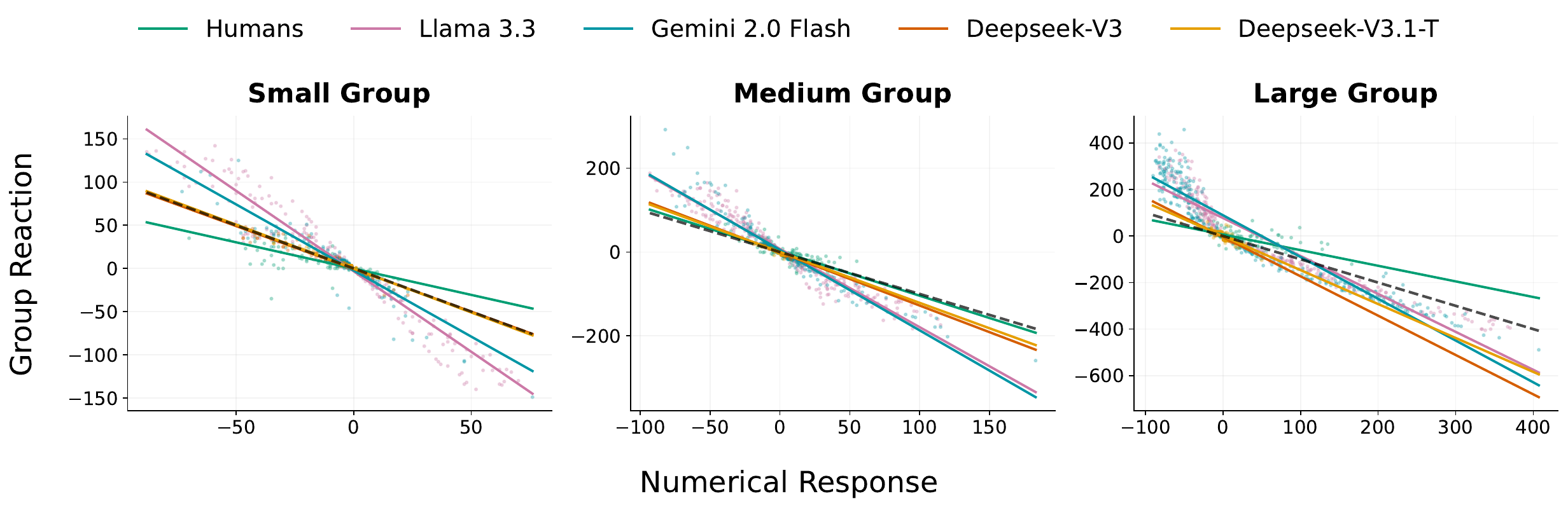}
\caption{Group reaction to numerical feedback under zero-shot CoT prompting. Each dot denotes the aggregate adjustment made by a group after the previous round's  feedback. The dotted line indicates the optimal collective correction, and the solid line shows the fitted group reaction. Human groups remain closer to stable underreaction, whereas LLM groups more often respond with steeper collective updates.}
\label{fig:numerical-group-reaction-cot}
\end{figure*}

Analysis of group reaction to feedback provides further insight into coordination dynamics. Humans tend to adapt their reactivity based on feedback, generally underreacting but modulating their adjustments, decreasing reactivity as they approach the target or when feedback changes direction. In contrast, LLMs adjust in the correct direction but often overreact (Fig.~\ref{fig:numerical-group-reaction-cot}; see also Fig.~\ref{fig:numerical-group-reaction-zs} for Zero-shot prompt).

To quantify this, we compared the slopes of best-fit lines for group reactions across conditions (small, medium, and large groups; zero-shot and zero-shot CoT prompts). Human slopes averaged -.767, indicating underreaction (less negative than -1). LLM slopes, averaged across models per condition, were consistently more negative, with a mean of -1.386, indicating overreaction. In all 6 conditions, LLMs exhibited more negative slopes. Considering individual models, 19 of 21 numerical-feedback LLM slopes were more negative than the corresponding human slope, with the 2 exceptions being Deepseek-V3 with zero-shot prompting for large (Deepseek-V3: -0.517, human: -0.663) and medium groups (Deepseek-V3: -0.827, human -1.050). All per-condition slopes are reported in Table~\ref{tab:reaction_slopes_numerical}.
    
This overreactivity in LLMs may contribute to their coordination challenges, especially in larger groups, where excessive adjustments can lead to oscillations around the target (see Appendix Sec. \ref{appendix:Gameplay}). Humans' underreaction, while suboptimal for immediate correction, may stabilize group progress by reducing noise. This difference suggests opportunities for future research on human-AI teaming, where LLM overreactivity could complement human caution, potentially optimizing collective adjustments in mixed groups.

\subsection{Switching Behavior Analysis} 
\label{sec:switching}
\begin{figure*}
    \centering
        \begin{subfigure}[b]{0.30\textwidth}
            \centering
            \includegraphics[width=1.0\textwidth]{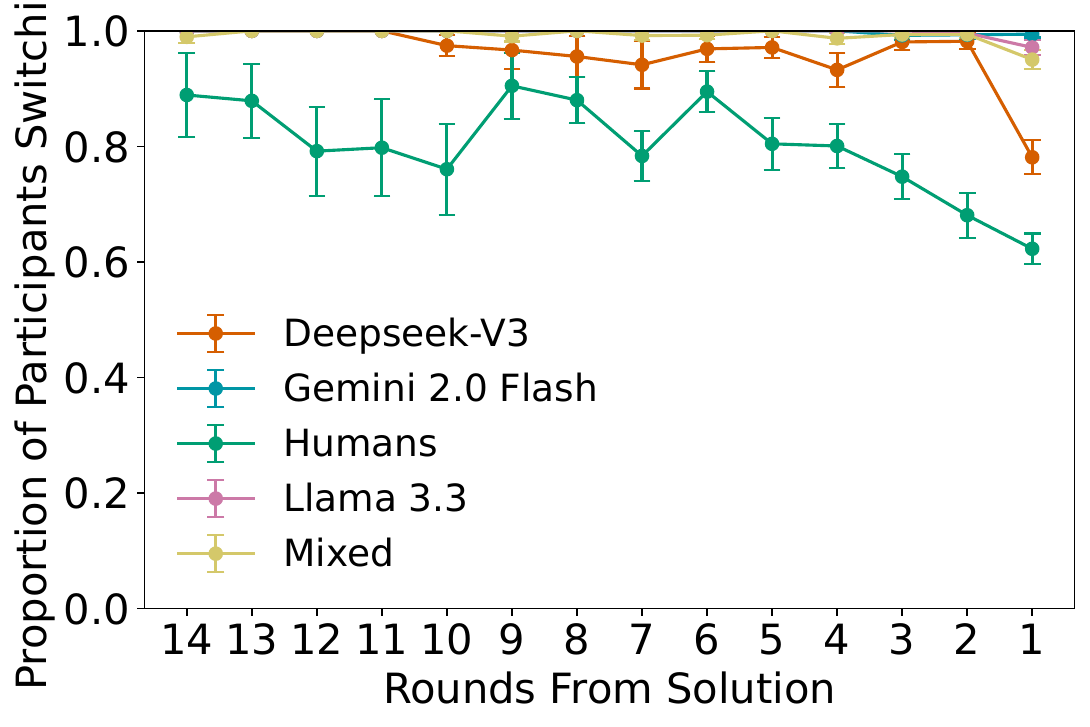}
            \caption{Small-group, zero-shot}
            \label{fig:2-player-zero-shot-proportion-change}
        \end{subfigure}
        \begin{subfigure}[b]{0.30\textwidth}
            \centering
            \includegraphics[width=1.0\textwidth]{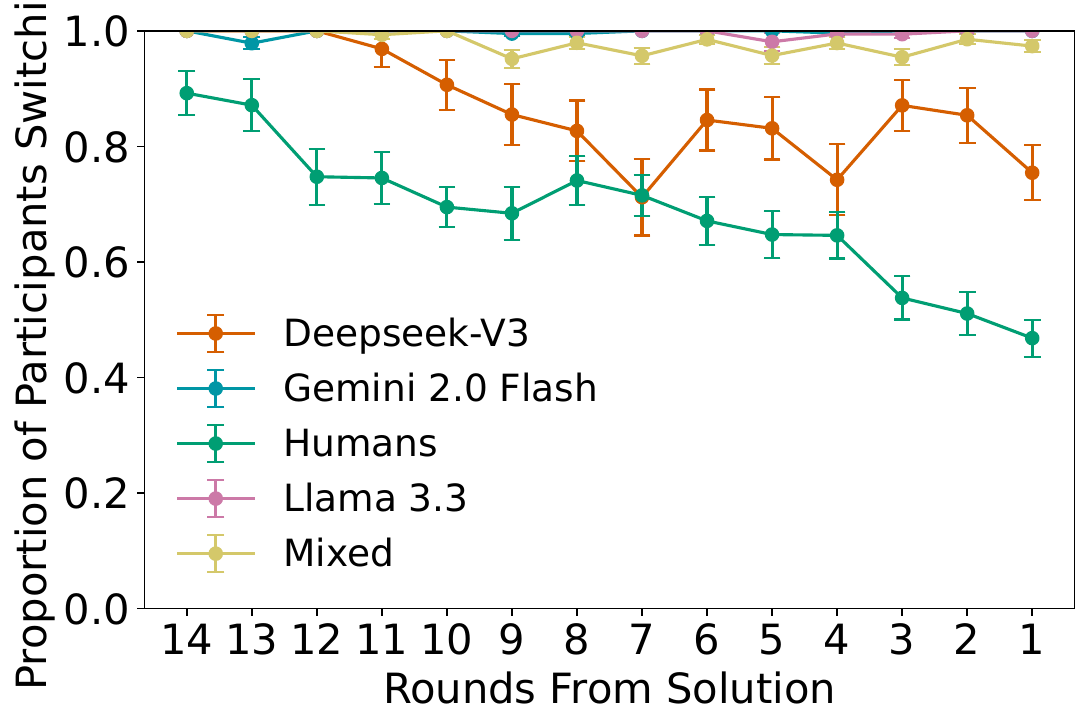}
            \caption{Medium-group, zero-shot}
            \label{fig:2-player-zero-shot-CoT-proportion-change}
        \end{subfigure}
        \begin{subfigure}[b]{0.30\textwidth}
            \centering
            \includegraphics[width=1.0\textwidth]{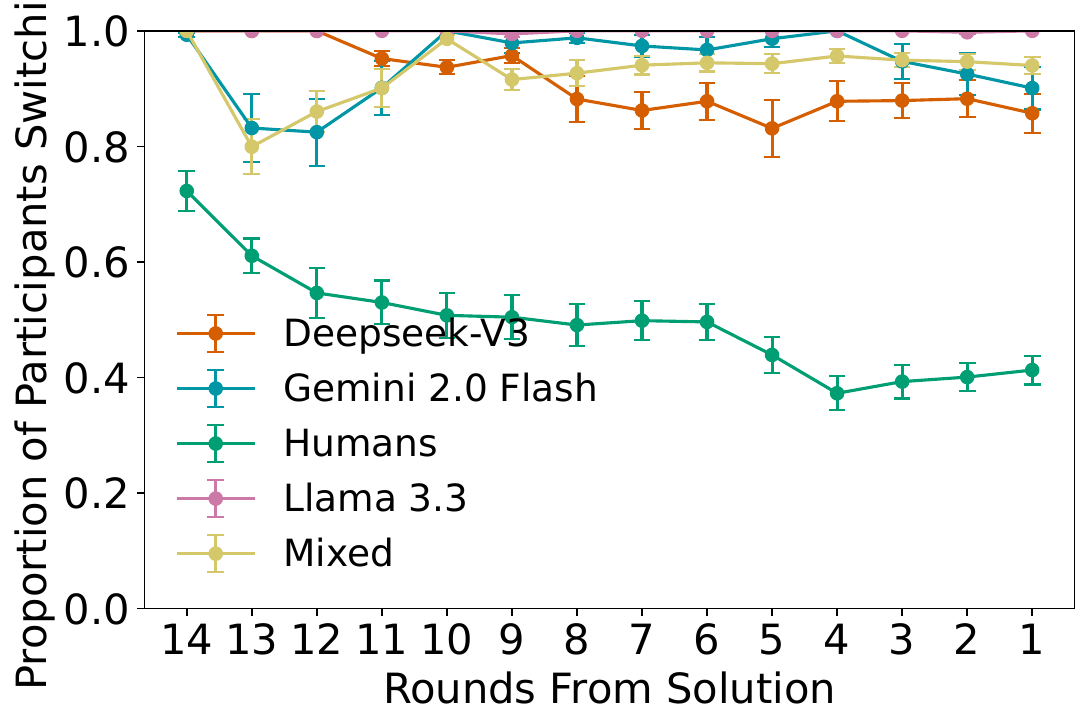}
            \caption{Large-group, zero-shot}
            \label{fig:2-player-multi-llm-zero-shot-proportion-change}
        \end{subfigure}
        
        \vspace{0.4cm}
        
        \begin{subfigure}[b]{0.30\textwidth}
            \centering
            \includegraphics[width=1.0\textwidth]{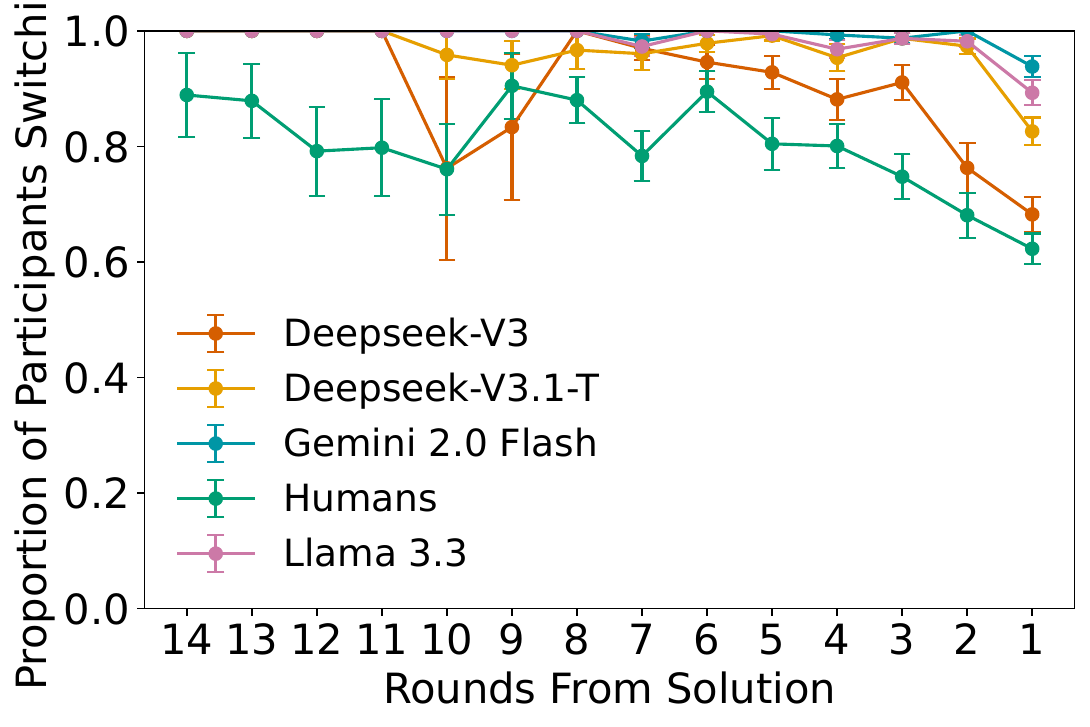}
            \caption{Small-group, ZS-CoT}
            \label{fig:6-player-zero-shot-proportion-change}
        \end{subfigure}
        \begin{subfigure}[b]{0.30\textwidth}
            \centering
            \includegraphics[width=1.0\textwidth]{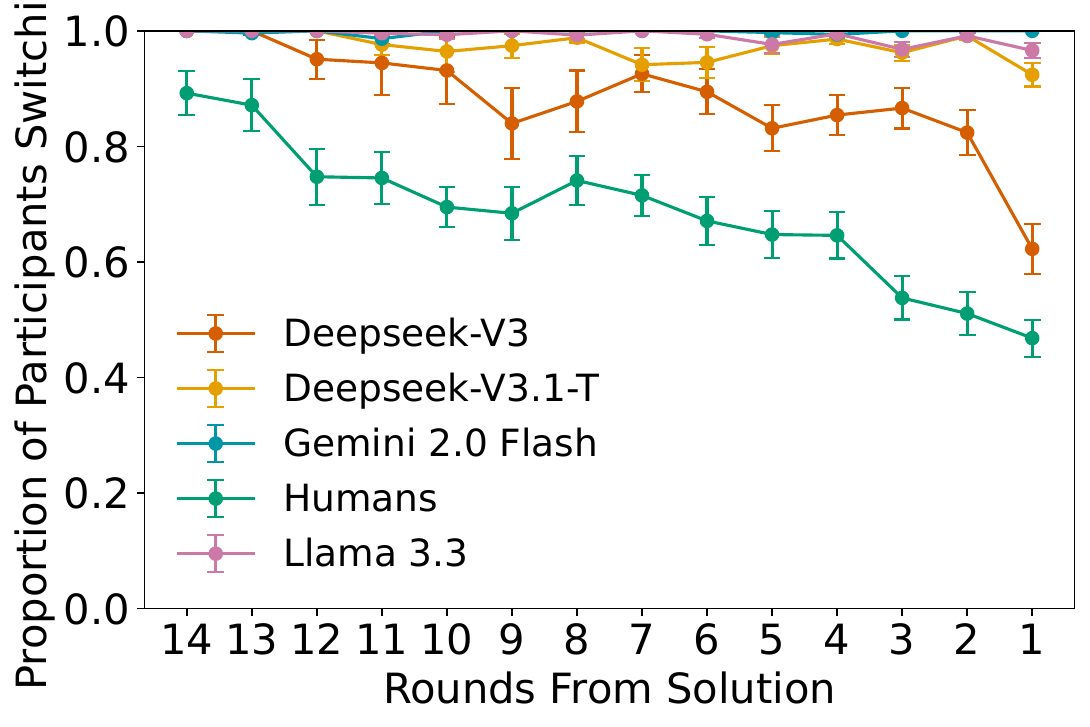}
            \caption{Medium-group, ZS-CoT}
            \label{fig:6-player-zero-shot-CoT-proportion-change}
        \end{subfigure}
        \begin{subfigure}[b]{0.30\textwidth}
            \centering
            \includegraphics[width=1.0\textwidth]{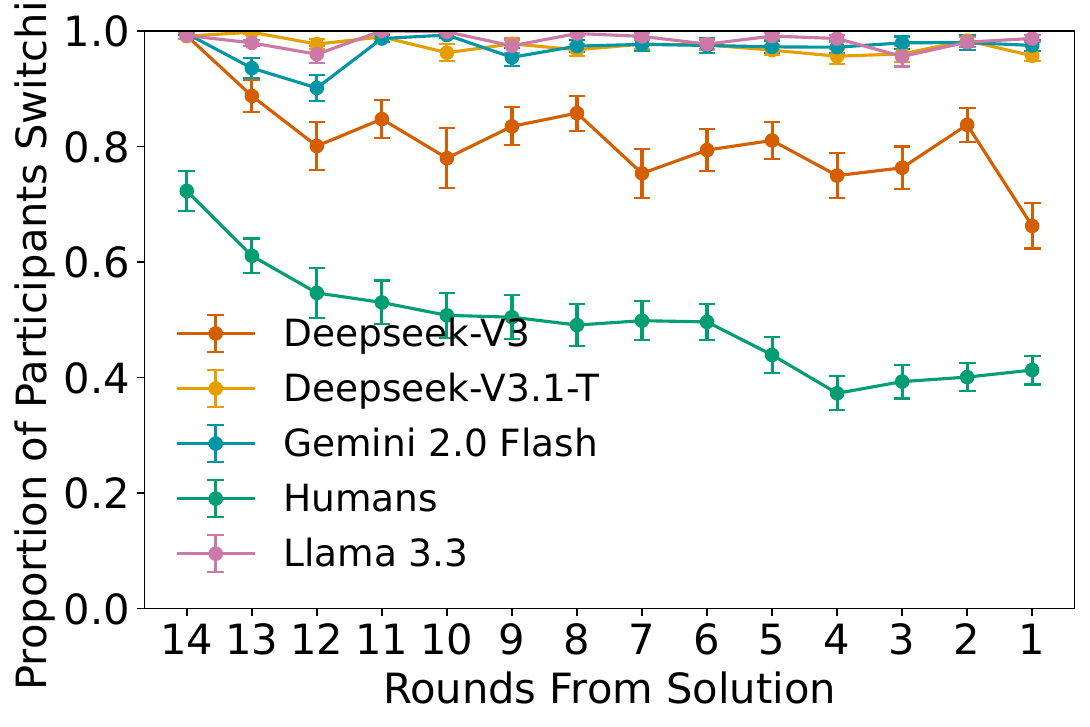}
            \caption{Large-group, ZS-CoT}
            \label{fig:6-player-multi-llm-zero-shot-proportion-change}
        \end{subfigure}

    \caption{Average proportion of players switching their guess from the previous round as the group approaches the end of the game (either by termination or by finding a solution), across different experimental conditions: small-group vs medium-group vs large-group and zero-shot vs. zero-shot CoT prompts. Error bars represent standard errors of the mean across contributing games.}
    \label{fig:combined-proportion-change}
\end{figure*}

We examined the proportion of agents that switched (i.e., changed) their submitted number from the previous round as the game progressed. LLMs had a remarkably stronger switching tendency across all group sizes and both feedback conditions (Fig.~\ref{fig:combined-proportion-change}). In human groups, particularly medium and large groups, individuals tend to strongly decrease their tendency to change the response as the group converges on the target, potentially reflecting role stabilization or an effort to reduce noise. 

In general, LLM groups show little reduction in switching and maintain much higher switching rates throughout the game. The distributions of switch magnitudes (Appendix Sec.~\ref{appendix:histograms_decision_switch_magnitude}) further confirm this: human players show a pronounced peak at zero change, reflecting frequent decisions to stay, whereas LLM distributions are more spread across non-zero magnitudes. This persistent reactivity can contribute to oscillation of the group sum and slow convergence. Neither zero-shot CoT prompting nor mixed-model composition qualitatively changed this pattern.

\begin{figure*}[ht!]
    \centering
    \begin{subfigure}[b]{0.48\textwidth}
        \centering
        
        \includegraphics[width=0.775\textwidth]{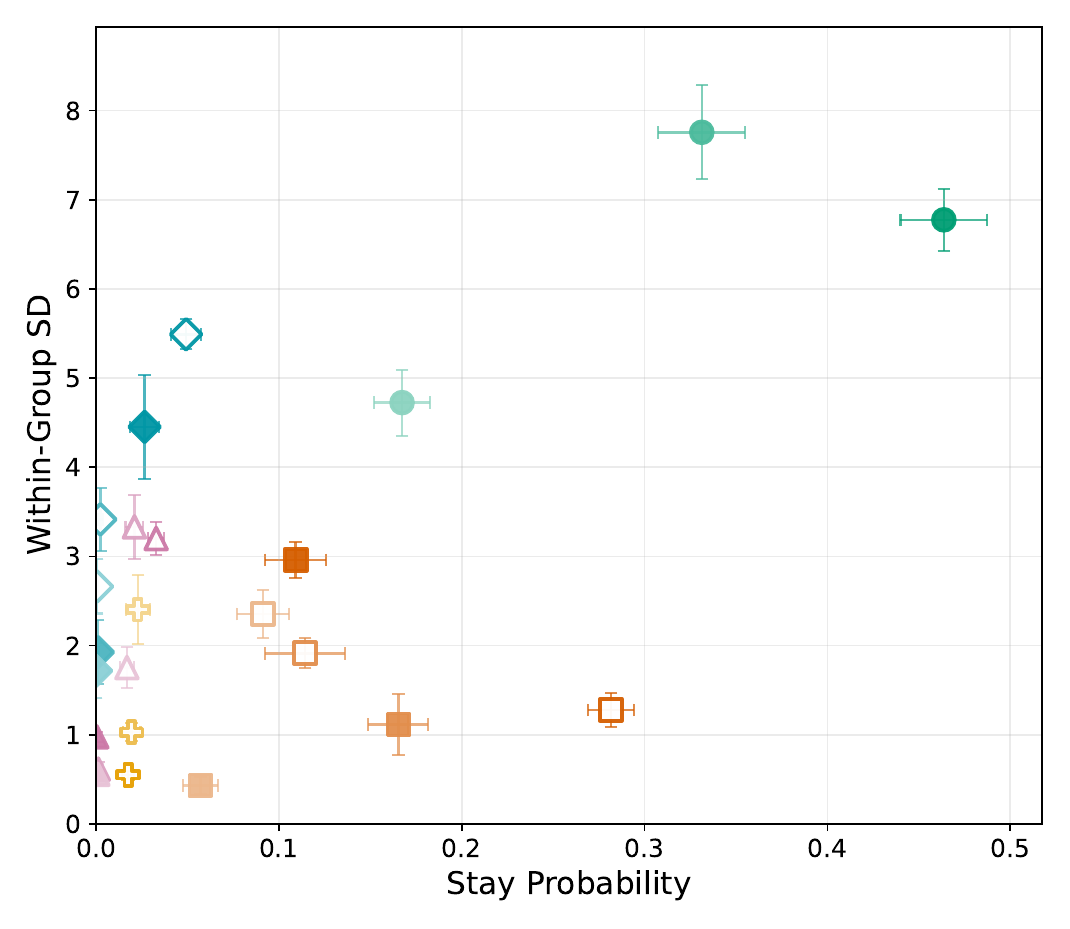}
        \caption{Directional feedback}
        \label{fig:coordination-signature-directional}
    \end{subfigure}
    \hfill
    \begin{subfigure}[b]{0.48\textwidth}
        \centering
        \includegraphics[width=\textwidth]{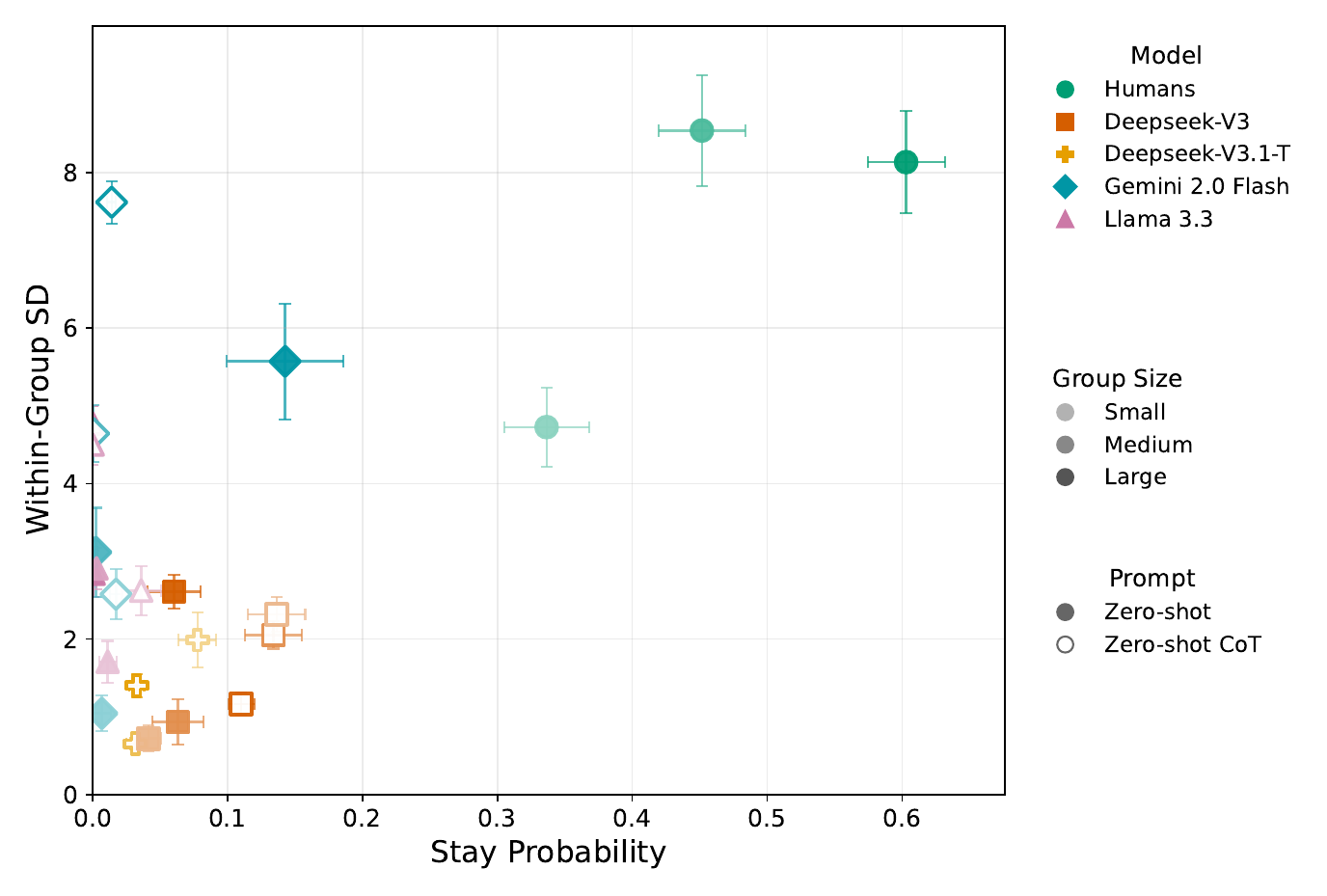}
        \caption{Numerical feedback}
        \label{fig:coordination-signature-numerical}
    \end{subfigure}
    \caption{Coordination signatures by model and group size. Horizontal and vertical bars summarize across-game variability within the condition. Across both feedback types, human conditions occupy a high-stability, high-dispersion region, while most LLM conditions cluster in a relatively lower-stability, lower-dispersion regime.}
    \label{fig:coordination-signature}
\end{figure*}

The coordination signatures in Fig.~\ref{fig:coordination-signature} show the same difference from a complementary perspective. Human conditions occupy a high-stability, high-dispersion region, whereas most LLM conditions cluster in a lower-stability, lower-dispersion regime. Selected numerical sessions in Fig.~\ref{fig:selected-stay-profiles} provide qualitative illustrations of that difference: in the human examples, switch rate is much lower and variability across players is larger, with some players choosing to keep their chosen number throughout each game, especially in sessions with a large number of players. 

Fig.~\ref{fig:stay-extremes} quantifies the same pattern at the game level. Human participants sometimes kept the same guess throughout an entire numerical game, corresponding to a per-game stay probability of 1. The proportion of such fully stable players increased with group size, from 0.04 in small groups to 0.12 in medium groups and 0.19 in large groups. No LLM condition showed players with stay probability of 1. At the opposite extreme, LLM groups were much more likely to contain players with stay probability of 0, meaning that they changed their guess on every post-initial round of a game. This pattern suggests that human groups sometimes spontaneously reduce the effective coordination burden by having a subset of members to remain fixed.

\subsection{Analysis of Reasoning in LLM Outputs}
\label{sec:reasoning}
While playing the games, LLMs evaluated with zero-shot CoT prompts had a tendency to output a large number of language tokens that described different strategies. The most elaborate outputs often came from Deepseek-V3.1-T compared to the other models in our evaluations. While in some cases the models' described strategy focused only on their own guesses and the feedback, in many cases the textual justifications incorporated the inferred strategy of other player(s) and adapted accordingly (see Appendix Sec. \ref{appendix: response examples} for examples of model outputs; See Supplemental Material for all model responses). Zero-shot CoT often improved performance relative to zero-shot prompting, with the clearest gains for Deepseek-V3 and gains for Llama 3.3 in most, but not all, size/feedback conditions (Table~\ref{tab:group_sizes_feedback}).

\subsection{Modifying Action Bias with GRPO}
\begin{figure*}[ht!]
    \centering
    \includegraphics[width=0.68\textwidth]{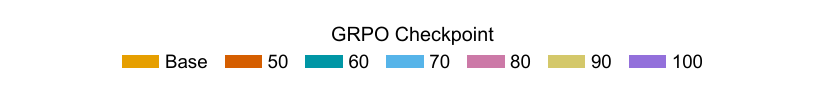}

    \begin{subfigure}[b]{0.31\textwidth}
        \centering
        \includegraphics[width=\textwidth]{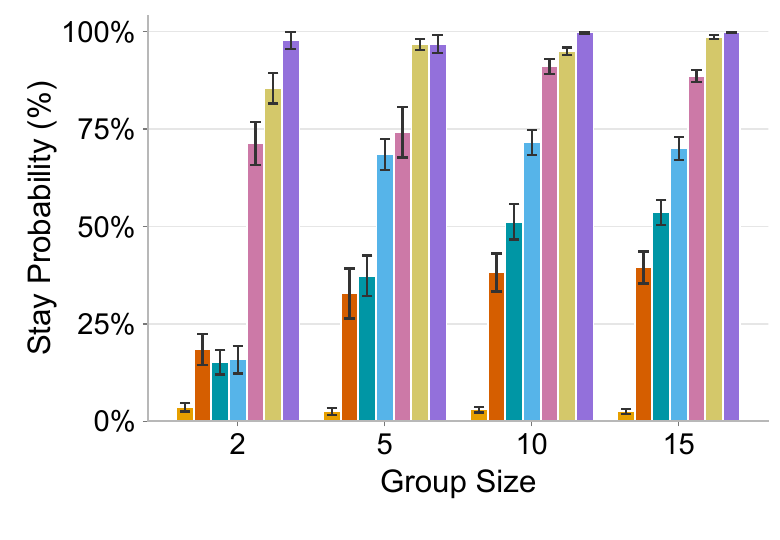}
        \caption{Stay probability}
        \label{fig:grpo-stay-probability}
    \end{subfigure}
    \hfill
    \begin{subfigure}[b]{0.31\textwidth}
        \centering
        \includegraphics[width=\textwidth]{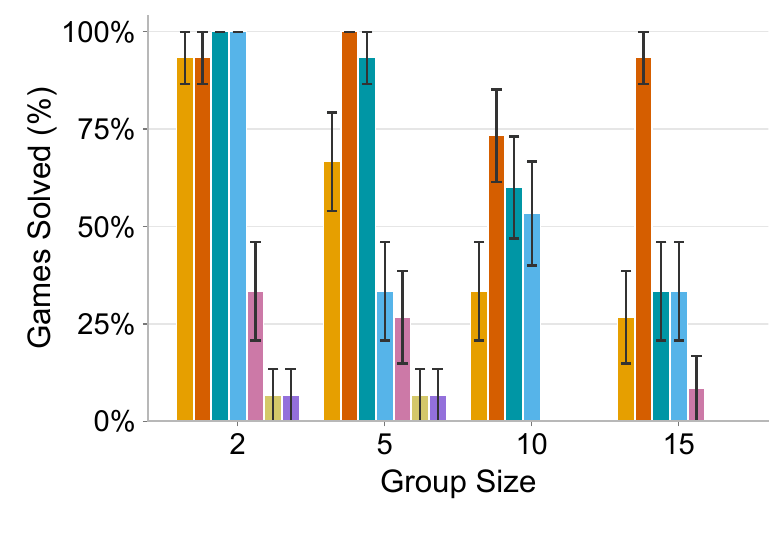}
        \caption{Games solved}
        \label{fig:grpo-games-solved}
    \end{subfigure}
    \hfill
    \begin{subfigure}[b]{0.31\textwidth}
        \centering
        \includegraphics[width=\textwidth]{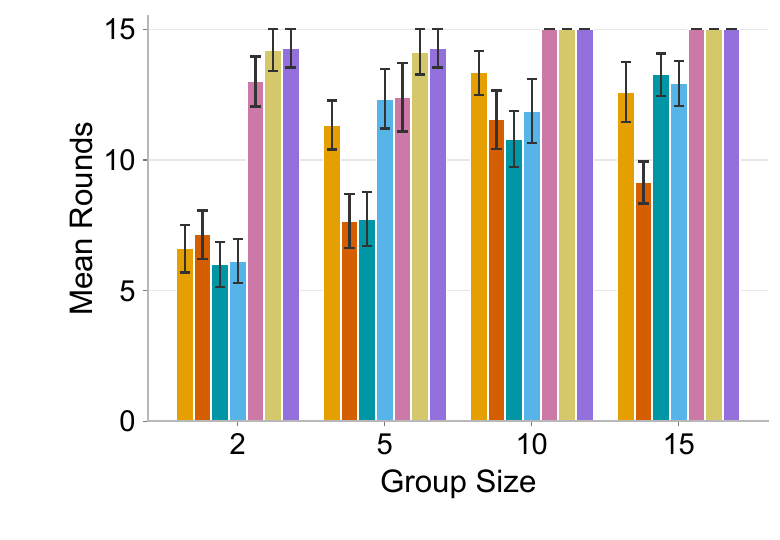}
        \caption{Mean rounds}
        \label{fig:grpo-mean-rounds}
    \end{subfigure}

    \caption{Numerical-feedback GBS performance across the base Qwen2.5-14B-Instruct model and a set of GRPO checkpoints. Panels show stay probability by group size (notice the systematic change), percentage of games solved, and mean rounds played (unsolved games are capped at 15). Error bars show standard errors across 15 games.}
    \label{fig:grpo-numerical-performance}
\end{figure*}

GRPO systematically increased stay probability in numerical-feedback GBS. The base model had stay probability near 0, which then gradually increased to near 100\% for the last checkpoint (Fig.~\ref{fig:grpo-stay-probability}) with intermediate checkpoints spanning a wide range  in between. This impacted the group performance, and groups with intermediate stay probabilities often solved more games than others (Fig.~\ref{fig:grpo-games-solved}) with fewest number of rounds to solution (Fig.~\ref{fig:grpo-mean-rounds}). Representative trajectories are shown in Appendix Fig.~\ref{fig:grpo-sample-game} and illustrate how the stay probability systematically differs across the checkpoints.

\section{Discussion}
\label{sec:discuss}
We investigated human and LLM coordination in GBS, a common-interest game with imperfect monitoring. LLM groups generally underperformed humans, especially with numerical feedback and in larger groups. Behavioral diagnostics linked slower LLM convergence to overreaction, persistent switching, and weak cross-game adaptation. These findings identify a task-specific coordination gap between the evaluated LLM and human groups, and provide behavioral diagnostics for future multi-agent LLM evaluations and interventions towards closing the coordination gap. Our results with GRPO fine-tuning provide an illustration of one such targeted intervention, demonstrating that the training can modulate the observed switch rate. %

\subsection{Coordination Deteriorates in Larger Groups}
Human and LLM performance degrades with group size in GBS, but the effect was often more pronounced for LLM groups (Table~\ref{tab:group_sizes_feedback}). Humans can partially mitigate these scaling challenges through adaptation across multiple timescales, such as learning better strategies across games and developing within-game adjustments like role differentiation, where some individuals become less reactive to reduce noise \citep{roberts2011adaptive}. Our analysis of switching behavior (Fig.~\ref{fig:combined-proportion-change}) suggests LLMs largely fail to adopt such dynamic strategies effectively; the proportion of agents changing their guesses remained high even as they neared the solution, in a sharp contrast to human behavior. Human players sometimes kept the same guess across all the rounds of a single game. The proportion of such players increased with the group size. This behavior effectively reduced the number of players that needed to coordinate, making the task easier. However, no LLM agent ever kept the same choices across all the rounds. While structured multi-agent networks with explicit communication may support collaboration at scale \citep{qian2025scaling}, our results show that under this imperfect monitoring setup, where agents receive only aggregate group feedback, the evaluated LLM groups commonly exhibited higher switching near the target than human groups, and, unlike the human groups, no LLM condition contained players with a per-game stay probability of 1.

\subsection{Learning from Experience is Challenging in LLM groups}
An important difference compared to human performance was the general lack of learning or strategy adaptation by LLMs over successive games (Fig.~\ref{fig:rounds_to_solution_numerical_cot} and Fig.~\ref{fig:rounds_to_solution_rest}). Human groups reliably improve their performance over repeated GBS games. In contrast, LLM performance typically remained flat and sometimes even degraded over the 10 games within a session. This lack of improvement suggests that under our setup, the evaluated LLMs are not effectively updating their strategies based on cumulative experience within the task session. While context window limitations might contribute to degradation in very long interactions \citep{liu2024deepseek}, the fundamental lack of improvement points to a deeper limitation in adaptive learning within this interactive, multi-agent setting. This is consistent with recent findings that LLMs exhibit significant performance degradation in multi-turn settings, making premature commitments and failing to recover from early errors \citep{laban2025llms}. Recent work has also argued that many LLM ToM benchmarks measure literal prediction of other agents rather than functional adaptation to partners over time \citep{riemer2025position}. The present GBS task is closer to this functional setting, since agents must adjust to the evolving behavior of unseen partners from shared feedback alone. 

\subsection{Behavioral Homogeneity in Homogeneous Groups}
An interesting observation was the high degree of behavioral similarity, sometimes resulting in identical guesses, between agents of the same LLM type within a group. This homogeneity arises naturally  from  sampling  the same model given  identical inputs (task description, history, feedback). While introducing heterogeneity by mixing LLM types did increase behavioral diversity, it did not lead to improved performance. This suggests that simply having different response patterns is insufficient for effective coordination; rather, agents need to \emph{adaptively} leverage their differences to develop complementary strategies, which neither homogeneous nor heterogeneous LLM groups achieved effectively here. Unlike human groups, where diversity can enhance performance \citep{page2007difference}, simple heterogeneity in LLM groups under our experimental setup offered no coordination benefit.

\subsection{Feedback Utilization: Human Groups Undershoot, LLM Groups Overshoot}
In this task, humans appear more adept at using precise error magnitude to modulate their responses effectively. In particular, when provided numerical feedback, LLM groups, while adjusting in the correct direction, tended to overreact, unlike human groups who were mostly under-reactive (Fig.~\ref{fig:numerical-group-reaction-cot}). This suggests a potential difference in how current LLMs and humans process and utilize quantitative feedback for fine-grained behavioral adjustment in dynamic tasks.

\subsection{Higher Baseline Action Rate in LLMs}
Beyond the dynamics of how switching behavior changes over rounds, we observed another notable difference: LLMs maintained a remarkably higher baseline proportion of switching compared to human participants throughout the game. While human groups, especially larger ones, start with a certain level of switching that then decreases substantially, LLM agents often started near or at a 100\% switching rate and, despite sometimes decreasing, generally finished at a much higher switching proportion than human players (Fig.~\ref{fig:combined-proportion-change}). We refer to this descriptive pattern as an action-bias-like tendency: the evaluated LLMs frequently, significantly more than human participants, changed their answers even when stability could have been useful. LLMs are biased towards changing their output or taking an action in each round, perhaps implicitly associating change with progress towards the goal, even when strategic inaction or stability might be more beneficial for group coordination. Understanding and mitigating this behavior could be important for developing LLMs that coordinate more effectively and in a more human-like manner. Through GRPO training we demonstrated that such mitigation could be achievable (Fig.~\ref{fig:grpo-stay-probability}).

\subsection{Scopes and Limitations}
Our results should be interpreted as evidence about LLM group coordination under imperfect monitoring setting, not as a general claim about all coordination settings. Our contribution is primarily in providing a behaviorally grounded, cognitive-science inspired diagnostic for LLM group coordination relative to human groups in a controlled, imperfect monitoring setup. The differences we identify are not impossibility results, but concrete coordination mechanisms (overreactivity, persistent switching, and weak cross-game adaptation) that future prompts, memory systems, or agent architectures could consider addressing.

\subsubsection*{Acknowledgments}
 We thank Dr. Michael Roberts for helpful correspondence regarding the human experimental data from \citet{roberts2011adaptive}. This research was supported in part by Lilly Endowment, Inc., through its support for the Indiana University Pervasive Technology Institute.

\bibliography{references}

\begin{thebibliography}{29}
\providecommand{\natexlab}[1]{#1}
\providecommand{\url}[1]{\texttt{#1}}
\expandafter\ifx\csname urlstyle\endcsname\relax
  \providecommand{\doi}[1]{doi: #1}\else
  \providecommand{\doi}{doi: \begingroup \urlstyle{rm}\Url}\fi

\bibitem[Agashe et~al.(2025)Agashe, Fan, Reyna, and Wang]{agashe-etal-2025-llm}
Saaket Agashe, Yue Fan, Anthony Reyna, and Xin~Eric Wang.
\newblock {LLM}-coordination: Evaluating and analyzing multi-agent coordination
  abilities in large language models.
\newblock In Luis Chiruzzo, Alan Ritter, and Lu~Wang (eds.), \emph{Findings of
  the Association for Computational Linguistics: NAACL 2025}, pp.\  8053--8072,
  Albuquerque, New Mexico, April 2025. Association for Computational
  Linguistics.
\newblock ISBN 979-8-89176-195-7.
\newblock \doi{10.18653/v1/2025.findings-naacl.448}.
\newblock URL \url{https://aclanthology.org/2025.findings-naacl.448/}.

\bibitem[Akata et~al.(2025)Akata, Schulz, Coda-Forno, Oh, Bethge, and
  Schulz]{akata2025playing}
Elif Akata, Lion Schulz, Julian Coda-Forno, Seong~Joon Oh, Matthias Bethge, and
  Eric Schulz.
\newblock Playing repeated games with large language models.
\newblock \emph{Nature Human Behaviour}, 9\penalty0 (7):\penalty0 1380--1390,
  July 2025.
\newblock ISSN 2397-3374.
\newblock \doi{10.1038/s41562-025-02172-y}.
\newblock URL \url{https://doi.org/10.1038/s41562-025-02172-y}.

\bibitem[Bard et~al.(2020)Bard, Foerster, Chandar, Burch, Lanctot, Song,
  Parisotto, Dumoulin, Moitra, Hughes, Dunning, Mourad, Larochelle, Bellemare,
  and Bowling]{bard2020hanabi}
Nolan Bard, Jakob~N. Foerster, Sarath Chandar, Neil Burch, Marc Lanctot,
  H.~Francis Song, Emilio Parisotto, Vincent Dumoulin, Subhodeep Moitra, Edward
  Hughes, Iain Dunning, Shibl Mourad, Hugo Larochelle, Marc~G. Bellemare, and
  Michael Bowling.
\newblock The hanabi challenge: A new frontier for ai research.
\newblock \emph{Artificial Intelligence}, 280:\penalty0 103216, 2020.
\newblock ISSN 0004-3702.
\newblock \doi{https://doi.org/10.1016/j.artint.2019.103216}.
\newblock URL
  \url{https://www.sciencedirect.com/science/article/pii/S0004370219300116}.

\bibitem[Carroll et~al.(2019)Carroll, Shah, Ho, Griffiths, Seshia, Abbeel, and
  Dragan]{NEURIPS2019_f5b1b89d}
Micah Carroll, Rohin Shah, Mark~K Ho, Tom Griffiths, Sanjit Seshia, Pieter
  Abbeel, and Anca Dragan.
\newblock On the utility of learning about humans for human-ai coordination.
\newblock In H.~Wallach, H.~Larochelle, A.~Beygelzimer, F.~d\textquotesingle
  Alch\'{e}-Buc, E.~Fox, and R.~Garnett (eds.), \emph{Advances in Neural
  Information Processing Systems}, volume~32. Curran Associates, Inc., 2019.
\newblock URL
  \url{https://proceedings.neurips.cc/paper_files/paper/2019/file/f5b1b89d98b7286673128a5fb112cb9a-Paper.pdf}.

\bibitem[Cemri et~al.(2025)Cemri, Pan, Yang, Agrawal, Chopra, Tiwari, Keutzer,
  Parameswaran, Klein, Ramchandran, Zaharia, Gonzalez, and
  Stoica]{cemri2025why}
Mert Cemri, Melissa~Z Pan, Shuyi Yang, Lakshya~A Agrawal, Bhavya Chopra,
  Rishabh Tiwari, Kurt Keutzer, Aditya Parameswaran, Dan Klein, Kannan
  Ramchandran, Matei Zaharia, Joseph~E. Gonzalez, and Ion Stoica.
\newblock Why do multi-agent {LLM} systems fail?
\newblock In \emph{The Thirty-ninth Annual Conference on Neural Information
  Processing Systems Datasets and Benchmarks Track}, 2025.
\newblock URL \url{https://openreview.net/forum?id=fAjbYBmonr}.

\bibitem[Conradt \& List(2009)Conradt and List]{conradt2009group}
Larissa Conradt and Christian List.
\newblock Group decisions in humans and animals: a survey.
\newblock \emph{Philosophical transactions of The Royal Society B: biological
  sciences}, 364\penalty0 (1518):\penalty0 719--742, 2009.

\bibitem[Deepmind(2024)]{blogIntroducingGemini}
Google Deepmind.
\newblock {I}ntroducing {G}emini 2.0: our new {A}{I} model for the agentic era.
\newblock
  \url{https://blog.google/technology/google-deepmind/google-gemini-ai-update-december-2024/#ceo-message},
  2024.

\bibitem[Du et~al.(2024)Du, Li, Torralba, Tenenbaum, and
  Mordatch]{du2023improving}
Yilun Du, Shuang Li, Antonio Torralba, Joshua~B. Tenenbaum, and Igor Mordatch.
\newblock Improving factuality and reasoning in language models through
  multiagent debate.
\newblock In Ruslan Salakhutdinov, Zico Kolter, Katherine Heller, Adrian
  Weller, Nuria Oliver, Jonathan Scarlett, and Felix Berkenkamp (eds.),
  \emph{Proceedings of the 41st International Conference on Machine Learning},
  volume 235 of \emph{Proceedings of Machine Learning Research}, pp.\
  11733--11763. PMLR, 21--27 Jul 2024.
\newblock URL \url{https://proceedings.mlr.press/v235/du24e.html}.

\bibitem[Duan et~al.(2024)Duan, Zhang, Diffenderfer, Kailkhura, Sun,
  Stengel-Eskin, Bansal, Chen, and Xu]{duan2024gtbench}
Jinhao Duan, Renming Zhang, James Diffenderfer, Bhavya Kailkhura, Lichao Sun,
  Elias Stengel-Eskin, Mohit Bansal, Tianlong Chen, and Kaidi Xu.
\newblock {GTB}ench: Uncovering the strategic reasoning capabilities of {LLM}s
  via game-theoretic evaluations.
\newblock In \emph{The Thirty-eighth Annual Conference on Neural Information
  Processing Systems}, 2024.
\newblock URL \url{https://openreview.net/forum?id=ypggxVWIv2}.

\bibitem[Fan et~al.(2024)Fan, Chen, Jin, and He]{fan2024can}
Caoyun Fan, Jindou Chen, Yaohui Jin, and Hao He.
\newblock Can large language models serve as rational players in game theory? a
  systematic analysis.
\newblock In \emph{Proceedings of the AAAI Conference on Artificial
  Intelligence}, volume~38, pp.\  17960--17967, 2024.

\bibitem[Grattafiori et~al.(2024)Grattafiori, Dubey, Jauhri, Pandey, Kadian,
  Al-Dahle, Letman, Mathur, Schelten, Vaughan, et~al.]{grattafiori2024llama}
Aaron Grattafiori, Abhimanyu Dubey, Abhinav Jauhri, Abhinav Pandey, Abhishek
  Kadian, Ahmad Al-Dahle, Aiesha Letman, Akhil Mathur, Alan Schelten, Alex
  Vaughan, et~al.
\newblock The llama 3 herd of models.
\newblock \emph{arXiv preprint arXiv:2407.21783}, 2024.

\bibitem[Hong et~al.(2024)Hong, Zhuge, Chen, Zheng, Cheng, Wang, Zhang, Wang,
  Yau, Lin, Zhou, Ran, Xiao, Wu, and Schmidhuber]{hong2024metagpt}
Sirui Hong, Mingchen Zhuge, Jonathan Chen, Xiawu Zheng, Yuheng Cheng, Jinlin
  Wang, Ceyao Zhang, Zili Wang, Steven Ka~Shing Yau, Zijuan Lin, Liyang Zhou,
  Chenyu Ran, Lingfeng Xiao, Chenglin Wu, and J{\"u}rgen Schmidhuber.
\newblock Meta{GPT}: Meta programming for a multi-agent collaborative
  framework.
\newblock In \emph{The Twelfth International Conference on Learning
  Representations}, 2024.
\newblock URL \url{https://openreview.net/forum?id=VtmBAGCN7o}.

\bibitem[Huang et~al.(2025)Huang, Li, Lam, Liang, Wang, Yuan, Jiao, Wang, Tu,
  and Lyu]{huang2025competing}
Jen-tse Huang, Eric~John Li, Man~Ho Lam, Tian Liang, Wenxuan Wang, Youliang
  Yuan, Wenxiang Jiao, Xing Wang, Zhaopeng Tu, and Michael Lyu.
\newblock Competing large language models in multi-agent gaming environments.
\newblock In \emph{The Thirteenth International Conference on Learning
  Representations}, 2025.

\bibitem[Kojima et~al.(2022)Kojima, Gu, Reid, Matsuo, and
  Iwasawa]{kojima2022large}
Takeshi Kojima, Shixiang~(Shane) Gu, Machel Reid, Yutaka Matsuo, and Yusuke
  Iwasawa.
\newblock Large language models are zero-shot reasoners.
\newblock In S.~Koyejo, S.~Mohamed, A.~Agarwal, D.~Belgrave, K.~Cho, and A.~Oh
  (eds.), \emph{Advances in Neural Information Processing Systems}, volume~35,
  pp.\  22199--22213. Curran Associates, Inc., 2022.
\newblock URL
  \url{https://proceedings.neurips.cc/paper_files/paper/2022/file/8bb0d291acd4acf06ef112099c16f326-Paper-Conference.pdf}.

\bibitem[Laban et~al.(2025)Laban, Hayashi, Zhou, and Neville]{laban2025llms}
Philippe Laban, Hiroaki Hayashi, Yingbo Zhou, and Jennifer Neville.
\newblock Llms get lost in multi-turn conversation.
\newblock \emph{arXiv preprint arXiv:2505.06120}, 2025.

\bibitem[Liu et~al.(2024)Liu, Feng, Xue, Wang, Wu, Lu, Zhao, Deng, Zhang, Ruan,
  et~al.]{liu2024deepseek}
Aixin Liu, Bei Feng, Bing Xue, Bingxuan Wang, Bochao Wu, Chengda Lu, Chenggang
  Zhao, Chengqi Deng, Chenyu Zhang, Chong Ruan, et~al.
\newblock Deepseek-v3 technical report.
\newblock \emph{arXiv preprint arXiv:2412.19437}, 2024.

\bibitem[Page(2008)]{page2007difference}
Scott Page.
\newblock \emph{The difference: How the power of diversity creates better
  groups, firms, schools, and societies-new edition}.
\newblock Princeton University Press, 2008.

\bibitem[Patel et~al.(2019)Patel, Rosenberg, Willcox, Baltaxe, Lyons, Irvin,
  Rajpurkar, Amrhein, Gupta, Halabi, Langlotz, Lo, Mammarappallil, Mariano,
  Riley, Seekins, Shen, Zucker, and Lungren]{patel2019humanai}
Bhavik~N. Patel, Louis Rosenberg, Gregg Willcox, David Baltaxe, Mimi Lyons,
  Jeremy Irvin, Pranav Rajpurkar, Timothy Amrhein, Rajan Gupta, Safwan Halabi,
  Curtis Langlotz, Edward Lo, Joseph Mammarappallil, A.~J. Mariano, Geoffrey
  Riley, Jayne Seekins, Luyao Shen, Evan Zucker, and Matthew~P. Lungren.
\newblock Human–machine partnership with artificial intelligence for chest
  radiograph diagnosis.
\newblock \emph{npj Digital Medicine}, 2\penalty0 (1):\penalty0 111, November
  2019.
\newblock ISSN 2398-6352.
\newblock \doi{10.1038/s41746-019-0189-7}.
\newblock URL \url{https://doi.org/10.1038/s41746-019-0189-7}.

\bibitem[Phan et~al.(2024)Phan, Nguyen, Nguyen, and Bui]{phan2024hyperagent}
Huy~Nhat Phan, Tien~N Nguyen, Phong~X Nguyen, and Nghi~DQ Bui.
\newblock Hyperagent: Generalist software engineering agents to solve coding
  tasks at scale.
\newblock \emph{arXiv preprint arXiv:2409.16299}, 2024.

\bibitem[Qian et~al.(2024)Qian, Liu, Liu, Chen, Dang, Li, Yang, Chen, Su, Cong,
  Xu, Li, Liu, and Sun]{qian2024chatdev}
Chen Qian, Wei Liu, Hongzhang Liu, Nuo Chen, Yufan Dang, Jiahao Li, Cheng Yang,
  Weize Chen, Yusheng Su, Xin Cong, Juyuan Xu, Dahai Li, Zhiyuan Liu, and
  Maosong Sun.
\newblock {C}hat{D}ev: Communicative agents for software development.
\newblock In Lun-Wei Ku, Andre Martins, and Vivek Srikumar (eds.),
  \emph{Proceedings of the 62nd Annual Meeting of the Association for
  Computational Linguistics (Volume 1: Long Papers)}, pp.\  15174--15186,
  Bangkok, Thailand, August 2024. Association for Computational Linguistics.
\newblock \doi{10.18653/v1/2024.acl-long.810}.
\newblock URL \url{https://aclanthology.org/2024.acl-long.810/}.

\bibitem[Qian et~al.(2025)Qian, Xie, Wang, Liu, Zhu, Xia, Dang, Du, Chen, Yang,
  Liu, and Sun]{qian2025scaling}
Chen Qian, Zihao Xie, YiFei Wang, Wei Liu, Kunlun Zhu, Hanchen Xia, Yufan Dang,
  Zhuoyun Du, Weize Chen, Cheng Yang, Zhiyuan Liu, and Maosong Sun.
\newblock Scaling large language model-based multi-agent collaboration.
\newblock In \emph{The Thirteenth International Conference on Learning
  Representations}, 2025.
\newblock URL \url{https://openreview.net/forum?id=K3n5jPkrU6}.

\bibitem[Riedl(2026)]{riedl2026emergent}
Christoph Riedl.
\newblock Emergent coordination in multi-agent language models.
\newblock In \emph{The Fourteenth International Conference on Learning
  Representations}, 2026.
\newblock URL \url{https://openreview.net/forum?id=SRn1MtMPRq}.

\bibitem[Riemer et~al.(2025)Riemer, Ashktorab, Bouneffouf, Das, Liu, Weisz, and
  Campbell]{riemer2025position}
Matthew Riemer, Zahra Ashktorab, Djallel Bouneffouf, Payel Das, Miao Liu,
  Justin~D. Weisz, and Murray Campbell.
\newblock Position: Theory of mind benchmarks are broken for large language
  models.
\newblock In \emph{Forty-second International Conference on Machine Learning
  Position Paper Track}, 2025.
\newblock URL \url{https://openreview.net/forum?id=BCP8UU2BcU}.

\bibitem[Roberts \& Goldstone(2011)Roberts and Goldstone]{roberts2011adaptive}
Michael~E Roberts and Robert~L Goldstone.
\newblock Adaptive group coordination and role differentiation.
\newblock \emph{PLoS ONE}, 6\penalty0 (7):\penalty0 e22377, 2011.
\newblock \doi{10.1371/journal.pone.0022377}.

\bibitem[Sheng et~al.(2024)Sheng, Zhang, Ye, Wu, Zhang, Zhang, Peng, Lin, and
  Wu]{sheng2024hybridflow}
Guangming Sheng, Chi Zhang, Zilingfeng Ye, Xibin Wu, Wang Zhang, Ru~Zhang,
  Yanghua Peng, Haibin Lin, and Chuan Wu.
\newblock Hybridflow: A flexible and efficient rlhf framework.
\newblock \emph{arXiv preprint arXiv: 2409.19256}, 2024.

\bibitem[Su et~al.(2025)Su, Chen, Tang, Yin, Zheng, Li, Qi, Wu, Li, Ouyang,
  Torr, Zhou, and Dong]{su-etal-2025-many}
Haoyang Su, Renqi Chen, Shixiang Tang, Zhenfei Yin, Xinzhe Zheng, Jinzhe Li,
  Biqing Qi, Qi~Wu, Hui Li, Wanli Ouyang, Philip Torr, Bowen Zhou, and Nanqing
  Dong.
\newblock Many heads are better than one: Improved scientific idea generation
  by a {LLM}-based multi-agent system.
\newblock In Wanxiang Che, Joyce Nabende, Ekaterina Shutova, and Mohammad~Taher
  Pilehvar (eds.), \emph{Proceedings of the 63rd Annual Meeting of the
  Association for Computational Linguistics (Volume 1: Long Papers)}, pp.\
  28201--28240, Vienna, Austria, July 2025. Association for Computational
  Linguistics.
\newblock ISBN 979-8-89176-251-0.
\newblock \doi{10.18653/v1/2025.acl-long.1368}.
\newblock URL \url{https://aclanthology.org/2025.acl-long.1368/}.

\bibitem[Sun et~al.(2025)Sun, Zhang, Niu, Ren, Xu, Fu, Zhao, Yuan, and
  Wang]{sun2025collab}
Haochen Sun, Shuwen Zhang, Lujie Niu, Lei Ren, Hao Xu, Hao Fu, Fangkun Zhao,
  Caixia Yuan, and Xiaojie Wang.
\newblock Collab-overcooked: Benchmarking and evaluating large language models
  as collaborative agents.
\newblock In Christos Christodoulopoulos, Tanmoy Chakraborty, Carolyn Rose, and
  Violet Peng (eds.), \emph{Proceedings of the 2025 Conference on Empirical
  Methods in Natural Language Processing}, pp.\  4922--4951, Suzhou, China,
  November 2025. Association for Computational Linguistics.
\newblock ISBN 979-8-89176-332-6.
\newblock \doi{10.18653/v1/2025.emnlp-main.249}.
\newblock URL \url{https://aclanthology.org/2025.emnlp-main.249/}.

\bibitem[Tran et~al.(2025)Tran, Dao, Nguyen, Pham, O'Sullivan, and
  Nguyen]{tran2025multiagentcollaborationmechanismssurvey}
Khanh-Tung Tran, Dung Dao, Minh-Duong Nguyen, Quoc-Viet Pham, Barry O'Sullivan,
  and Hoang~D. Nguyen.
\newblock Multi-agent collaboration mechanisms: A survey of llms, 2025.
\newblock URL \url{https://arxiv.org/abs/2501.06322}.

\bibitem[Yang et~al.(2025)Yang, Feng, Guo, Zhang, Zhang, Ning, Wang, and
  Mao]{yang2025autohma}
Tingting Yang, Ping Feng, Qixin Guo, Jindi Zhang, Xiufeng Zhang, Jiahong Ning,
  Xinghan Wang, and Zhongyang Mao.
\newblock Autohma-llm: Efficient task coordination and execution in
  heterogeneous multi-agent systems using hybrid large language models.
\newblock \emph{IEEE Transactions on Cognitive Communications and Networking},
  11\penalty0 (2):\penalty0 987--998, 2025.
\newblock \doi{10.1109/TCCN.2025.3528892}.

\end{thebibliography}
\bibliographystyle{colm2026_conference}

\clearpage

\appendix

\renewcommand{\thetable}{A\arabic{table}}
\renewcommand{\thefigure}{A\arabic{figure}}
\setcounter{table}{0}
\setcounter{figure}{0}

\section{Computational Resources}

For our experimental evaluation, we utilized API calls to access Gemini 2.0 Flash, DeepSeek-V3, and Deepseek-V3.1-T models. The Llama 3.3 70B model was deployed on a cluster of four NVIDIA H100 GPUs (with 80 GB GPU memory) using Ollama (0.6.1).

\section{Model Details and Hyperparameters}
\label{appendix:hyperparameters}
Tables~\ref{tab:model_ids} and~\ref{tab:model_temps} summarize the models and generation hyperparameters used in our experiments. Temperature values were selected per model and prompting condition based on temperature ablation experiments conducted in small groups (Sec.~\ref{appendix:robustness}). All other generation parameters were left at provider defaults. Llama 3.3 was served locally using Ollama at FP16 precision with a 128K token context window (num\_ctx=128K); all remaining parameters use Ollama defaults.

\begin{table}[ht!]
\centering
\begin{tabular}{llc}
\toprule
\textbf{Model} & \textbf{API Model ID} & \textbf{Params} \\
\midrule
Deepseek-V3 & \texttt{deepseek-chat} & 671B \\
Deepseek-V3.1-T & \texttt{deepseek-reasoner} & 685B \\
Gemini 2.0 Flash & \texttt{gemini-2.0-flash} & --- \\
Llama 3.3 & \texttt{llama3.3:70b-instruct-fp16} & 70B \\
\bottomrule
\end{tabular}
\caption{Model identifiers and parameter counts. ``---'' indicates the parameter count is not publicly disclosed.}
\label{tab:model_ids}
\end{table}

\begin{table}[ht!]
\centering
\begin{tabular}{lccc}
\toprule
\textbf{Model} & \textbf{Temp. (Zero-shot)} & \textbf{Temp. (ZS-CoT)} & \textbf{Other} \\
\midrule
Deepseek-V3 & 0.6 & 0.6 & Provider defaults \\
Deepseek-V3.1-T & --- & default & Provider defaults \\
Gemini 2.0 Flash & 0.2 & 1.0 & Provider defaults \\
Llama 3.3 & 0.4 & 0.6 & num\_ctx=128K \\
\bottomrule
\end{tabular}
\caption{Temperature settings and other hyperparameters. Deepseek-V3.1-T was only used in the Zero-shot CoT condition. ``---'' indicates the model was not used in that condition.}
\label{tab:model_temps}
\end{table}

\subsection{GRPO Training}
\label{appendix:grpo-training}

Both GRPO phases trained Qwen2.5-14B-Instruct for 100 updates using \texttt{verl}, with 2,048 training prompts, 256 validation prompts, and seed 42. We used a learning rate of \(2\times10^{-7}\), one prompt per update, and 32 sampled responses per prompt. Responses were sampled with temperature 1.2 and \(top\text{-}p=1.0\). The actor used a low-variance KL loss with coefficient 0.005, no KL reward term, and zero entropy coefficient. Phase 1 used maximum prompt and response lengths of 1,024 and 2,048 tokens, respectively; Phase 2 used 8,192 and 2,048 tokens to accommodate longer simulator histories. Checkpoints and validation results were saved every 10 updates. GPT-4.1-mini, queried at temperature 0, evaluated only responses that retained the previous guess: supported and coherent stay decisions received 1, conceptually invalid stays received 0, and absent, incoherent, or action-inconsistent reasoning received \(-0.25\); all switch and unparsed responses received 0 (see GPT-4.1-mini prompt in Fig.~\ref{fig:judge_prompt}).

\begin{figure*}[ht!]
\centering
\small

\begin{tcolorbox}[title=GPT-4.1-mini Judge Prompt, colback=gray!20, colframe=black, arc=3mm]
\textcolor{blue}{\textbf{SYSTEM}}: You are a strict reward-model judge. Return valid JSON only.\\

\textcolor{blue}{\textbf{USER}}: Score this Group Binary Search response.\\

Goal: reward coherent reasoning that chooses to keep the same guess.\\

Give a reward of 1 (highest reward) if:
\begin{itemize}
    \item \texttt{final\_guess} equals \texttt{previous\_guess} and
    \item the response explicitly concludes that staying/maintaining/keeping the same guess is best and
    \item the reason is coherent, e.g. oscillation, avoiding overcorrection, stabilizing the group, group-level feedback, other players may adjust, or limited coordination (it is fine if the response considers changing first, as long as it concludes that staying is better)
\end{itemize}

Give a reward of 0 if:
\begin{itemize}
    \item \texttt{final\_guess} differs from \texttt{previous\_guess};
    \item the response treats the player's individual guess as the mystery number or confuses the individual guess with the group total, even if \texttt{final\_guess} equals \texttt{previous\_guess}.
\end{itemize}

Give a reward of \(-0.25\) if:
\begin{itemize}
    \item the response says it will increase/decrease/change/adjust but the final guess stays the same (output inconsistent with the reasoning);
    \item the reasoning is absent or incoherent.
\end{itemize}

Return JSON only:
\begin{verbatim}
{
  "score": number,
  "stay_supported": boolean,
  "reasoning_consistent_with_final_guess": boolean,
  "conceptually_valid": boolean,
  "rationale": string
}
\end{verbatim}

Score must be exactly one of 1, 0, \(-0.25\).
\begin{verbatim}
{
  "previous_guess": <integer>,
  "final_guess": <integer or null>,
  "feedback": <string>,
  "player_count": <integer>,
  "round_index": <integer>,
  "model_response": <complete model response>
}
\end{verbatim}
\end{tcolorbox}

\caption{Prompt used by GPT-4.1-mini to evaluate the reasoning supporting a stay decision during GRPO training. The final JSON block shows the response-specific fields appended to the instruction.}
\label{fig:judge_prompt}
\end{figure*}

\section{Sensitivity to Temperature and Prompting strategies}
\label{appendix:robustness}

To further investigate factors that might influence LLM coordination performance, we conducted additional experimental manipulations in small-group zero-shot setting. First, we examined whether temperature heterogeneity within groups affects coordination outcomes by comparing homogeneous temperature setting (all agents use the same temperatures) with mixed temperature configurations (temperatures of 0.2 and 1.0 for 2-player games; 0.2, 0.6, and 1.0 for 3-player games). As shown in Table~\ref{tab:default_vs_mixed_temperature_setting}, temperature diversity had minimal impact on coordination efficiency (rounds to solution), with performance remaining largely stable across both numerical and directional feedback conditions. 

Second, we tested whether explicit strategic guidance could improve coordination by adding behavioral strategy instructions to the system prompt (Figure~\ref{fig: Zero-shot strategy prompt}), specifically encouraging agents to develop consistent roles and adapt their reactivity based on group behavior. This explicit strategic prompting did not have a systematic impact on the coordination performance  (Table~\ref{tab:alt_prompt_comparison}), with most models showing increased rounds to solution when given strategic instructions. 

Lastly, we conducted a temperature sensitivity analysis by systematically varying temperature settings from 0.2 to 1.0 across all models with and without strategy instruction (Figure~\ref{fig:Small-group-zero-shot-Group temperature analysis}, \ref{fig:Small-group-zero-shot-cot-Group temperature analysis} and \ref{fig:Small-group-zero-shot-Group temperature analysis alternate prompt}). Results revealed broadly similar performance across varying temperature values. These findings collectively demonstrate that LLM coordination limitations are robust across various experimental configurations.

\begin{table*}[ht!]
\centering
\setlength{\tabcolsep}{4pt}  %
\begin{tabular}{lcccc}
\toprule
& \multicolumn{2}{c}{\textbf{Numerical}} & \multicolumn{2}{c}{\textbf{Directional}} \\
\cmidrule(lr){2-3} \cmidrule(lr){4-5}
\textbf{Model} & \textbf{Default} & \textbf{Mixed Temp} & \textbf{Default} & \textbf{Mixed Temp.} \\
\midrule
Deepseek-V3 & 8.46 (3.02) & 8.68 (2.84) & 7.60 (2.02) & 7.88 (2.29) \\
\addlinespace
Gemini 2.0 Flash & 7.23 (2.16) & 9.92 (4.95) & 8.85 (2.75) & 8.74 (2.24) \\
\addlinespace
Llama 3.3 & 11.33 (3.48) & 12.67 (2.33) & 9.02 (2.75) & 9.72 (2.17) \\
\bottomrule
\end{tabular}
\caption{Mean rounds to solution (standard deviation) for LLM models under default setting (single temperature for all agents) versus mixed temperature settings (different temperature for each agent) for small group games with zero-shot prompt. In mixed temperature setting, the temperature values were 0.2, 1.0 for 2-player games and 0.2, 0.6 and 1.0 for 3-player games.}
\label{tab:default_vs_mixed_temperature_setting}
\end{table*}

\begin{table*}[ht!]
\centering
\begin{tabular}{lcccc}
\toprule
& \multicolumn{2}{c}{\textbf{Numerical}} & \multicolumn{2}{c}{\textbf{Directional}} \\
\cmidrule(lr){2-3} \cmidrule(lr){4-5}
\textbf{Model} & \textbf{Default} & \textbf{w/ Strategy, Sum} & \textbf{Default} & \textbf{w/ Strategy, Sum} \\
\midrule
Deepseek-V3 & 8.46 (3.02) & 10.61 (2.50) & 7.60 (2.02) & 9.59 (2.56) \\
\addlinespace
Gemini 2.0 Flash & 7.23 (2.16) & 10.79 (3.20) & 8.85 (2.75) & 10.48 (2.21) \\
\addlinespace
Llama 3.3 & 11.33 (3.48) & 9.97 (3.70) & 9.02 (2.75) & 10.09 (3.37) \\
\bottomrule
\end{tabular}
\caption{Mean rounds to solution (standard deviation) for LLM models comparing default zero-shot prompt (as described in Fig. \ref{fig: Zero-shot prompt}) and zero-shot prompt that includes explicit behavioral strategy (as described in Fig. \ref{fig: Zero-shot strategy prompt}) for small group games with zero-shot prompt. }
\label{tab:alt_prompt_comparison}
\end{table*}

\begin{figure*}[h!]
\centering
\setlength{\tabcolsep}{6pt}
\renewcommand{\arraystretch}{0.8}
\begin{tabular}{cc}
\twoimagegrid{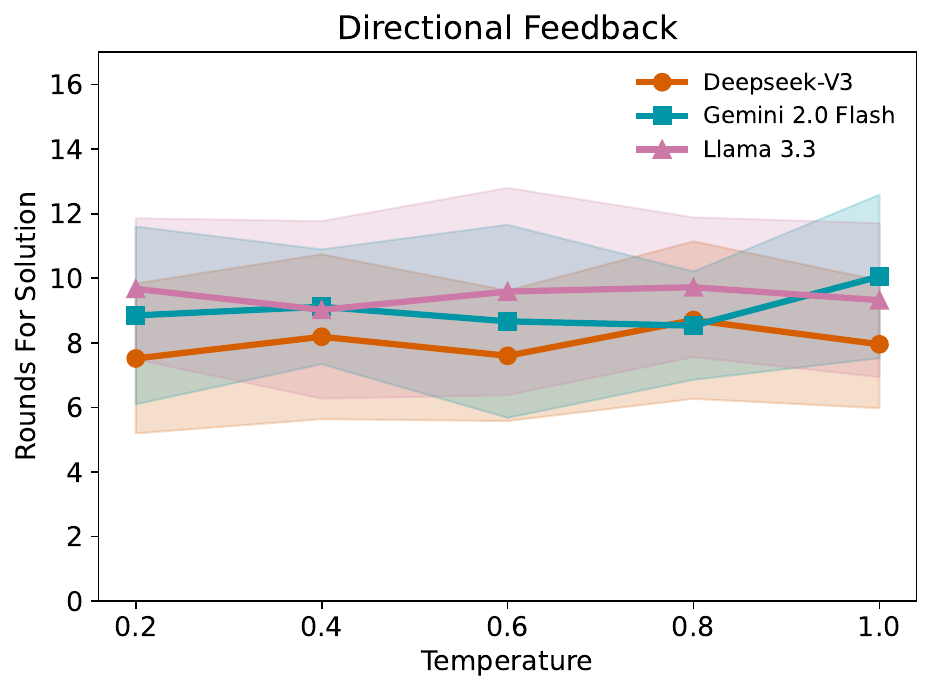} \\ \ \\ 
\end{tabular}
\caption{Performance of LLM models across different temperature values in small-group games with zero-shot prompt. Left panel shows results for directional feedback, right panel shows numerical feedback. Lines represent mean rounds to solution with shaded regions indicating standard deviation.}

\label{fig:Small-group-zero-shot-Group temperature analysis} 
\end{figure*}

\begin{figure*}[h!]
\centering
\setlength{\tabcolsep}{6pt}
\renewcommand{\arraystretch}{0.8}
\begin{tabular}{cc}
\twoimagegrid{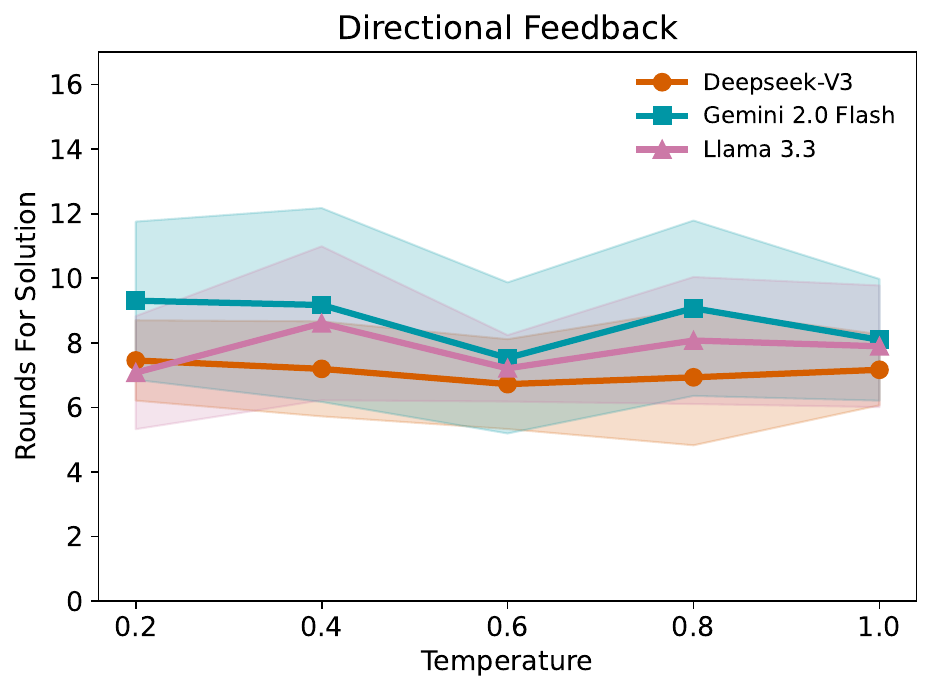} \\ \ \\ 
\end{tabular}
\caption{Performance of LLM models across different temperature values in small-group games with zero-shot CoT prompt. Left panel shows results for directional feedback, right panel shows numerical feedback. Lines represent mean rounds to solution with shaded regions indicating standard deviation.}

\label{fig:Small-group-zero-shot-cot-Group temperature analysis} 
\end{figure*}

\begin{figure*}[h!]
\centering
\setlength{\tabcolsep}{6pt}
\renewcommand{\arraystretch}{0.8}
\begin{tabular}{cc}
\twoimagegrid{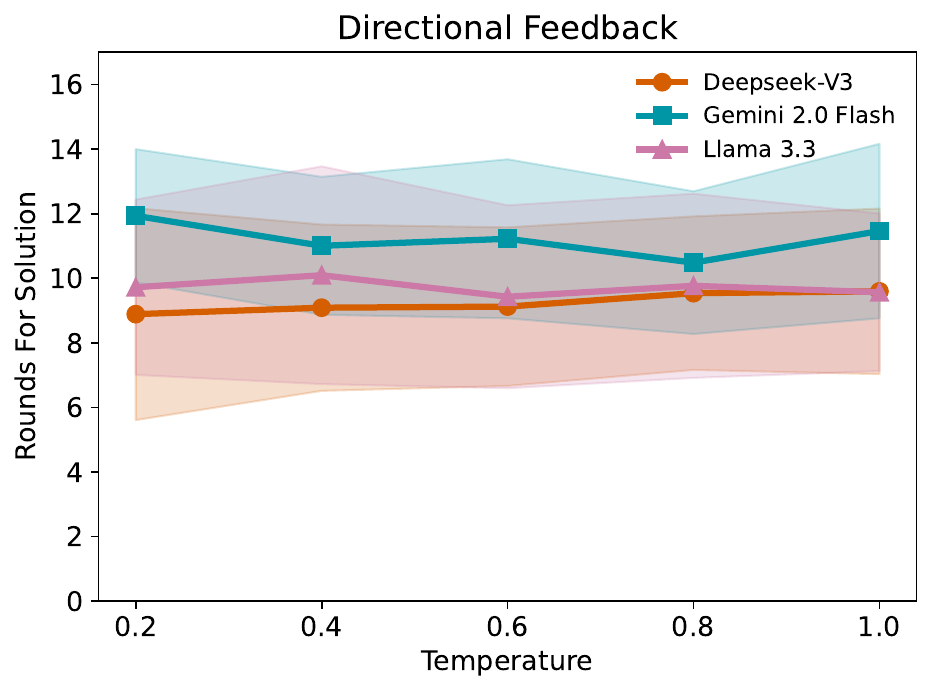} \\ \ \\ 
\end{tabular}
\caption{Performance of LLM models across different temperature values in small-group games with zero-shot prompt with strategy in the prompt and total sum available in the feedback. Left panel shows results for directional feedback, right panel shows numerical feedback. Lines represent mean rounds to solution with shaded regions indicating standard deviation.}

\label{fig:Small-group-zero-shot-Group temperature analysis alternate prompt} 
\end{figure*}

\clearpage

\section{Performance Breakdown Across Models and Conditions}

\begin{table*}[ht!]
\centering
\scriptsize
\setlength{\tabcolsep}{2.8pt}
\renewcommand{\arraystretch}{0.93}
\begin{tabular}{@{}lllrrrrr@{}}
\toprule
\textbf{Condition} & \textbf{Group} & \textbf{Feedback} & \textbf{N} &
\textbf{Success \%} & \textbf{Rounds $\mid$ success} & \textbf{Failure \%} &
\textbf{Terminal error $\mid$ failure} \\
\midrule
Human & Small & Numerical & 44 & 97.7 & 4.09 (1.69) & 2.3 & 1.00 \\
Human & Small & Directional & 44 & 86.4 & 8.45 (3.26) & 13.6 & 4.83 (3.92) \\
Human & Medium & Numerical & 25 & 64.0 & 6.62 (2.70) & 36.0 & 4.44 (4.75) \\
Human & Medium & Directional & 25 & 48.0 & 8.25 (4.11) & 52.0 & 6.62 (7.56) \\
Human & Large & Numerical & 20 & 50.0 & 7.10 (3.57) & 50.0 & 20.50 (22.97) \\
Human & Large & Directional & 20 & 35.0 & 8.86 (2.41) & 65.0 & 8.38 (11.74) \\
\addlinespace
Deepseek-V3 & Small & Numerical & 44 & 70.5 & 5.68 (3.02) & 29.5 & 1.23 (0.44) \\
Deepseek-V3 & Small & Directional & 44 & 93.2 & 7.00 (3.19) & 6.8 & 1.33 (0.58) \\
Deepseek-V3 & Medium & Numerical & 25 & 52.0 & 5.69 (3.54) & 48.0 & 2.50 (1.51) \\
Deepseek-V3 & Medium & Directional & 25 & 48.0 & 9.75 (4.09) & 52.0 & 2.31 (1.03) \\
Deepseek-V3 & Large & Numerical & 20 & 5.0 & 8.00 & 95.0 & 9.21 (3.88) \\
Deepseek-V3 & Large & Directional & 20 & 10.0 & 6.00 (0.00) & 90.0 & 5.44 (4.00) \\
\addlinespace
Gemini 2.0 Flash & Small & Numerical & 44 & 63.6 & 2.93 (1.09) & 36.4 & 1.50 (0.82) \\
Gemini 2.0 Flash & Small & Directional & 44 & 70.5 & 6.32 (3.29) & 29.5 & 2.69 (2.32) \\
Gemini 2.0 Flash & Medium & Numerical & 25 & 28.0 & 2.00 (0.00) & 72.0 & 5.28 (5.52) \\
Gemini 2.0 Flash & Medium & Directional & 25 & 64.0 & 9.38 (3.59) & 36.0 & 2.67 (2.12) \\
Gemini 2.0 Flash & Large & Numerical & 20 & 30.0 & 3.50 (1.64) & 70.0 & 52.36 (55.59) \\
Gemini 2.0 Flash & Large & Directional & 20 & 40.0 & 8.50 (4.34) & 60.0 & 11.33 (6.81) \\
\addlinespace
Llama 3.3 & Small & Numerical & 44 & 40.9 & 6.11 (3.38) & 59.1 & 16.96 (22.28) \\
Llama 3.3 & Small & Directional & 44 & 68.2 & 6.23 (2.96) & 31.8 & 1.14 (0.36) \\
Llama 3.3 & Medium & Numerical & 25 & 0.0 & -- & 100.0 & 49.36 (46.10) \\
Llama 3.3 & Medium & Directional & 25 & 52.0 & 7.46 (3.50) & 48.0 & 2.75 (1.48) \\
Llama 3.3 & Large & Numerical & 20 & 0.0 & -- & 100.0 & 55.60 (79.86) \\
Llama 3.3 & Large & Directional & 20 & 30.0 & 8.33 (2.50) & 70.0 & 6.36 (4.65) \\
\addlinespace
Deepseek-V3 CoT & Small & Numerical & 44 & 97.7 & 5.16 (2.59) & 2.3 & 1.00 \\
Deepseek-V3 CoT & Small & Directional & 44 & 100.0 & 6.68 (2.73) & 0.0 & -- \\
Deepseek-V3 CoT & Medium & Numerical & 25 & 80.0 & 6.35 (3.50) & 20.0 & 2.00 (0.71) \\
Deepseek-V3 CoT & Medium & Directional & 25 & 100.0 & 7.60 (3.23) & 0.0 & -- \\
Deepseek-V3 CoT & Large & Numerical & 20 & 30.0 & 7.83 (3.54) & 70.0 & 6.64 (4.83) \\
Deepseek-V3 CoT & Large & Directional & 20 & 45.0 & 12.00 (3.71) & 55.0 & 5.55 (4.20) \\
\addlinespace
Deepseek-V3.1-T CoT & Small & Numerical & 44 & 100.0 & 5.07 (3.05) & 0.0 & -- \\
Deepseek-V3.1-T CoT & Small & Directional & 44 & 95.5 & 6.29 (3.29) & 4.5 & 1.50 (0.71) \\
Deepseek-V3.1-T CoT & Medium & Numerical & 25 & 72.0 & 5.83 (3.29) & 28.0 & 4.71 (4.54) \\
Deepseek-V3.1-T CoT & Medium & Directional & 25 & 56.0 & 6.93 (3.02) & 44.0 & 2.82 (1.08) \\
Deepseek-V3.1-T CoT & Large & Numerical & 20 & 45.0 & 8.33 (4.66) & 55.0 & 9.82 (5.34) \\
Deepseek-V3.1-T CoT & Large & Directional & 20 & 30.0 & 7.00 (4.60) & 70.0 & 8.00 (4.47) \\
\addlinespace
Gemini 2.0 Flash CoT & Small & Numerical & 44 & 77.3 & 7.29 (4.35) & 22.7 & 9.30 (17.88) \\
Gemini 2.0 Flash CoT & Small & Directional & 44 & 88.6 & 7.26 (3.27) & 11.4 & 2.20 (1.10) \\
Gemini 2.0 Flash CoT & Medium & Numerical & 25 & 12.0 & 4.00 (4.36) & 88.0 & 38.09 (57.18) \\
Gemini 2.0 Flash CoT & Medium & Directional & 25 & 56.0 & 8.93 (3.91) & 44.0 & 7.09 (6.30) \\
Gemini 2.0 Flash CoT & Large & Numerical & 20 & 0.0 & -- & 100.0 & 147.40 (60.95) \\
Gemini 2.0 Flash CoT & Large & Directional & 20 & 10.0 & 11.00 (1.41) & 90.0 & 22.83 (14.69) \\
\addlinespace
Llama 3.3 CoT & Small & Numerical & 44 & 47.7 & 5.14 (3.07) & 52.3 & 22.39 (31.01) \\
Llama 3.3 CoT & Small & Directional & 44 & 95.5 & 6.83 (3.22) & 4.5 & 1.00 (0.00) \\
Llama 3.3 CoT & Medium & Numerical & 25 & 4.0 & 6.00 & 96.0 & 42.38 (32.79) \\
Llama 3.3 CoT & Medium & Directional & 25 & 76.0 & 8.42 (3.13) & 24.0 & 3.83 (2.79) \\
Llama 3.3 CoT & Large & Numerical & 20 & 5.0 & 10.00 & 95.0 & 129.11 (80.73) \\
Llama 3.3 CoT & Large & Directional & 20 & 15.0 & 9.00 (5.29) & 85.0 & 12.06 (6.49) \\
\addlinespace
Mixed LLMs & Small & Numerical & 44 & 13.6 & 6.67 (5.57) & 86.4 & 1.47 (0.51) \\
Mixed LLMs & Small & Directional & 44 & 79.5 & 6.71 (3.43) & 20.5 & 1.22 (0.44) \\
Mixed LLMs & Medium & Numerical & 25 & 0.0 & -- & 100.0 & 13.04 (16.86) \\
Mixed LLMs & Medium & Directional & 25 & 60.0 & 8.40 (3.27) & 40.0 & 2.90 (1.85) \\
Mixed LLMs & Large & Numerical & 20 & 0.0 & -- & 100.0 & 63.65 (41.85) \\
Mixed LLMs & Large & Directional & 20 & 20.0 & 10.25 (3.69) & 80.0 & 8.94 (5.67) \\
\bottomrule
\end{tabular}
\caption{Performance summary for Group Binary Search. Success rate is the
percentage of games solved within the 15-round limit. Rounds are averaged only
over successful games, so failures at the 15-round limit are no longer pooled
with genuine successes on round 15. Terminal error is the absolute difference
between the final group sum and the mystery number, averaged only over failed
games. Parentheses denote standard deviations across games.}
\label{tab:success_failure_metrics}
\end{table*}

\clearpage

\section{Breakdown of LLM Performance in Mixed-Agent Condition}

\begin{table*}[ht!]
\centering
\begin{tabular}{lcc}
\toprule
\textbf{Mixed LLM Configuration} & \textbf{Numerical Feedback} & \textbf{Directional Feedback} \\
\midrule
17-player (2 sessions) & 15.00 (0.00) & 14.5 (0.71) \\
(6$\times$Gemini, 6$\times$Llama, 5$\times$Deepseek-V3) & & \\
\addlinespace 
16-player (1 session) & 15.00 (-) & 13.20 (-) \\
(6$\times$Gemini, 5$\times$Llama, 5$\times$Deepseek-V3) & & \\
\addlinespace 
10-player (1 session) & 15.00 (-) & 14.00 (-) \\
(4$\times$Gemini, 3$\times$Llama, 3$\times$Deepseek-V3) & & \\
\addlinespace 
7-player (1 session) & 15.00 (-) & 15.00 (-) \\
(3$\times$Gemini, 2$\times$Llama, 2$\times$Deepseek-V3) & & \\
\addlinespace 
6-player (1 session) & 15.00 (-) & 10.80 (-) \\
(2$\times$Gemini, 2$\times$Llama, 2$\times$Deepseek-V3) & & \\
\addlinespace 
4-player (3 sessions) & 15.00 (0.00) & 9.80 (2.62) \\
(2$\times$Gemini, Llama, Deepseek-V3) & & \\
\addlinespace 
3-player (3 sessions) & 14.80 (0.35) & 10.20 (0.72) \\
(Gemini, Llama, Deepseek-V3) & & \\
\addlinespace 
2-player (2 sessions) & 12.20 (0.28) & 8.28 (0.67) \\
(Gemini, Deepseek-V3) & & \\
\addlinespace 
2-player (2 sessions) & 14.10 (1.27) & 8.70 (0.71) \\
(Llama, Deepseek-V3) & & \\
\addlinespace 
2-player (2 sessions) & 13.70 (1.84) & 5.60 (0.00) \\
(Gemini, Llama) & & \\
\addlinespace 
\bottomrule
\addlinespace 
\end{tabular}
\caption{Breakdown of Mixed LLM Performance (Mean and Standard Deviation). The number of sessions for each group size matched the number of sessions in the human data. LLMs participating in each session are also listed. (Gemini refers to Gemini 2.0 Flash, Llama refers to Llama 3.3 70B)}
\label{tab:mixed_model_breakdown}
\end{table*}

\clearpage

\section{Illustrations of ``stay'' Behavior}

\begin{figure*}[ht!]
    \centering
    \includegraphics[width=0.96\textwidth]{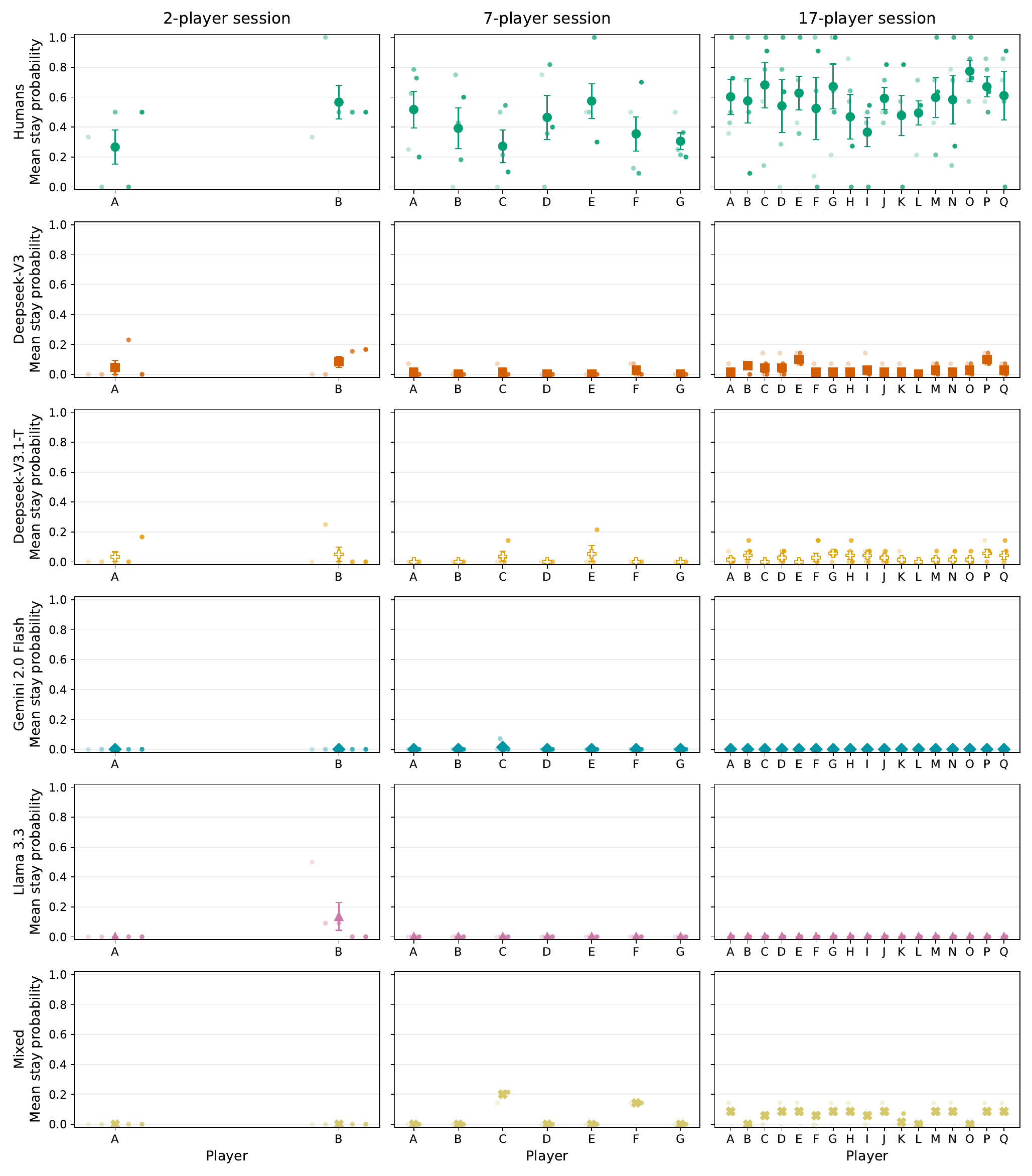}
    \caption{Representative player-level stay profiles across five numerical games. Columns show one 2-player session (day 10), one 7-player session (day 1), and one 17-player session (day 15). Rows show humans, Deepseek-V3, Deepseek-V3.1-T, Gemini 2.0 Flash, Llama 3.3, and mixed-model groups. Faint points denote per-game stay probabilities for each player, shaded from light to dark for games 1 through 5. Large markers and error bars show mean stay probability and variability across the five numerical games. In the human examples, players are more separated from one another than in the LLM examples, which are typically lower and more compressed. The remaining stay profiles are provided as a part of the Supplemental Material.}
    \label{fig:selected-stay-profiles}
\end{figure*}

\begin{figure*}[ht!]
    \centering
    \includegraphics[width=0.94\textwidth]{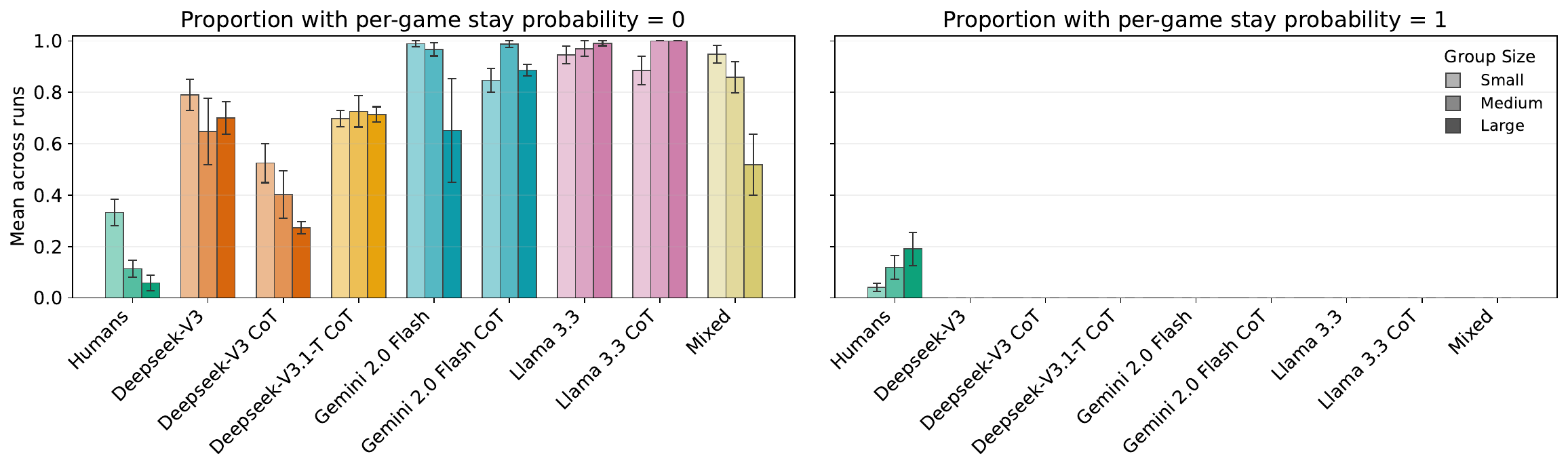}
    \caption{Extreme stay behavior in numerical games. For each player and game, stay probability is the fraction of post-initial rounds in which the player repeated the previous guess. The left panel shows the proportion of players with stay probability 0, meaning they changed their guess on every post-initial round of a game. The right panel shows the proportion of players with stay probability 1, meaning they kept the same guess throughout the game. Bars show means across runs and error bars show standard errors across runs. The right panel highlights a qualitative difference between humans and LLMs: fully stable players appear only in human groups and become more common as group size increases.}
    \label{fig:stay-extremes}
\end{figure*}

\clearpage

\section{Prompt Templates and Examples of LLM Responses During the Gameplay}
\label{appendix: response examples}

\begin{figure*}[h!]
\centering
\small

\begin{tcolorbox}[title= Zero-Shot Prompt, colback=gray!20, colframe=black, arc=3mm]
\textcolor{blue}{\textbf{SYSTEM}}: You will now play a game with a group of players. You are player A, and 
you will be playing with player B. You will play 10 games, where each game 
contains 15 rounds after which the game ends. Each game will have a different 
mystery number between 51 and 100. In each round, each player submits their 
own number. All of the players' numbers are summed together and compared to 
the mystery number that has a value between 51 and 100. All of the players 
are given identical feedback on whether their group's total sum was too low, 
too high, or just right, and each player decides for themselves whether and 
how to adjust their number for the next round. Your goal as a member of the 
group is to help the group converge to the mystery number as soon as possible 
in each game. You will be provided with your own guesses from previous rounds and the corresponding group feedback for those rounds. 
Provide the chosen integer between 0 and 50. \\

\textcolor{blue}{\textbf{USER}}: This is Game 1 Round 1. There is no history yet. Please provide your 
answer in the specified format. The output should be formatted as a JSON 
instance that conforms to the JSON schema below.\\

As an example, for the schema \{"properties": \{"foo": \{"title": "Foo", 
"description": "a list of strings", "type": "array", "items": \{"type": 
"string"\}\}\}, "required": ["foo"]\}
the object \{"foo": ["bar", "baz"]\} is a well-formatted instance of the schema. 
The object \{"properties": \{"foo": ["bar", "baz"]\}\} is not well-formatted.\\

Here is the output schema:

\begin{verbatim}
{"description": "The player's chosen number for the guessing 
game.", "properties": {"chosen_number": {"description": "The 
player's chosen number for the next round (between 0 and 
50)", "title": "Chosen Number", "type": "integer"}}, 
"required": ["chosen_number"]}
\end{verbatim}

\textcolor{darkgreen}{\textbf{Assistant}}: \{"chosen\_number":25\} \\

\textcolor{blue}{\textbf{USER}}: In the previous round your choice was 25 and the total sum of guesses 
by all players was too low. This is Game 1 Round 2. You need to 
choose a number to help your group converge to the mystery number. Provide 
your answer in the specified format. The output should be formatted as a JSON 
instance that conforms to the JSON schema below.\\

As an example, for the schema \{"properties": \{"foo": \{"title": "Foo", 
"description": "a list of strings", "type": "array", "items": \{"type": 
"string"\}\}\}, "required": ["foo"]\}
the object \{"foo": ["bar", "baz"]\} is a well-formatted instance of the schema. 
The object \{"properties": \{"foo": ["bar", "baz"]\}\} is not well-formatted.\\

Here is the output schema:

\begin{verbatim}
{"description": "The player's chosen number for the guessing 
game.", "properties": {"chosen_number": {"description": "The 
player's chosen number for the next round (between 0 and 50)"
, "title": "Chosen Number", "type": "integer"}}, "required":
["chosen_number"]}
\end{verbatim}

\end{tcolorbox}
\caption{Zero-Shot prompt provided to each agent during the game in the first two rounds of the experiment.}
\label{fig: Zero-shot prompt}
\end{figure*}

\begin{figure*}[h!]
\centering
\small

\begin{tcolorbox}[title= Zero-Shot CoT Prompt, colback=gray!20, colframe=black, arc=3mm]
\textcolor{blue}{\textbf{SYSTEM}}: You will now play a game with a group of players. You are player A, and 
you will be playing with player B. You will play 10 games, where each game 
contains 15 rounds after which the game ends. Each game will have a different 
mystery number between 51 and 100. In each round, each player submits their 
own number. All of the players' numbers are summed together and compared to 
the mystery number that has a value between 51 and 100. All of the players 
are given identical feedback on whether their group's total sum was too low, 
too high, or just right, and each player decides for themselves whether and 
how to adjust their number for the next round. Your goal as a member of the 
group is to help the group converge to the mystery number as soon as possible 
in each game. You will be provided with your own guesses from previous rounds and the corresponding group feedback for those rounds. 
Provide the chosen integer between 0 and 50. \\

\textcolor{blue}{\textbf{USER}}: This is Game 1 Round 1. There is no history yet. Please provide your answer in the specified format. The output should be formatted as a JSON instance that conforms to the JSON schema below.\\

As an example, for the schema \{"properties": \{"foo": \{"title": "Foo", 
"description": "a list of strings", "type": "array", "items": \{"type": 
"string"\}\}\}, "required": ["foo"]\} the object \{"foo": ["bar", "baz"]\} is a well-formatted instance of the schema. The object \{"properties": \{"foo": ["bar", "baz"]\}\} is not well-formatted. \\

Here is the output schema:

\begin{verbatim}
{"description": "The player's chosen number for the guessing 
game.", "properties": {"chosen_number": {"description": "The 
player's chosen number for the next round (between 0 and 
50)", "title": "Chosen Number", "type": "integer"}}, 
"required": ["chosen_number"]}
\end{verbatim}

\textcolor{darkgreen}{\textbf{Assistant}}: \{"chosen\_number":25\} \\

\textcolor{blue}{\textbf{USER}}: In the previous round your choice was 25 and the total sum of guesses 
by all players was too low. This is Game 1 Round 2. You need to 
choose a number to help your group converge to the mystery number. Provide 
your answer in the specified format. The output should be formatted as a JSON instance that conforms to the JSON schema below.\\ 

As an example, for the schema \{"properties": \{"foo": \{"title": "Foo", "description": "a list of strings", "type": "array", "items": \{"type": "string"\}\}\}, "required": ["foo"]\} the object \{"foo": ["bar", "baz"]\} is a well-formatted instance of the schema. The object \{"properties": \{"foo": ["bar", "baz"]\}\} is not well-formatted.\\ 

Here is the output schema:

\begin{verbatim}
{"description": "The player's chosen number for the guessing 
game.", "properties": {"chosen_number": {"description": "The 
player's chosen number for the next round (between 0 and 50)"
, "title": "Chosen Number", "type": "integer"}}, "required":
["chosen_number"]}
\end{verbatim}

\textcolor{darkgreen}{\textbf{Assistant}}: Let's think step by step:

\end{tcolorbox}

\caption{Zero-Shot CoT prompt provided to each agent during the game in the first two rounds of the experiment.}
\label{fig: Zero-shot CoT prompt}

\end{figure*}

\clearpage

\begin{figure*}[h!]
\centering
\small
\begin{tcolorbox}[title= Zero-Shot Prompt with Strategy and Group Sum from Previous Round as Part of the Prompt, colback=gray!20, colframe=black, arc=3mm]
\textcolor{blue}{\textbf{SYSTEM}}: You will now play a game with a group of players. You are player A, and 
you will be playing with player B. You will play 10 games, where each game 
contains 15 rounds after which the game ends. Each game will have a different 
mystery number between 51 and 100. In each round, each player submits their 
own number. All of the players' numbers are summed together and compared to 
the mystery number that has a value between 51 and 100. All of the players 
are given identical feedback on whether their group's total sum was too low, 
too high, or just right, and each player decides for themselves whether and 
how to adjust their number for the next round. Your goal as a member of the 
group is to help the group converge to the mystery number as soon as possible 
in each game. You will be provided with your own guesses from previous rounds and the corresponding group feedback for those rounds. \textcolor{red}{It will help if you try, over successive rounds of play, to develop a consistent role in terms of how much you react to the feedback, while also trying to make your role unique compared to others in your group. For example, if you think that others in your group are reacting too much to the feedback (your group often guesses numbers that are too high and then too low), then you would want to react less. If others are reacting too little to the feedback (your group is always guessing number that are too high, or always guessing numbers that are too low) then you would want to react more.} Provide the chosen integer between 0 and 50. \\

\textcolor{blue}{\textbf{USER}}: This is Game 1 Round 1. There is no history yet. Please provide your 
answer in the specified format. The output should be formatted as a JSON 
instance that conforms to the JSON schema below.\\

As an example, for the schema ... \\
... \\

\textcolor{darkgreen}{\textbf{Assistant}}: \{"chosen\_number":25\} \\

\textcolor{blue}{\textbf{USER}}: In the previous round your choice was 25 \textcolor{red}{and the total sum of guesses 
by all players was 50} which was too low. This is Game 1 Round 2. You need to 
choose a number to help your group converge to the mystery number. Provide 
your answer in the specified format. The output should be formatted as a JSON 
instance that conforms to the JSON schema below.\\

As an example, for the schema \{"properties": \{"foo": \{"title": "Foo", 
"description": "a list of strings", "type": "array", "items": \{"type": 
"string"\}\}\}, "required": ["foo"]\}
the object \{"foo": ["bar", "baz"]\} is a well-formatted instance of the schema. 
The object \{"properties": \{"foo": ["bar", "baz"]\}\} is not well-formatted.\\

Here is the output schema:

\begin{verbatim}
{"description": "The player's chosen number for the guessing 
game.", "properties": {"chosen_number": {"description": "The 
player's chosen number for the next round (between 0 and 50)"
, "title": "Chosen Number", "type": "integer"}}, "required":
["chosen_number"]}
\end{verbatim}

\end{tcolorbox}
\caption{Zero-shot prompt with an explicit behavioral strategy in the system prompt and the raw group sum from the previous round in the user prompt (both highlighted in red). The raw group sum was provided only in this strategy-plus-sum ablation. In the main experiments, models received their own prior guesses and either directional feedback (too low, too high, or just right) or numerical error-magnitude feedback (too low or too high by $k$), but not the raw group sum.}
\label{fig: Zero-shot strategy prompt}
\end{figure*}

\clearpage

\begin{figure*}[h!]
\centering
\small

\begin{tcolorbox}[title= DeepSeek-V3.1-T Response in 2-Player Game (Game 10 Round 1), colback=gray!20, colframe=black, arc=3mm]

 First, this is Game 10 Round 1. I'm player B, and I need to choose a number between 0 and 50 to help the group converge to the mystery number, which is between 51 and 100.
Since it's the first round, I have no history from this game. But I can use what I've learned from previous games. In the earlier games, I often started with 25 as a mid-point since the mystery number is between 51 and 100, and there are two players (A and B), so the sum of our guesses should aim for the mystery number.
Assuming there are two players: me and player A. The system message says "a group of players," and in the interactions, it's implied there's at least one other player, probably player A. I need to coordinate with them.

From previous games, I've seen that starting around 25 is common, and then adjusting based on feedback. In Game 1, I started with 25, and it was too high, so I adjusted down. In Game 2, I started with 25, and it was too low, so I adjusted up. Similarly in other games.
For Game 10 Round 1, I should start with a reasonable guess. The mystery number is between 51 and 100, so if there are two players, each guessing between 0 and 50, the sum could be up to 100, which matches the mystery number range.
To be efficient, I might start with a number that is half of the average mystery number. The average mystery number is around (51+100)/2 = 75.5, so half of that is about 37.75, but since I can only choose integers, maybe 38 or something. But in previous games, I often started with 25, which is more conservative.\\

Looking back at my responses:

- Game 1 Round 1: 25

- Game 2 Round 1: 25

- Game 3 Round 1: 25

- Game 4 Round 1: 25

- Game 5 Round 1: 25

- Game 6 Round 1: 25

- Game 7 Round 1: 25

- Game 8 Round 1: 25

- Game 9 Round 1: 25

- Now Game 10 Round 1: I should probably start with 25 again for consistency.\\

But in some games, when the sum was too low by a large margin, I increased my guess significantly, like in Game 2 Round 2, I went from 25 to 42 when it was too low by 34.\\

However, for the first round, I don't have any feedback, so I need an initial guess.\\

The output must be in JSON format with "chosen\_number" as an integer between 0 and 50.\\

I think sticking with 25 is a safe bet, as it's a common starting point in these simulations.\\

So for Game 10 Round 1, I'll choose 25.\\

\begin{verbatim}
```json
{"chosen_number": 25}
```
\end{verbatim}

\end{tcolorbox}

\caption{}
\label{fig: deepseek v31t response}

\end{figure*}

\begin{figure*}[h!]
\centering
\small
\begin{tcolorbox}[title= DeepSeek-V3 Zero-Shot CoT Response in 2-Player Game (Game 10 Round 1), colback=gray!20, colframe=black, arc=3mm]

1. Since this is the first round of a new game, I don't have any previous feedback to work with. \\
2. The mystery number is between 51-100, and I'm one of two players. \\
3. A reasonable starting point would be to choose a number around the middle of my possible range (0-50). \\
4. Choosing 25 gives room to adjust upward if needed while still leaving flexibility for my partner to contribute. \\

\begin{verbatim}
```json
{"chosen_number": 25}
```
\end{verbatim}

\end{tcolorbox}

\caption{}
\label{fig: deepseek v3 response}

\end{figure*}

\begin{figure*}[h!]
\centering
\small
\begin{tcolorbox}[title= Gemini 2.0 Flash Zero-Shot CoT Response in 2-Player Game (Game 10 Round 1), colback=gray!20, colframe=black, arc=3mm]

* **Initial Round:** Without any history, it's best to start with a number around the middle of the allowed range (0-50). Let's start with 30.

\begin{verbatim}
```json
{"chosen_number": 30}
```
\end{verbatim}

\end{tcolorbox}

\caption{}
\label{fig: gemini flash 2.0 response}

\end{figure*}

\begin{figure*}[h!]
\centering
\small
\begin{tcolorbox}[title= Llama 3.3 Zero-Shot CoT Response in 2-Player Game (Game 10 Round 1), colback=gray!20, colframe=black, arc=3mm]

This is the first round of Game 10, so there is no history yet. The goal is to choose a number that will help the group converge to the mystery number as soon as possible. \\

Given the constraints that the mystery number is between 51 and 100, and each player's guess must be between 0 and 50, it seems reasonable to start with a moderate value to allow for adjustments in either direction based on the feedback from the total sum of guesses. \\

A good starting point could be the middle of the range 0-50, which is 25. This choice allows for flexibility to increase or decrease the guess in subsequent rounds based on whether the total sum is too high or too low. \\

Therefore, the chosen number for Game 10 Round 1 should be 25. \\

\begin{verbatim}
```json
{"chosen_number": 25}
```
\end{verbatim}

\end{tcolorbox}

\caption{}
\label{fig: llama 3.3 response}

\end{figure*}

\clearpage

\section{Learning Across Games} \label{appendix:Analysis}

\begin{figure*}[h!]
\centering
\includegraphics[width=\textwidth]{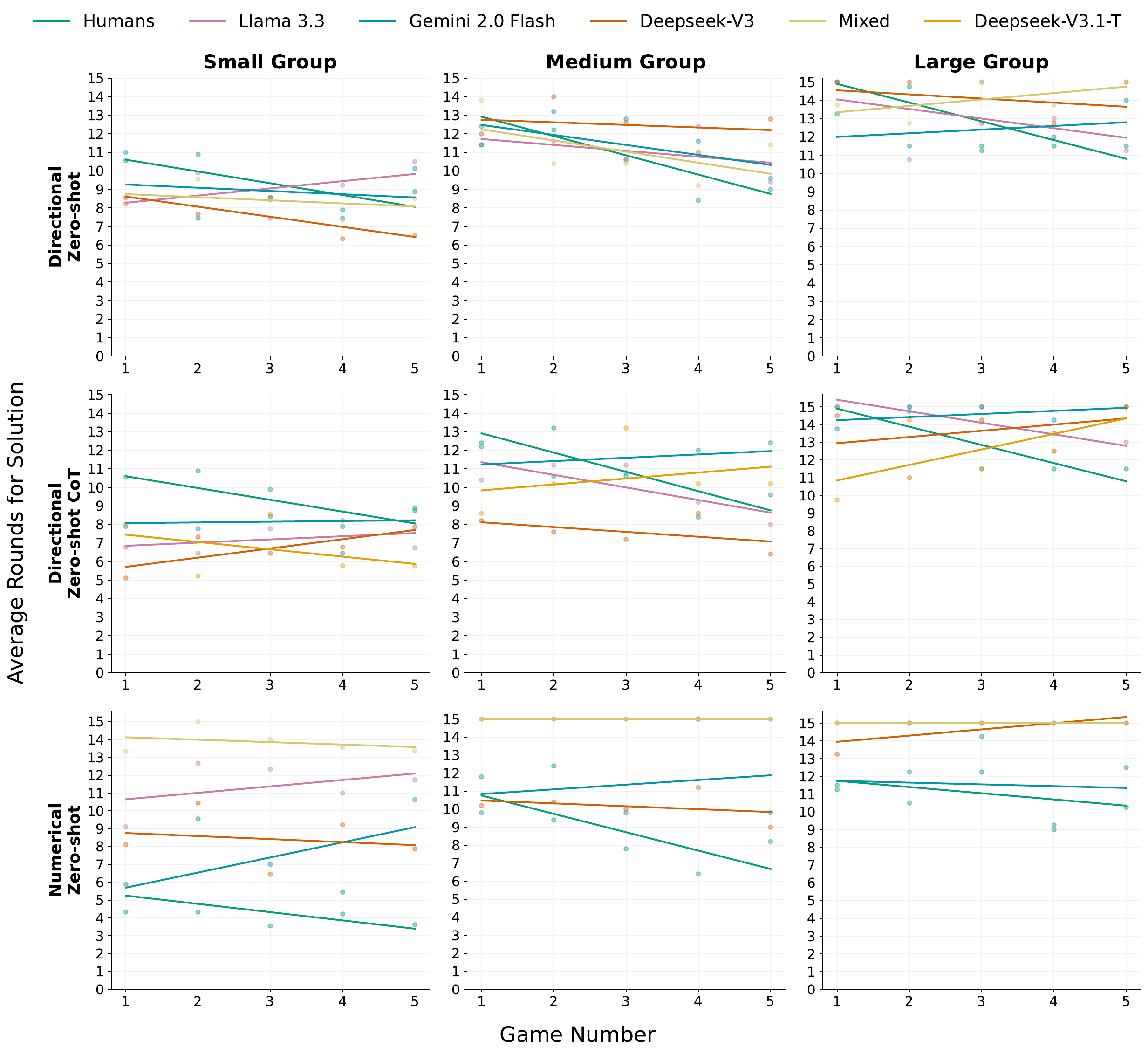}
\caption{Average number of rounds needed to finish the game across five games under zero-shot prompt and directional feedback (top), zero-shot CoT prompt and directional feedback (middle), and zero-shot prompt and numerical feedback (bottom) conditions.}
\label{fig:rounds_to_solution_rest}
\end{figure*}

\begin{table}[ht]
\centering
\small
\begin{tabular}{llrcll}
\toprule
Model & Feedback & Runs & Mean Slope & 95\% Bootstrap CI & \% Negative \\
\midrule
Humans           & Directional & 18 & $-0.91$ & $[-1.34, -0.43]$   & 78\% \\
Humans           & Numerical   & 18 & $-0.57$ & $[-1.12, -0.04]$   & 72\% \\
Deepseek-V3      & Directional & 18 & $-0.39$ & $[-0.97, 0.19]$    & 56\% \\
Deepseek-V3      & Numerical   & 18 & $0.02$  & $[-0.61, 0.61]$    & 33\% \\
Gemini 2.0 Flash & Directional & 18 & $-0.21$ & $[-0.94, 0.51]$    & 44\% \\
Gemini 2.0 Flash & Numerical   & 18 & $0.41$  & $[-0.17, 0.99]$    & 33\% \\
Llama 3.3        & Directional & 18 & $-0.07$ & $[-0.52, 0.42]$    & 50\% \\
Llama 3.3        & Numerical   & 18 & $0.16$  & $[-0.13, 0.48]$    & 6\%  \\
Mixed            & Directional & 18 & $-0.24$ & $[-0.82, 0.33]$    & 50\% \\
Mixed            & Numerical   & 18 & $0.02$  & $[-0.47, 0.58]$    & 11\% \\
\bottomrule
\end{tabular}
\caption{Mean slope of rounds-to-solution vs. game number (1–5) for Zero-shot prompts, pooled across group sizes (small, medium, large). A negative slope indicates fewer rounds needed in later games, reflecting cross-game learning. 95\% confidence intervals computed via bootstrap resampling (10,000 iterations). ``\% Negative'' reports the fraction of individual runs with a negative slope.}
\label{slope_rounds_to_solution_zs}
\end{table}

\begin{table}[ht]
\centering
\small
\begin{tabular}{llrcll}
\toprule
Model & Feedback & Runs & Mean Slope & 95\% Bootstrap CI & \% Negative \\
\midrule
Humans           & Directional & 18 & $-0.91$ & $[-1.34, -0.43]$   & 78\% \\
Humans           & Numerical   & 18 & $-0.57$ & $[-1.12, -0.04]$   & 72\% \\
Deepseek-V3      & Directional & 18 & $0.31$  & $[-0.09, 0.70]$    & 33\% \\
Deepseek-V3      & Numerical   & 18 & $0.27$  & $[-0.28, 0.84]$    & 39\% \\
Deepseek-V3.1-T  & Directional & 18 & $0.10$  & $[-0.31, 0.52]$    & 39\% \\
Deepseek-V3.1-T  & Numerical   & 18 & $-0.22$ & $[-0.77, 0.29]$    & 50\% \\
Gemini 2.0 Flash & Directional & 18 & $0.07$  & $[-0.54, 0.70]$    & 44\% \\
Gemini 2.0 Flash & Numerical   & 18 & $0.89$  & $[0.33, 1.52]$     & 11\% \\
Llama 3.3        & Directional & 18 & $-0.19$ & $[-0.73, 0.31]$    & 33\% \\
Llama 3.3        & Numerical   & 18 & $0.00$  & $[-0.67, 0.66]$    & 28\% \\
\bottomrule
\end{tabular}
\caption{Mean slope of rounds-to-solution vs. game number (1–5) for Zero-shot CoT prompts, pooled across group sizes (small, medium, large). A negative slope indicates fewer rounds needed in later games, reflecting cross-game learning. 95\% confidence intervals computed via bootstrap resampling (10,000 iterations). ``\% Negative'' reports the fraction of individual runs with a negative slope.}
\label{slope_rounds_to_solution_zs_cot}
\end{table}

\begin{figure*}[h!]
\centering
\includegraphics[width=\textwidth]{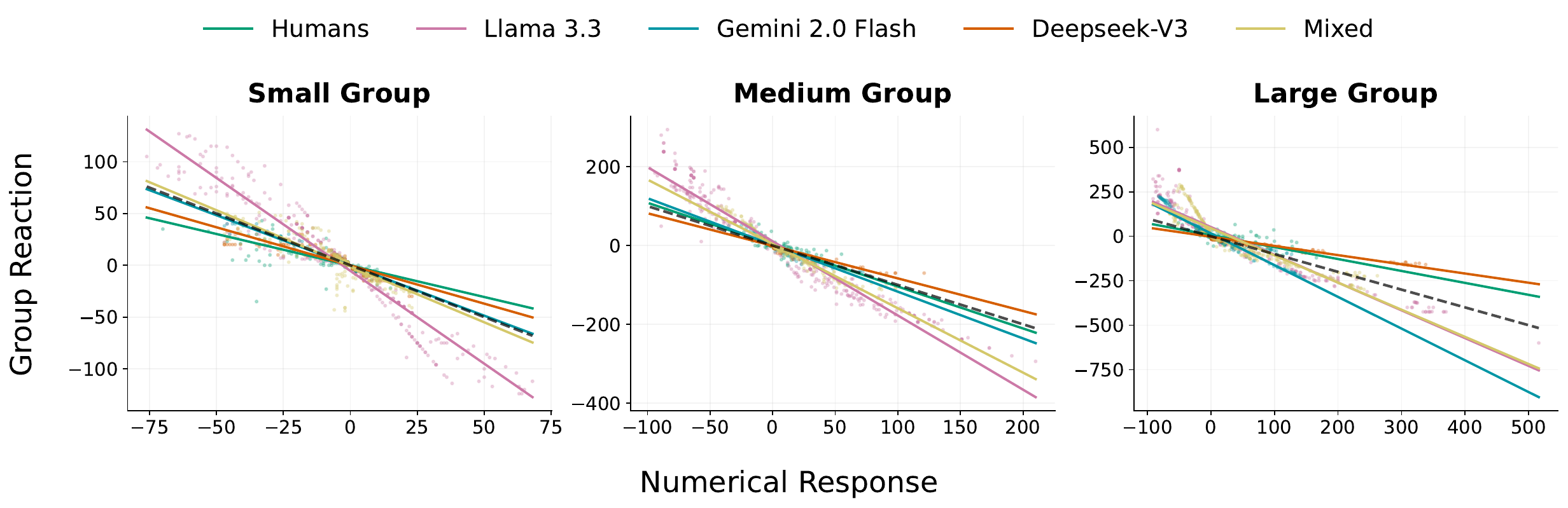}
\caption{Group reaction to numerical feedback under zero-shot prompting. Each dot denotes the aggregate adjustment made by a group after the previous round's numerical feedback. The dotted line indicates the optimal collective correction, and the solid line shows the fitted descriptive group reaction. }
\label{fig:numerical-group-reaction-zs}
\end{figure*}

\begin{table}[ht!]
\centering
\small
\begin{tabular}{llr}
\toprule
\textbf{Condition} & \textbf{Model} & \textbf{Slope} \\
\midrule
Small Group & Humans & $-0.589$ \\
\addlinespace
Small Group, Zero-shot & Deepseek-V3 & $-0.738$ \\
Small Group, Zero-shot & Gemini 2.0 Flash & $-0.969$ \\
Small Group, Zero-shot & Llama 3.3 & $-1.794$ \\
\addlinespace
Small Group, Zero-shot CoT & Deepseek-V3 & $-0.991$ \\
Small Group, Zero-shot CoT & Deepseek-V3.1-T & $-1.018$ \\
Small Group, Zero-shot CoT & Gemini 2.0 Flash & $-1.531$ \\
Small Group, Zero-shot CoT & Llama 3.3 & $-1.864$ \\
\midrule
Medium Group & Humans & $-1.050$ \\
\addlinespace
Medium Group, Zero-shot & Deepseek-V3 & $-0.827$ \\
Medium Group, Zero-shot & Gemini 2.0 Flash & $-1.186$ \\
Medium Group, Zero-shot & Llama 3.3 & $-1.883$ \\
\addlinespace
Medium Group, Zero-shot CoT & Deepseek-V3 & $-1.270$ \\
Medium Group, Zero-shot CoT & Deepseek-V3.1-T & $-1.214$ \\
Medium Group, Zero-shot CoT & Gemini 2.0 Flash & $-1.920$ \\
Medium Group, Zero-shot CoT & Llama 3.3 & $-1.866$ \\
\midrule
Large Group & Humans & $-0.663$ \\
\addlinespace
Large Group, Zero-shot & Deepseek-V3 & $-0.517$ \\
Large Group, Zero-shot & Gemini 2.0 Flash & $-1.785$ \\
Large Group, Zero-shot & Llama 3.3 & $-1.562$ \\
\addlinespace
Large Group, Zero-shot CoT & Deepseek-V3 & $-1.691$ \\
Large Group, Zero-shot CoT & Deepseek-V3.1-T & $-1.458$ \\
Large Group, Zero-shot CoT & Gemini 2.0 Flash & $-1.795$ \\
Large Group, Zero-shot CoT & Llama 3.3 & $-1.631$ \\
\bottomrule
\end{tabular}
\caption{Slopes of best-fit lines for group reaction (change in group sum) vs. numerical response (difference between group sum and target), computed across all observations within each condition. A slope of $-1.0$ indicates ideal correction; slopes more negative than $-1.0$ indicate overreaction, and slopes less negative than $-1.0$ indicate underreaction. Human slopes average $-0.767$ across conditions, indicating systematic underreaction. LLM slopes average $-1.386$ (computed as the mean of per-condition model averages), indicating systematic overreaction. In 19 of 21 condition–model pairs, LLMs exhibit more negative slopes than the corresponding human baseline.}
\label{tab:reaction_slopes_numerical}
\end{table}

\clearpage

\subsection{Gameplay dynamics for GRPO Checkpoints}
\begin{figure*}[ht!]
    \centering
    \includegraphics[width=\textwidth]{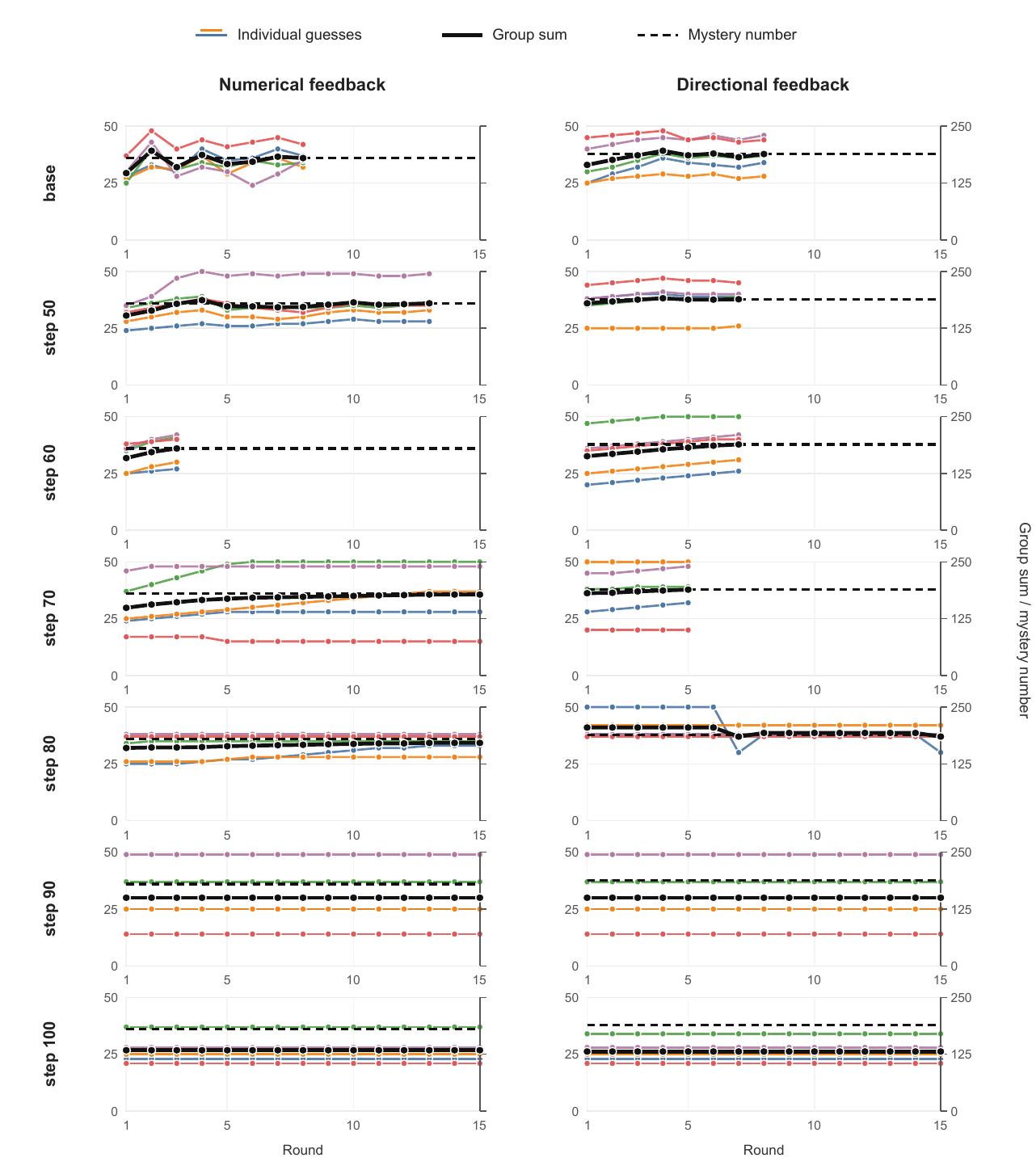}
    \caption{Representative five-player numerical- and directional-feedback GBS games across the base model and Phase 2 GRPO checkpoints. Colored lines show individual guesses, the solid black line the group sum, and the dashed black line the mystery number.}
    \label{fig:grpo-sample-game}
\end{figure*}

\clearpage

\section{Distribution of Player Decisions} 
\label{appendix:histograms_decisions}

\begin{figure*}[h!]
\centering
\includegraphics[width=\textwidth]{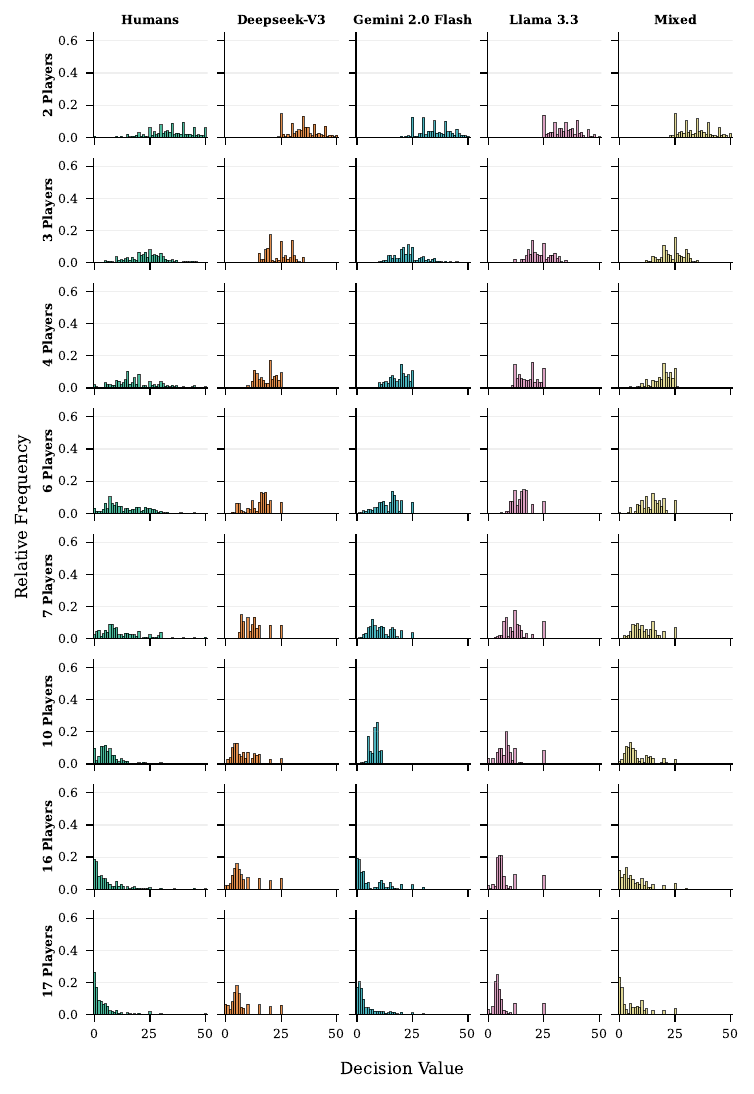}
\caption{Distribution of player decisions under the zero-shot condition with directional feedback.}
\label{zs-directional-decision-histogram}
\end{figure*}

\begin{figure*}[h!]
\centering
\includegraphics[width=\textwidth]{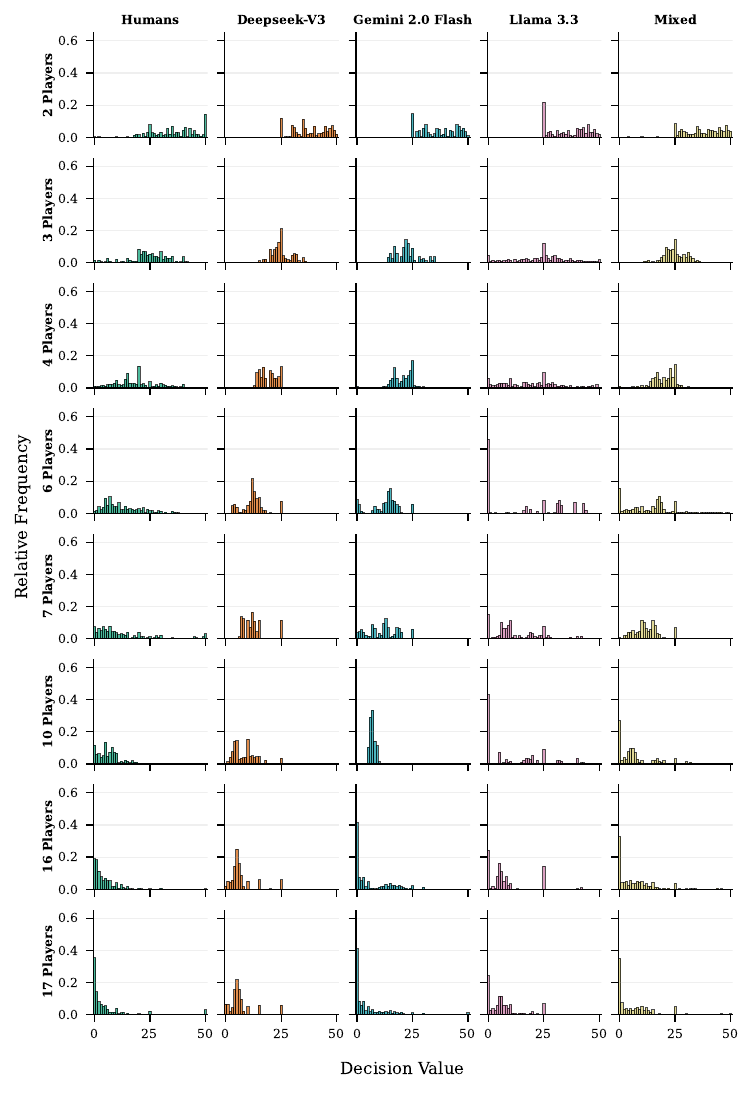}
\caption{Distribution of player decisions under the zero-shot condition with numerical feedback.}
\label{zs-numerical-decision-histogram}
\end{figure*}

\begin{figure*}[h!]
\centering
\includegraphics[width=\textwidth]{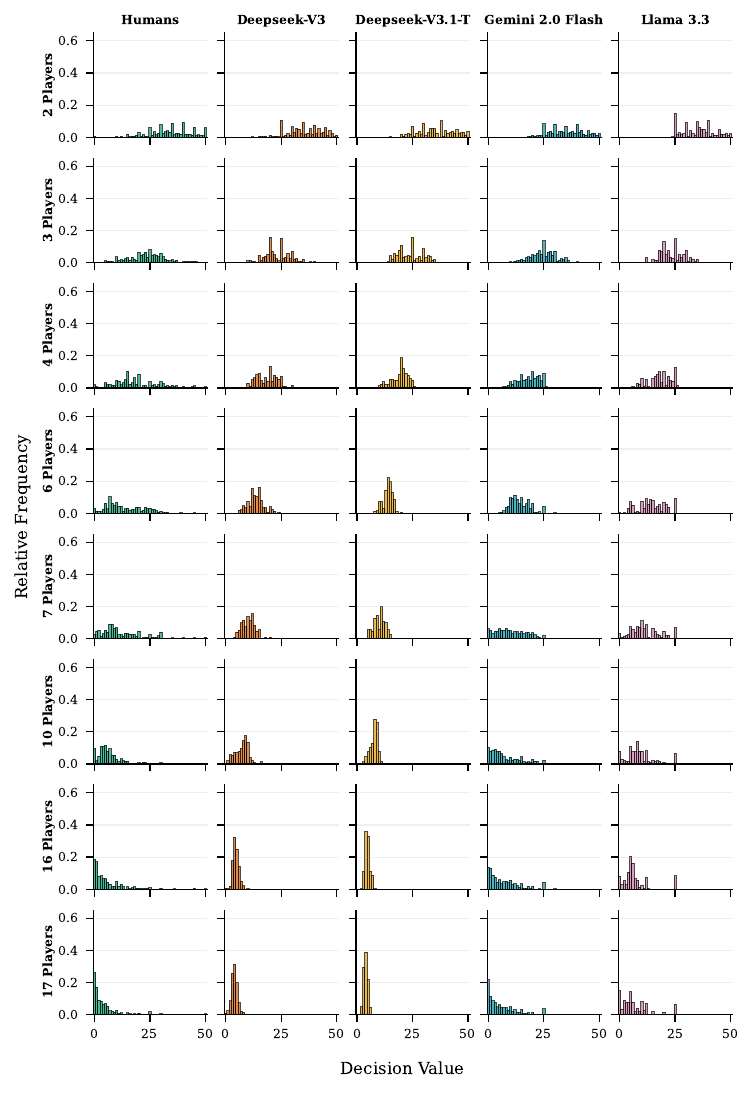}
\caption{Distribution of player decisions under the zero-shot CoT condition with directional feedback.}
\label{zs-CoT-directional-decision-histogram}
\end{figure*}

\begin{figure*}[h!]
\centering
\includegraphics[width=\textwidth]{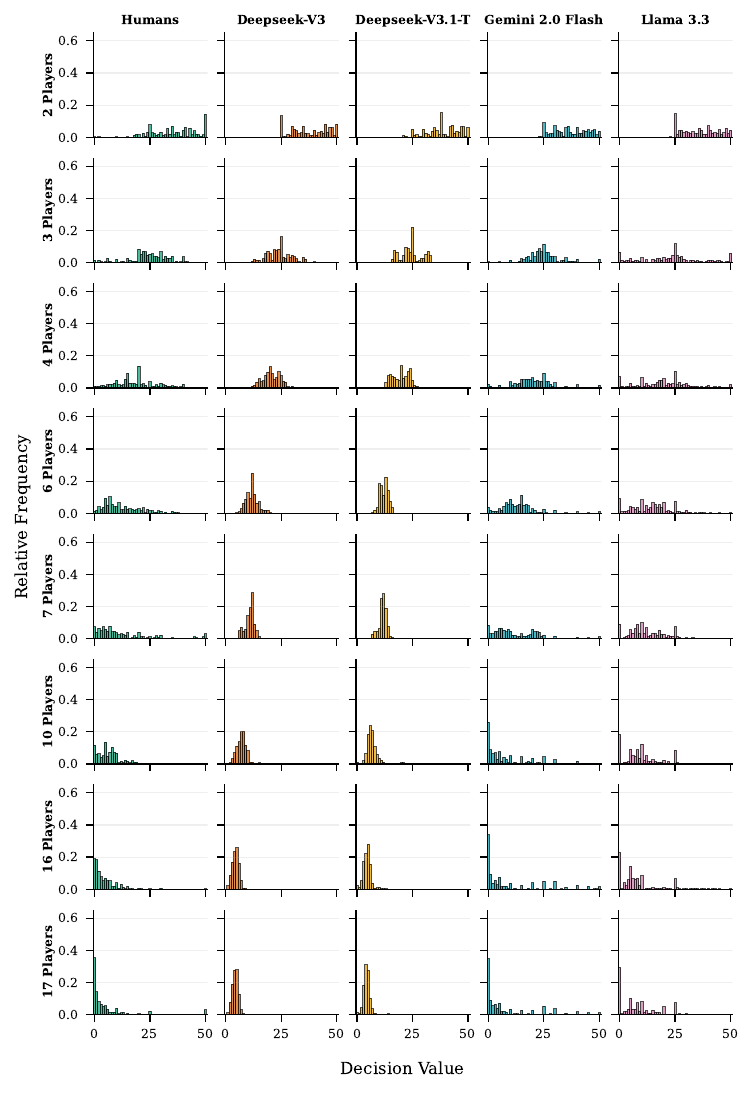}
\caption{Distribution of player decisions under the zero-shot CoT condition with numerical feedback.}
\label{zs-CoT-numerical-decision-histogram}
\end{figure*}

\clearpage

\section{Distribution of Magnitude of Decision Changes Between Rounds} 
\label{appendix:histograms_decision_switch_magnitude}

\begin{figure*}[h!]
\centering
\includegraphics[width=\textwidth]{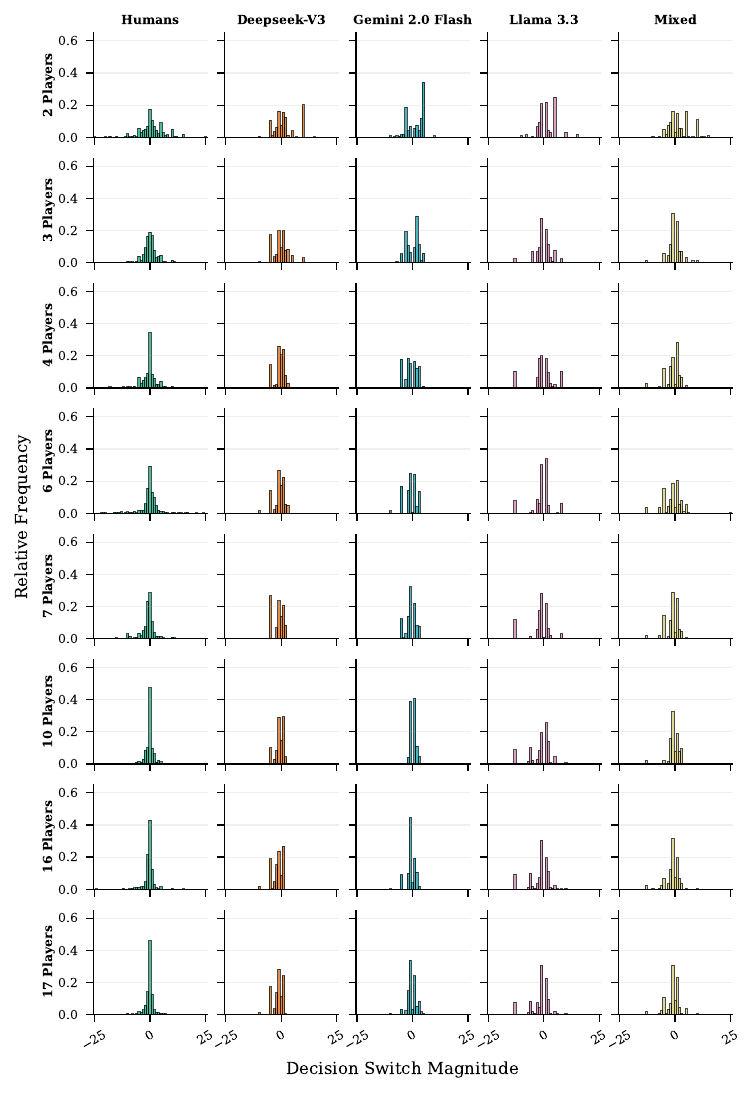}
\caption{Distribution of decision switch magnitudes (change in a player's guess between consecutive rounds) under the zero-shot condition with directional feedback.}
\label{zs-directional-decision-switch-histogram}
\end{figure*}

\begin{figure*}[h!]
\centering
\includegraphics[width=\textwidth]{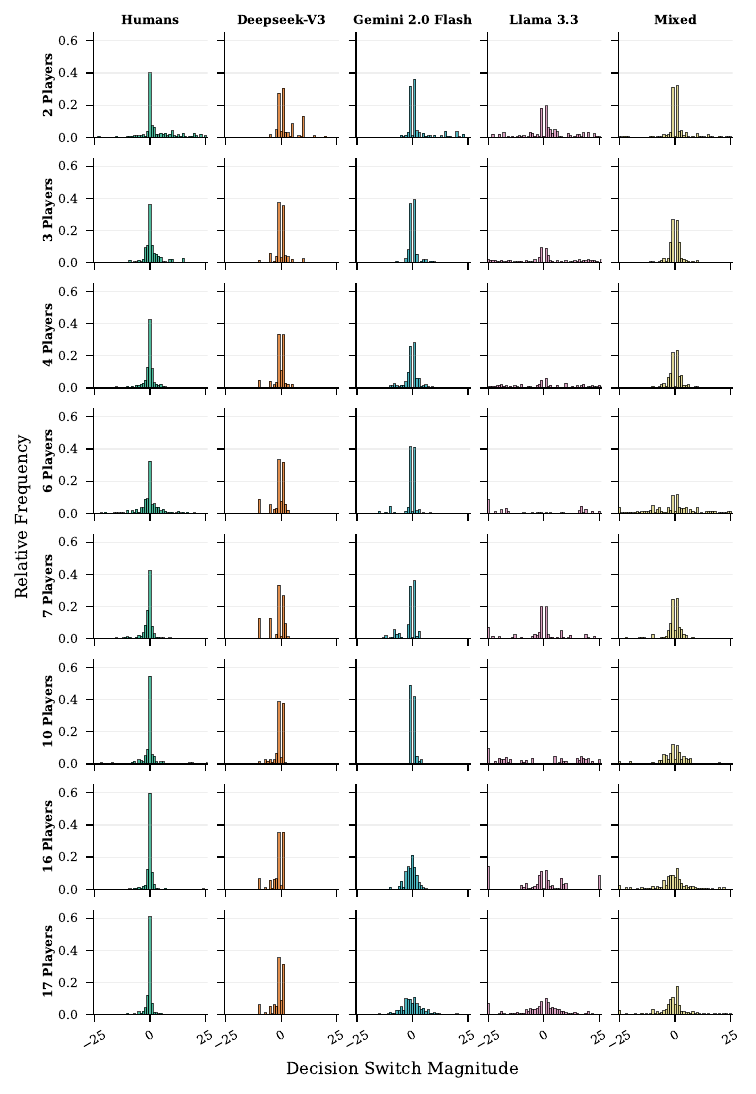}
\caption{Distribution of decision switch magnitudes (change in a player's guess between consecutive rounds) under the zero-shot condition with numerical feedback.}
\label{zs-numerical-decision-switch-histogram}
\end{figure*}

\begin{figure*}[h!]
\centering
\includegraphics[width=\textwidth]{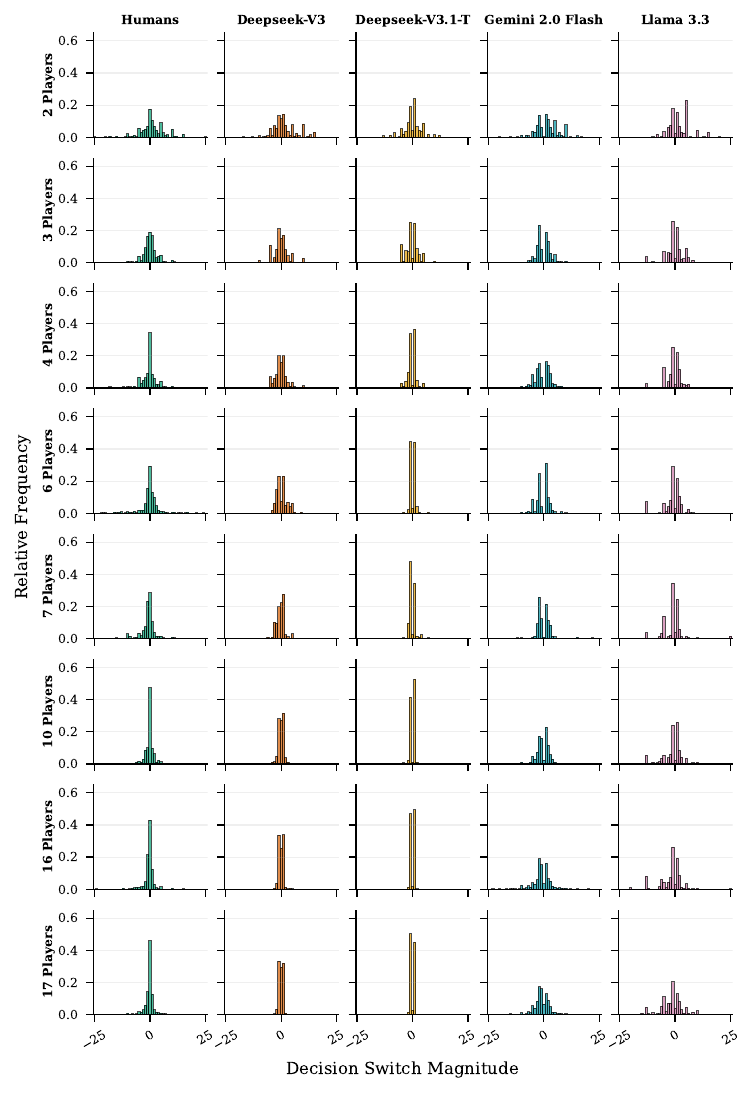}
\caption{Distribution of decision switch magnitudes (change in a player's guess between consecutive rounds) under the zero-shot CoT condition with directional feedback.}
\label{zs-CoT-directional-decision-switch-histogram}
\end{figure*}

\begin{figure*}[h!]
\centering
\includegraphics[width=\textwidth]{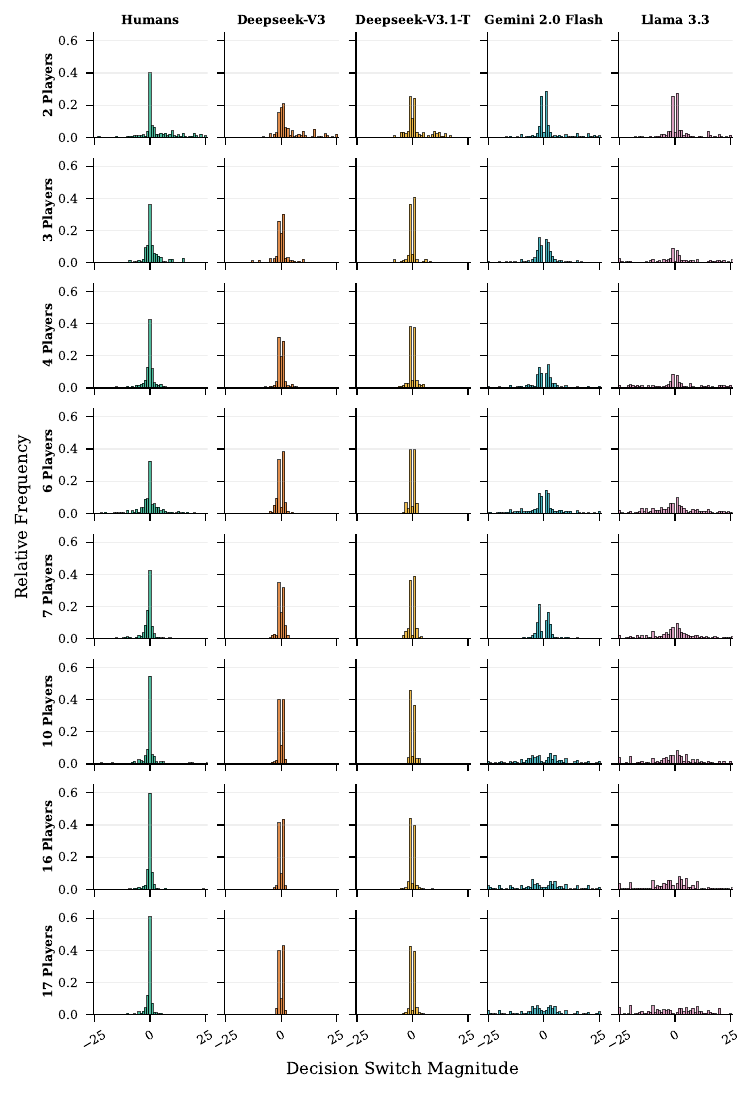}
\caption{Distribution of decision switch magnitudes (change in a player's guess between consecutive rounds) under the zero-shot CoT condition with numerical feedback.}
\label{zs-CoT-numerical-decision-switch-histogram}
\end{figure*}

\clearpage

\section{Visualization of Gameplay Dynamics} 
\label{appendix:Gameplay}

\begin{figure*}[h!]
\centering
\includegraphics[width=1.0\textwidth]{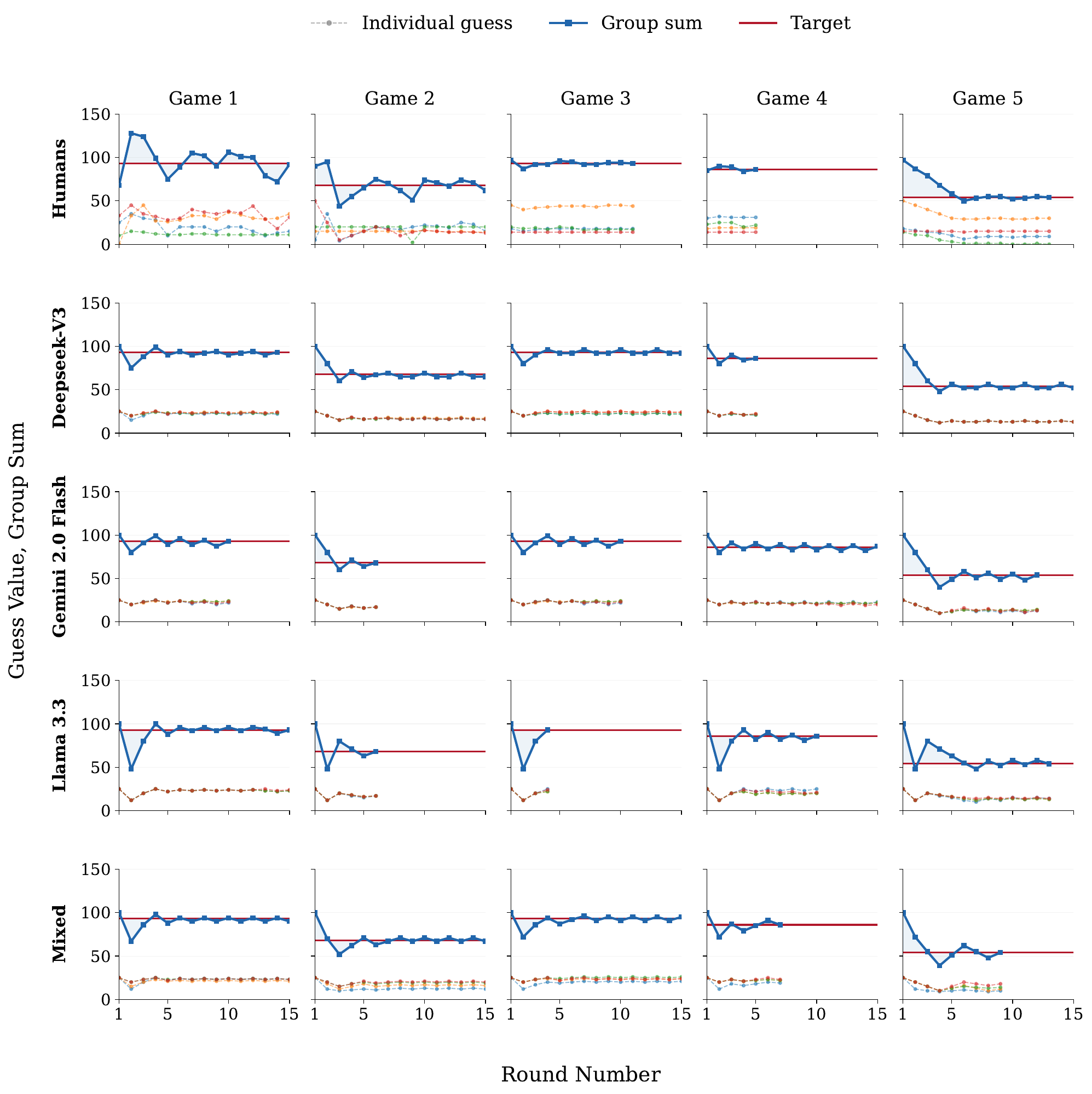}
\caption{Example of coordination in 4-player games with directional feedback and zero-shot prompts. The solid horizontal line indicates the mystery number, the other solid line indicates the sum of the group, and the dashed line represents the decisions of each agent in the group.}
\label{fig:4-player-zero-shot-directional-12}
\end{figure*}

\begin{figure*}[h!]
\centering
\includegraphics[width=1.0\textwidth]{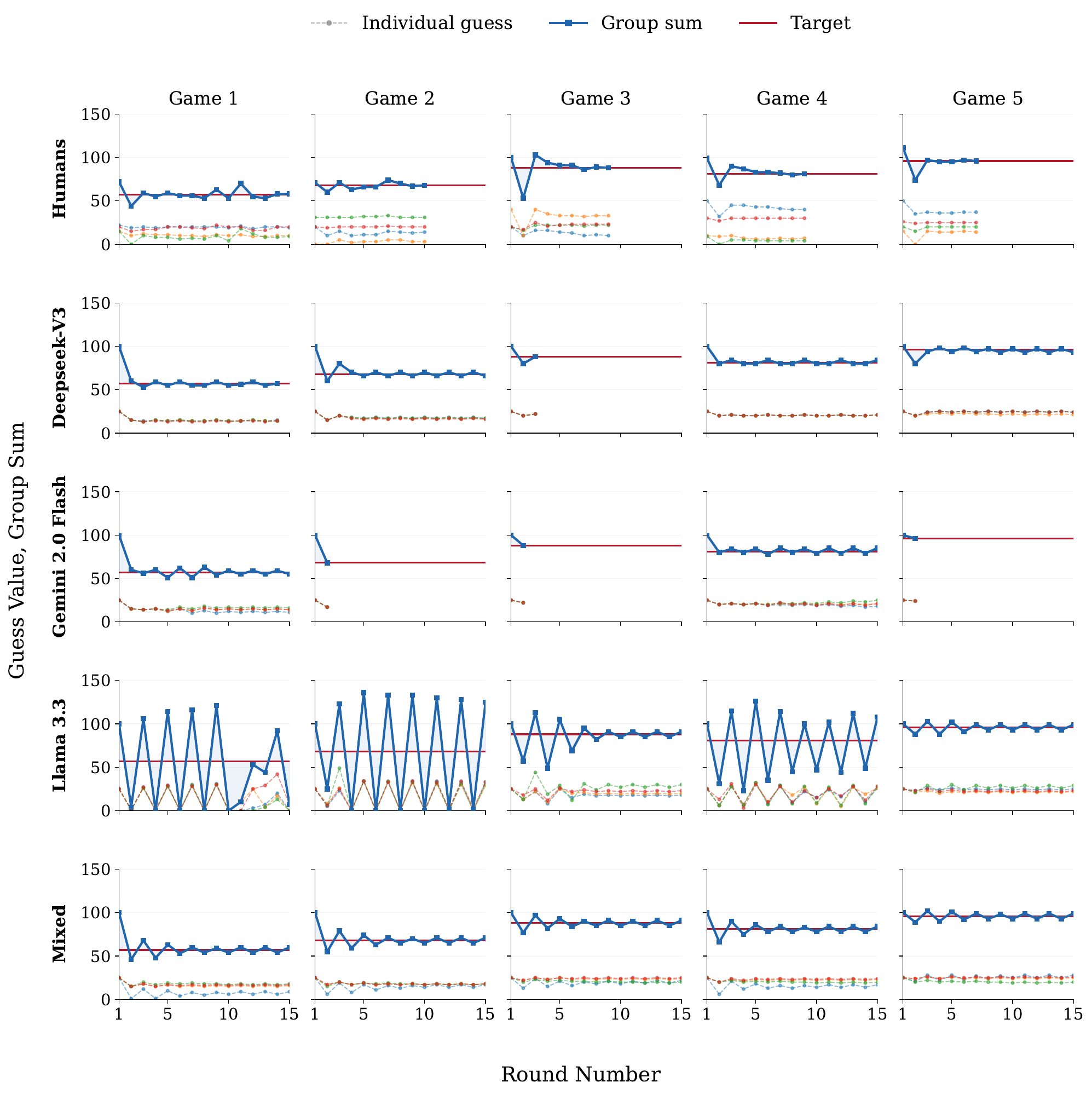}
\caption{Example of coordination in 4-player games with numerical feedback and zero-shot prompts. The solid horizontal line indicates the mystery number, the other solid line indicates the sum of the group, and the dashed line represents the decisions of each agent in the group.}
\label{fig:4-player-zero-shot-numerical-4}
\end{figure*}

\begin{figure*}[h!]
\centering
\includegraphics[width=1.0\textwidth]{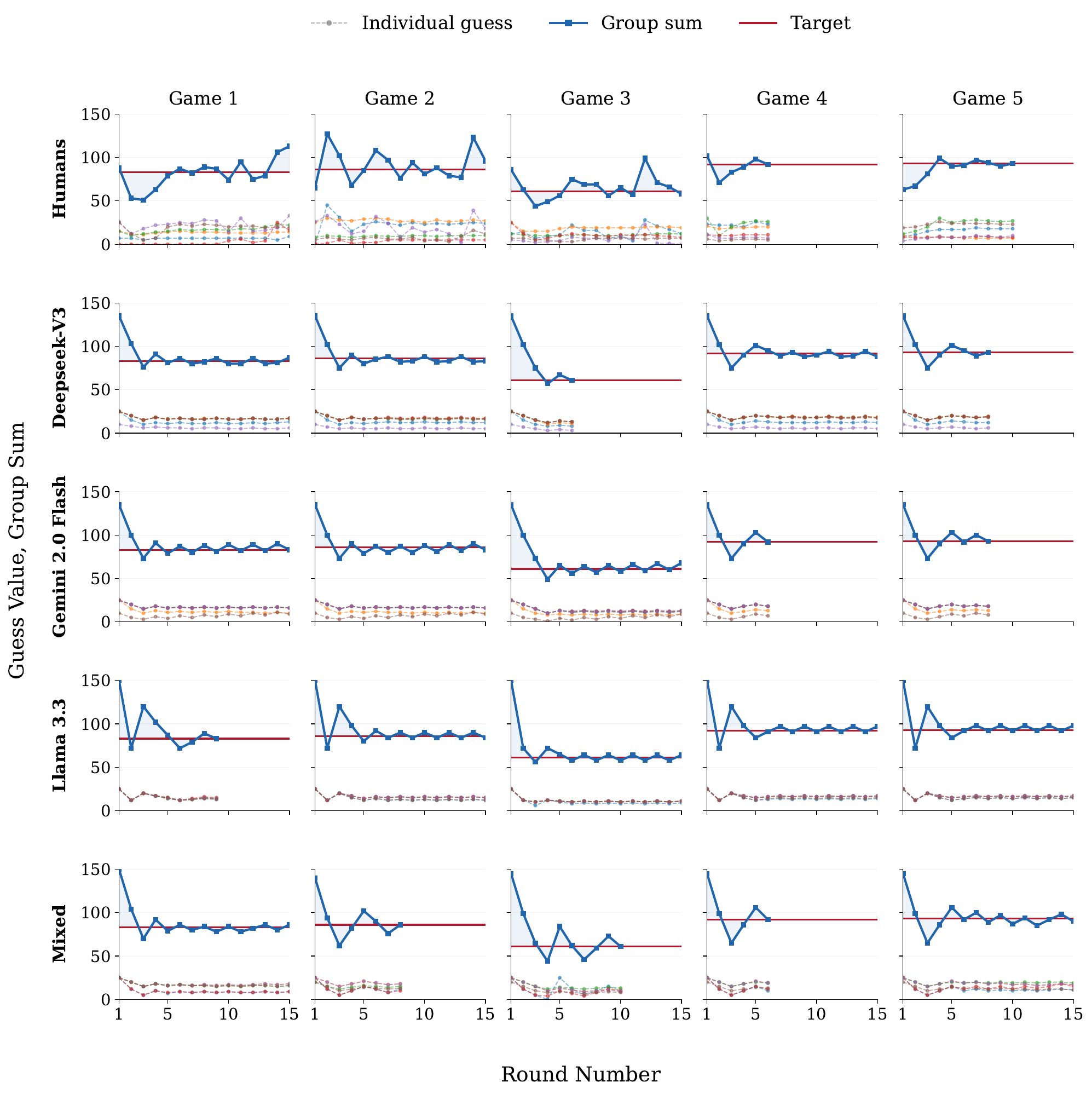}
\caption{Example of coordination in 6-player games with directional feedback and zero-shot prompts. The solid horizontal line indicates the mystery number, the other solid line indicates the sum of the group, and the dashed line represents the decisions of each agent in the group.}
\label{fig:6-player-zero-shot-directional-2}
\end{figure*}

\begin{figure*}[h!]
\centering
\includegraphics[width=1.0\textwidth]{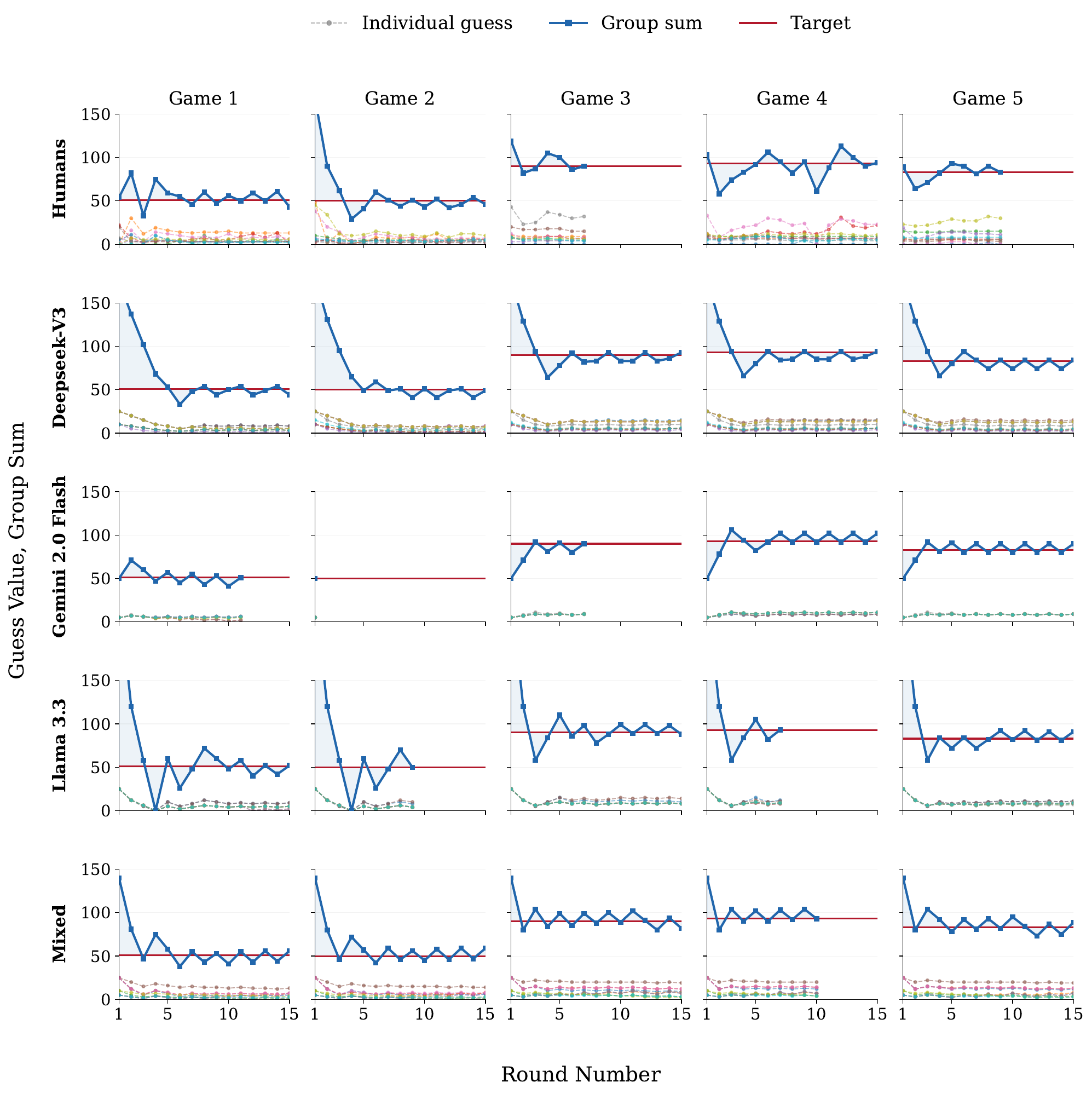}
\caption{Example of coordination in 10-player games with directional feedback and zero-shot prompts. The solid horizontal line indicates the mystery number, the other solid line indicates the sum of the group, and the dashed line represents the decisions of each agent in the group.}
\label{fig:10-player-zero-shot-directional-17}
\end{figure*}

\begin{figure*}[h!]
\centering
\includegraphics[width=1.0\textwidth]{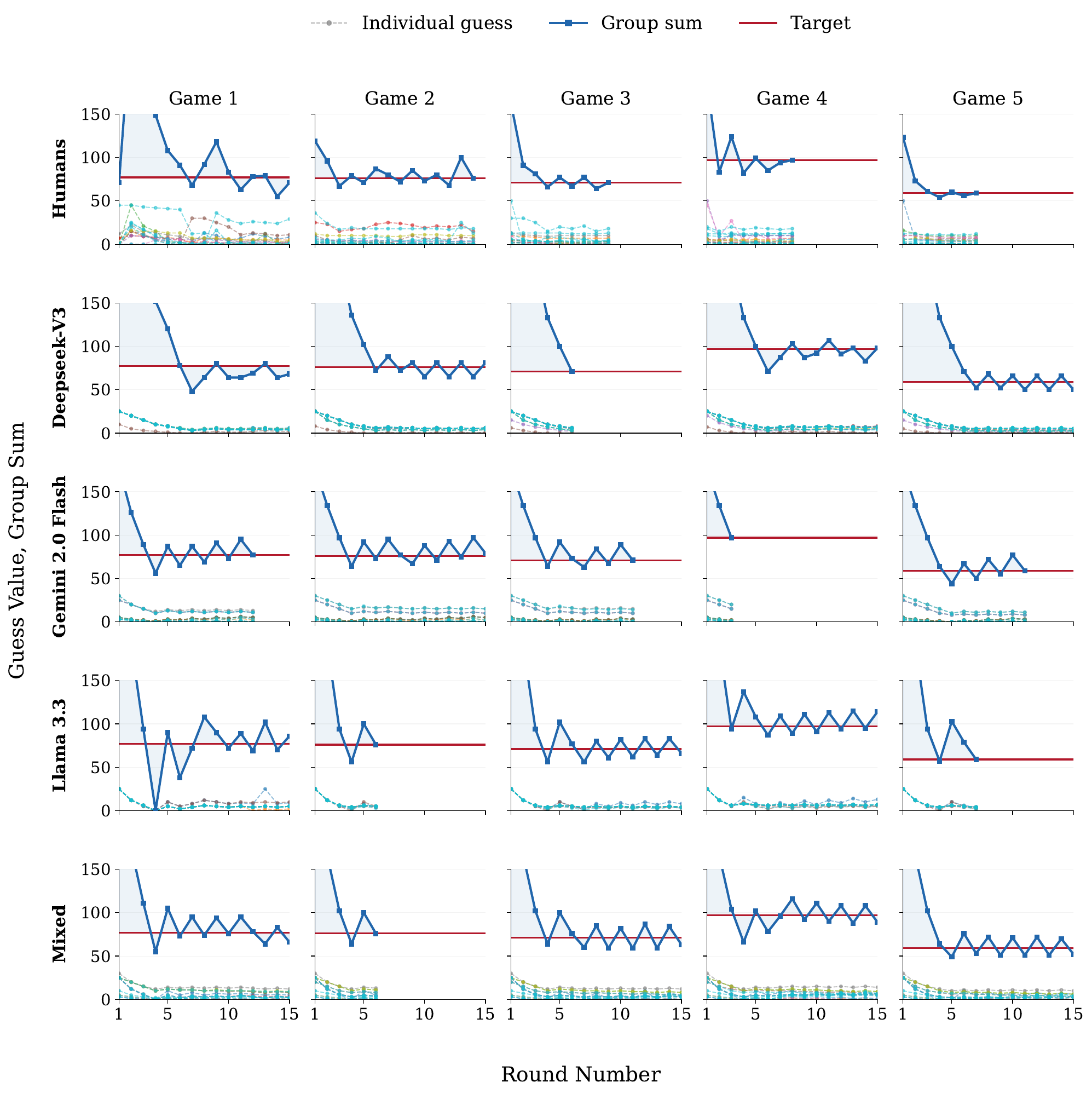}
\caption{Example of coordination in 16-player games with directional feedback and zero-shot prompts. The solid horizontal line indicates the mystery number, the other solid line indicates the sum of the group, and the dashed line represents the decisions of each agent in the group.}
\label{fig:16-player-zero-shot-directional-18}
\end{figure*}

\begin{figure*}[h!]
\centering
\includegraphics[width=1.0\textwidth]{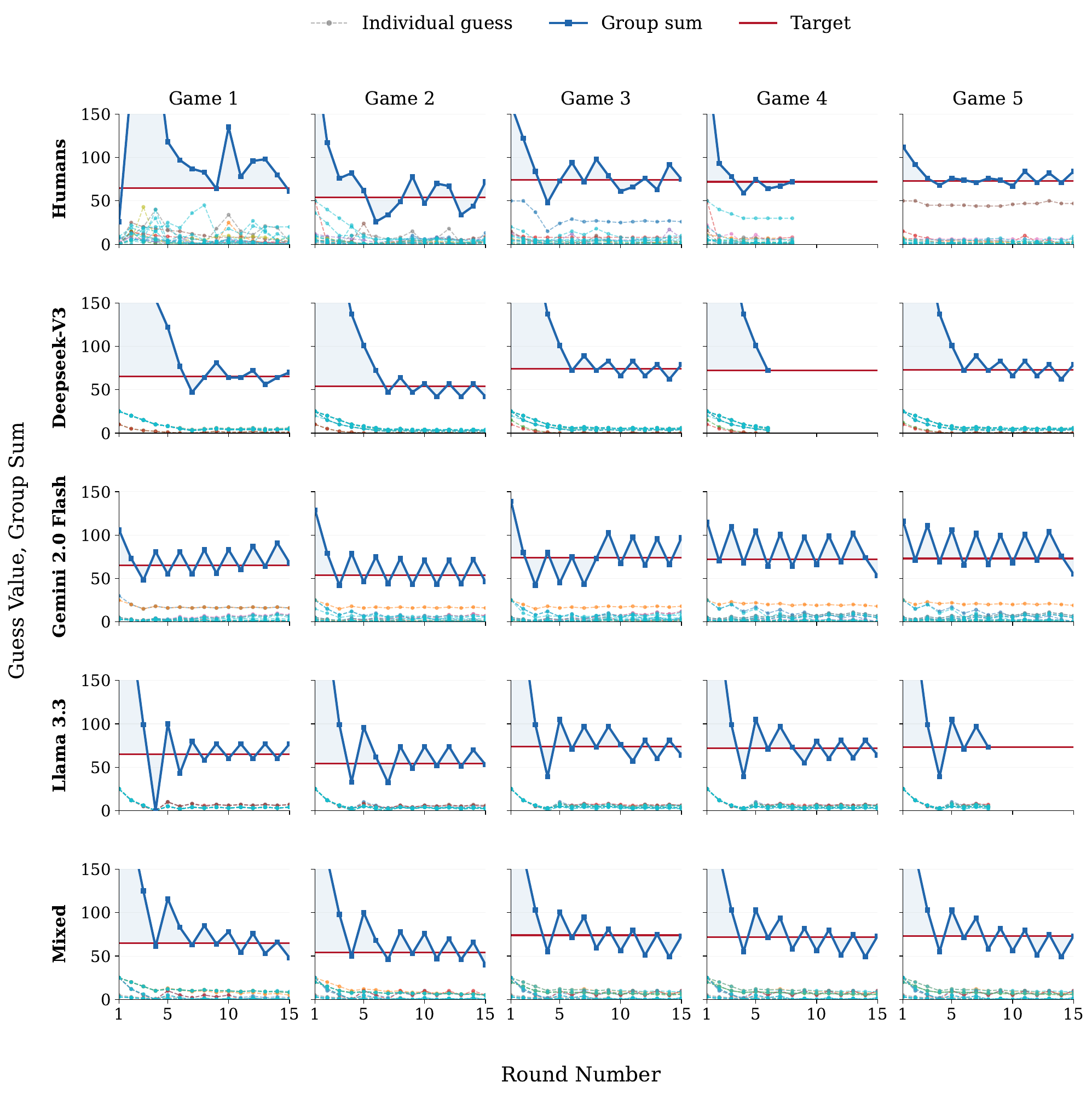}
\caption{Example of coordination in 17-player games with directional feedback and zero-shot prompts. The solid horizontal line indicates the mystery number, the other solid line indicates the sum of the group, and the dashed line represents the decisions of each agent in the group.}
\label{fig:17-player-zero-shot-directional-15}
\end{figure*}

\begin{figure*}[h!]
\centering
\includegraphics[width=1.0\textwidth]{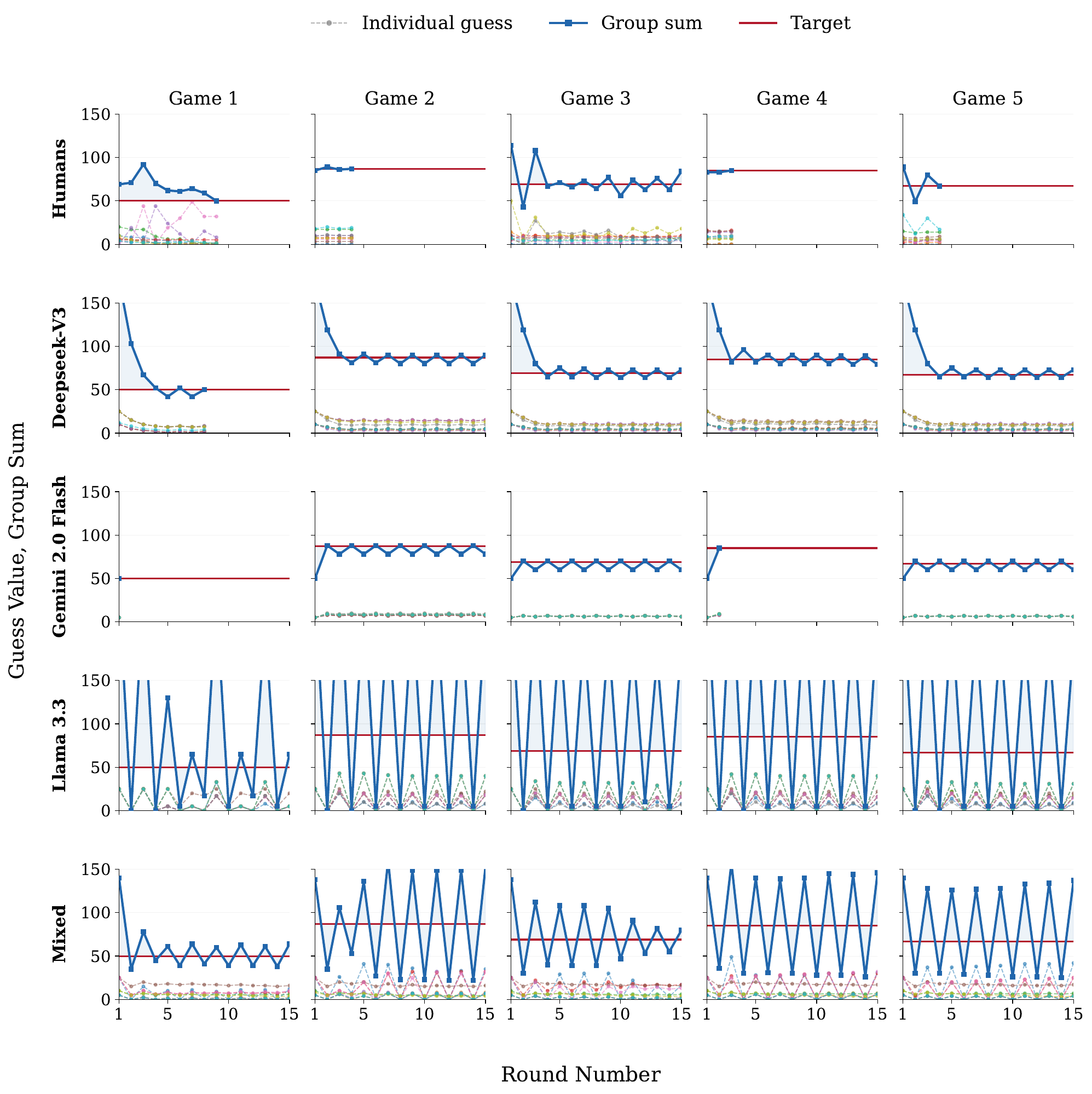}
\caption{Example of coordination in 10-player games with numerical feedback and zero-shot prompts. The solid horizontal line indicates the mystery number, the other solid line indicates the sum of the group, and the dashed line represents the decisions of each agent in the group.}
\label{fig:10-player-zero-shot-numerical-17}
\end{figure*}

\begin{figure*}[h!]
\centering
\includegraphics[width=1.0\textwidth]{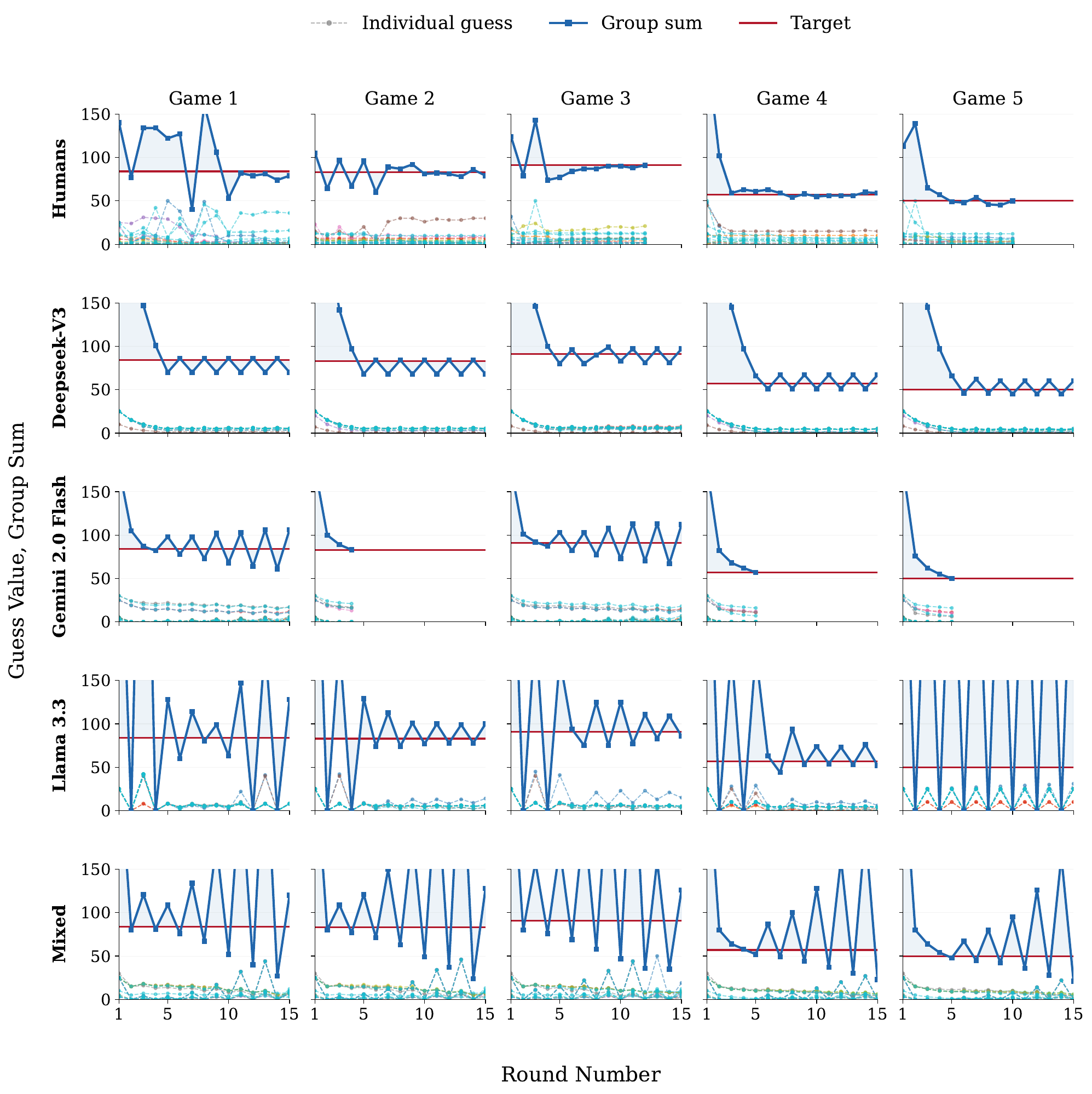}
\caption{Example of coordination in 16-player games with numerical feedback and zero-shot prompts. The solid horizontal line indicates the mystery number, the other solid line indicates the sum of the group, and the dashed line represents the decisions of each agent in the group.}
\label{fig:16-player-zero-shot-numerical-18}
\end{figure*}

\begin{figure*}[h!]
\centering
\includegraphics[width=1.0\textwidth]{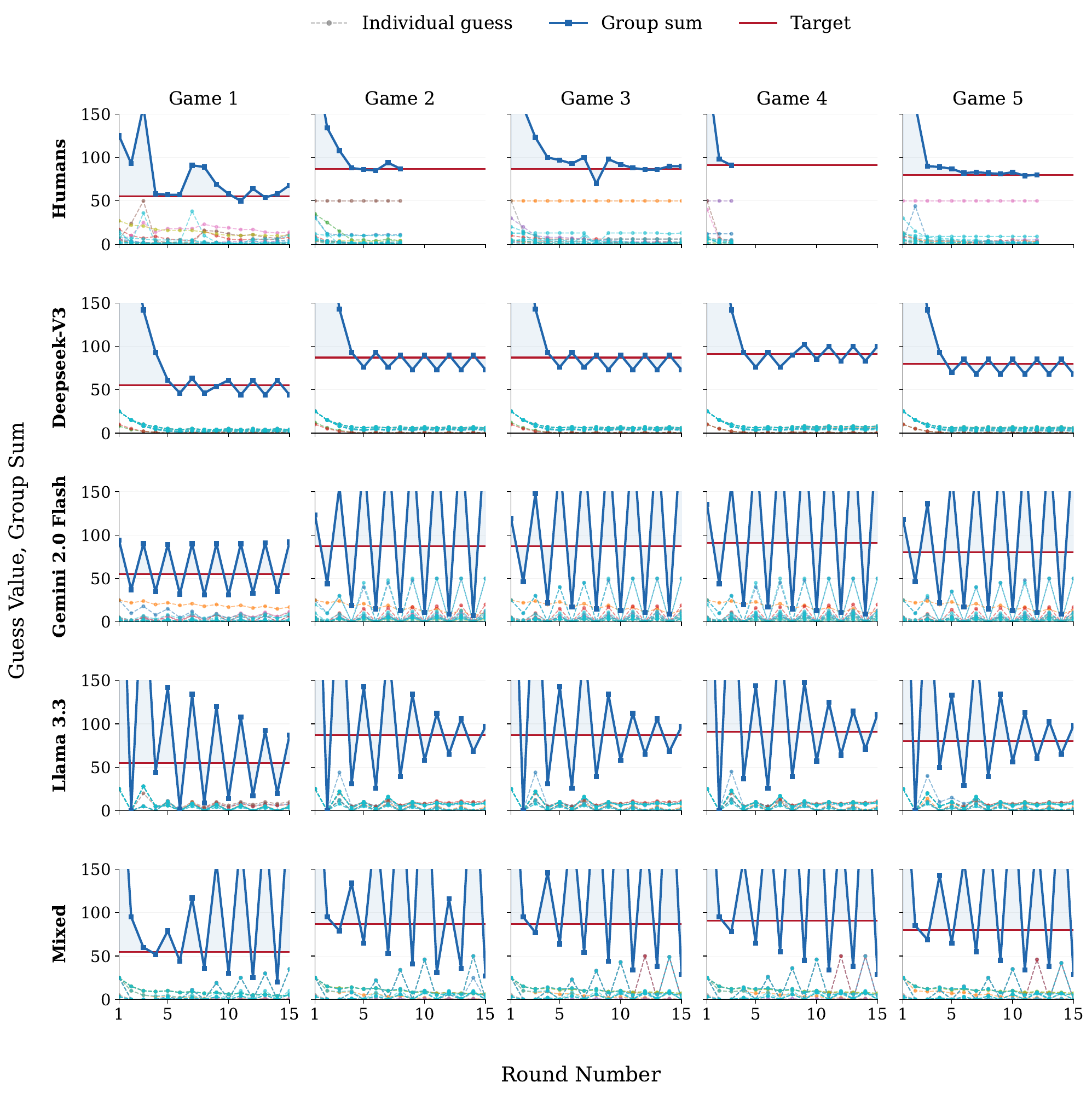}
\caption{Example of coordination in 17-player games with numerical feedback and zero-shot prompts. The solid horizontal line indicates the mystery number, the other solid line indicates the sum of the group, and the dashed line represents the decisions of each agent in the group.}
\label{fig:17-player-zero-shot-numerical-15}
\end{figure*}

\end{document}